\documentclass[
    reprint,
    superscriptaddress,
    amsmath,amssymb,
    aps,
    prx,
    floatfix,
]{revtex4-2}

\usepackage{graphicx} 
\usepackage{xcolor}
\usepackage{siunitx}
\usepackage{braket}
\usepackage{bm}
\usepackage[colorlinks=true,linkcolor=blue,citecolor=blue]{hyperref}
\usepackage[capitalise]{cleveref}

\usepackage{enumitem}
\graphicspath{{figures/}}
\newlist{outline}{enumerate}{3}
\setlist[outline]{label=\Roman*.}
\setlist[outline,1]{label=\Roman*.}
\setlist[outline,2]{label=\Alph*.}
\setlist[outline,3]{label=\arabic*.}
\newcommand{\bout}{\begin{outline}}
\newcommand{\eout}{\end{outline}}

\usepackage[
    textwidth=0.8in,
    textsize=footnotesize,
    color=red!25,
    colorinlistoftodos,
]{todonotes}

\begin{document}

\title{Fluxonium as a control qubit for bosonic quantum information}

\author{Ke Nie}
\author{J. Nofear Bradford}
\author{Supriya Mandal}
\author{Aayam Bista}
\affiliation{Department of Physics, University of Illinois at Urbana-Champaign, Urbana, IL 61801, USA}
\author{Wolfgang Pfaff}
\email{wpfaff@illinois.edu}
\affiliation{Department of Physics, University of Illinois at Urbana-Champaign, Urbana, IL 61801, USA}
\affiliation{Materials Research Laboratory, University of Illinois at Urbana-Champaign, Urbana, IL 61801, USA}
\author{Angela Kou}
\affiliation{Department of Physics, University of Illinois at Urbana-Champaign, Urbana, IL 61801, USA}
\affiliation{Materials Research Laboratory, University of Illinois at Urbana-Champaign, Urbana, IL 61801, USA}
\affiliation{Holonyak Micro and Nanotechnology Lab, University of Illinois at Urbana-Champaign, Urbana, IL 61801, USA}

\begin{abstract}
    Bosonic codes in superconducting resonators are a hardware-efficient avenue for quantum error correction and benefit from favorable error hierarchies provided by long-lived cavities compared to typical superconducting qubits. 
    The required coupling to an ancillary control qubit, however, can negate these benefits by inducing highly detrimental effects such as excess decoherence and undesired nonlinearities.
    An important question is thus whether a cavity-qubit coupling can be realized that offers readout and control capabilities without spoiling the cavity.
    Here, motivated by its long lifetime and design flexibility of its Hamiltonian, we experimentally investigate the fluxonium as a control qubit for superconducting cavities.
    We couple a fluxonium qubit to a superconducting resonator in the strong-dispersive regime and use it to measure the coherence and inherited nonlinearities of the resonator.
    We then demonstrate universal control by preparing and characterizing resonator Fock states and their superpositions, with fidelities limited by resonator decay in our planar prototype device. 
    Finally, we use the predictability of the resonator's inherited nonlinearities to show numerically that the fluxonium can reach cavity-coupling regimes that eliminate undesirable cavity nonlinearities. 
    These results demonstrate the potential of the fluxonium as a high-performance bosonic control qubit for superconducting cavities.
\end{abstract}
\maketitle

\section{Introduction}

Bosonic codes store quantum states in harmonic oscillators, which offer a hardware-efficient approach for realizing the large Hilbert space required for logical qubits~\cite{vitali_quantumstate_1998,gottesman_encoding_2001,zippilli_scheme_2003,michael_new_2016}.
In superconducting quantum circuits, bosonic encodings are a particularly appealing route toward fault-tolerant quantum computers~\cite{cai_bosonic_2021, ma_quantum_2021}.
For one, superconducting cavities can display quantum state storage times that outperform the best available superconducting qubits~\cite{reagor_quantum_2016,chakram_seamless_2021,milul_superconducting_2023}.
Second, errors in cavities can be strongly biased toward relaxation~\cite{reagorReaching10Ms2013,rosenblum_faulttolerant_2018,huang_fast_2025}, which reduces requirements on quantum error correction.
The challenge in utilizing bosonic encodings is to find strategies for quantum control of the oscillator that preserve these benefits.

Universal control of an oscillator's quantum state can be achieved by coupling it to an ancillary qubit~\cite{lawArbitraryControlQuantum1996,lloyd_quantum_1999,hofheinz_synthesizing_2009}.
In particular, a static dispersive coupling is a simple yet powerful means to achieve arbitrary state manipulation and tomography~\cite{krastanov_universal_2015,lutterbach_method_1997,bertet_direct_2002}.
In superconducting circuits, this coupling scheme has been investigated extensively with the transmon qubit and has enabled key bosonic code functionalities such as quantum state preparation and manipulation~\cite{vlastakisDeterministicallyEncodingQuantum2013, heeres_implementing_2017,chakram_multimode_2022}, readout and tomography~\cite{kirchmairObservationQuantumState2013,vlastakisDeterministicallyEncodingQuantum2013}, and quantum error correction~\cite{Ofek2016,Gertler2021,Sivak2023,ni_beating_2023}.

The introduction of the ancillary qubit, however, can spoil the oscillator: hybridization gives rise to new error channels, such as Purcell decay or qubit-induced dephasing~\cite{reagor_quantum_2016}, and spurious nonlinearites induced can dramatically reduce circuit performance~\cite{lescanne_exponential_2020,berdou_one_2023}.
Uncontrolled nonlinearities are unavoidable when using the transmon, which has led to alternative control strategies such as minimized or tuneable qubit-cavity coupling~\cite{eickbusch_fast_2022, milul_superconducting_2023, huang_fast_2025, amazon_bosonic} or parametric control schemes~\cite{vrajitoareaQuantumControlOscillator2020, lescanne_exponential_2020, grimm_stabilization_2020, ding_quantum_2024, réglade_quantum_2024, amazon_bosonic}; the downside of these approaches is that they can lead to slowed control speed, or added complexity in circuit design or operation.
It thus remains a compelling question whether an alternative ancillary qubit could enable a large dispersive coupling for fast and universal bosonic control, while eliminating inherited detrimental effects on the cavity.

Here, we consider the fluxonium~\cite{manucharyan_fluxonium_2009} as an ancillary qubit for superconducting cavities due to several possible benefits over the transmon.
First, it has been shown to have millisecond lifetimes~\cite{somoroff_millisecond_2023a, leon_high_2023, wang_high_2025} and could thus minimize qubit-induced cavity decoherence.
Second, its flux tunability enables \emph{in situ} tuning of qubit-cavity coupling~\cite{atanasova_situ_2025}.
Finally, the rich energy level structure of the fluxonium and the resulting interactions with the cavity it is coupled to can be modified through multiple different circuit parameters \cite{zhu_circuit_2013,smith_quantization_2016}.
This latter property suggests the possibility of tailoring the effective qubit-cavity Hamiltonian such that undesirable nonlinearities are minimized or eliminated.

In this work, we experimentally employ the fluxonium as an ancillary qubit for a superconducting resonator.
Operating in the strong-dispersive coupling regime, we use the qubit to characterize the resonator and measure the nonlinearities that the resonator inherits from the coupling, in quantitative agreement with simulations.
We demonstrate that the quantum control techniques pioneered with transmon qubits can be similarly used with the fluxonium to manipulate and reconstruct quantum states of the resonator.
With our simulations benchmarked against data, we use them to find parameters for a fluxonium that can be coupled to a cavity in the strong-dispersive regime while practically eliminating the self-Kerr nonlinearity of the cavity.
Our results establish fluxonium as a promising avenue for the control of bosonic quantum information.

\begin{figure}
    \centering
    \includegraphics{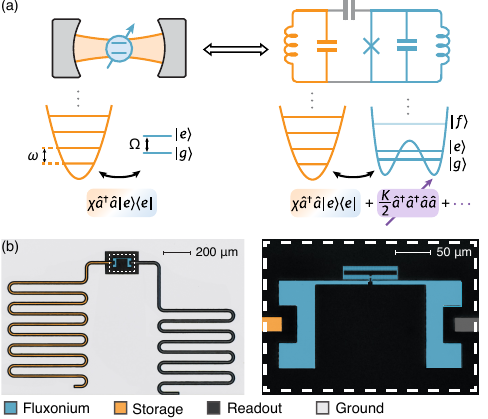}
    \caption{Fluxonium-resonator system.
    (a) Schematic comparison of an idealized cavity QED system versus a circuit implementation.
    Our target is a dispersive shift $\chi$ as the only interaction (left). 
    In practice, a resonator coupled to an artificial atom implemented by a superconducting circuit also inherits (at least) a self-Kerr nonlinearity $K$ arising from the qubit’s anharmonicity (right). 
    With sufficient control over circuit parameters, $K$ can be tuned and potentially suppressed.
    (b) False-colored optical images of the fabricated device. 
    A fluxonium qubit is capacitively coupled to storage and readout modes, implemented as coplanar waveguide (CPW) resonators.
    }
    \label{fig1:fluxonium-resonator}
\end{figure}

\section{Fluxonium-resonator system}

\begin{figure}
    \centering
    \includegraphics{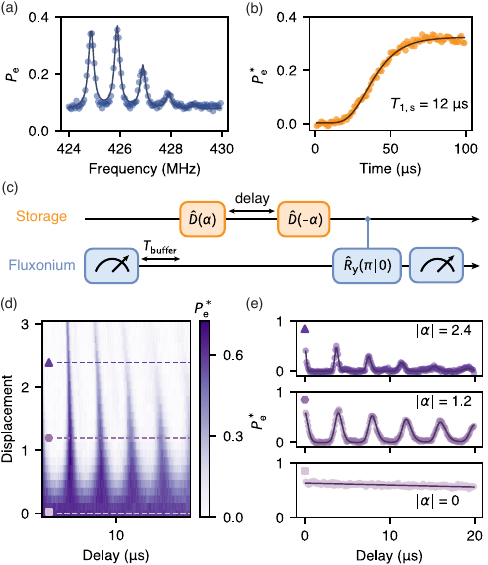}
    \caption{
    Characterization of the resonator and its coupling to the qubit.
    (a) Photon-number splitting in the fluxonium spectrum with the storage resonator prepared in $\ket{\alpha = 1}$.
    The nonzero background in the spectroscopy arises from imperfect qubit initialization (Appendix~\ref{supplement_qubit_initialization}).
    (b) Time-dependent vacuum state probability after an initial displacement.
    The energy relaxation rate $\kappa \equiv 1/T_{1,s}$ is determined from a fit to a function $\propto \exp(-|\alpha|^2 \exp(-\kappa t))$~\cite{reagor_quantum_2016}.
    Data are background corrected (indicated by an asterisk) for
    skewness arising from qubit decay and resonator crosstalk~\cite{vlastakisDeterministicallyEncodingQuantum2013}.
    (c) Pulse sequence for Ramsey interferometry experiment on the storage resonator.
    Initial fluxonium measurement serves to cool the qubit by postselection.  
    (d) Resonator Ramsey experiment as a function of displacement amplitude. 
    The Kerr nonlinearity of the storage resonator induces a amplitude-dependent shift in the detuning.
    (e) Data at different displacement amplitudes and fit to a master equation simulation.
    }  
    \label{fig2:half_flux}
\end{figure}

We consider a minimal circuit composed of a storage resonator and the fluxonium to test it as an ancillary control qubit for a bosonic mode.
An additional readout resonator serves to measure the qubit state (for details see~\cref{Spectroscopic Characterization}).
The fluxonium-storage Hamiltonian is given by
\begin{multline}\label{fluxonium_resonator_Hamiltonian}
    \hat{H}/\hbar =  4E_C\hat{n}^2 + \frac{1}{2}E_L\hat{\phi}^2 - E_J\cos{(\hat{\phi}-\phi_\mathrm{ext})} \\
     + \tilde{\omega}\hat{\tilde{a}}^\dagger\hat{\tilde{a}} - ig\hat{n}(\hat{\tilde{a}}- \hat{\tilde{a}}^\dagger).
\end{multline}
Here, $E_{C}$, $E_{L}$, $E_{J}$ represent the charging, inductive, and Josephson energies of the fluxonium.
The Cooper pair number and phase operator are denoted by $\hat{n}$ and $\hat{\phi}$. 
The external flux is represented by $\Phi_\mathrm{ext}$, and we introduce $\phi_\mathrm{ext}/2\pi = \Phi_\mathrm{ext}/\Phi_0$. 
The bare storage mode has a resonant frequency $\tilde{\omega}$ and annihilation operator $\hat{\tilde{a}}$, and its coupling strength to the fluxonium is given by $g$.

Starting from the circuit design (\cref{fluxonium_resonator_Hamiltonian}) we aim to relate circuit parameters to a description of the interaction between fluxonium and storage resonator in the dispersive regime of cavity quantum electrodynamics (cavity QED). 
Ideally, a dispersive shift of magnitude $\chi$ is the sole interaction (Fig.~\ref{fig1:fluxonium-resonator}(a)).
In practice, however, the resonator also inherits a nonlinearity from the interaction with the qubit.
The lowest-order nonlinearity is the self-Kerr, $K$, corresponding to a photon-number dependent shift that results in typically undesired state evolution that limits operation fidelities.
To this order, the effective Hamiltonian is
\begin{equation}\label{effective Hamiltonian}
    \hat{H}_\text{eff}/\hbar = \Omega\ket{e}\bra{e} + \omega\hat{a}^\dagger\hat{a} + \chi\hat{a}^\dagger\hat{a}\ket{e}\bra{e} + \frac{K}{2}\hat{a}^\dagger\hat{a}^\dagger\hat{a}\hat{a},
\end{equation}
where we model the fluxonium as a two-level system of transition frequency $\Omega$. 
The dressed frequency of the storage mode is $\omega$, while $\hat{a}$ is the dressed annihilation operator of the storage mode. 

To experimentally make the connection between circuit description and cavity QED model, we have designed a device as shown in (\cref{fig1:fluxonium-resonator}(b)).
We chose to implement it using a two-dimensional (2D) on-chip architecture because it offers precise control over fluxonium and resonator circuit parameters.
For our proof-of-principle experiment, the benefit of Hamiltonian predictability outweighs the tradeoff in resonator quality factor compared to a high-$Q$, three-dimensional (3D) implementation.
This circuit provides a simple yet effective platform for investigating the fluxonium-resonator interaction, as well as using the fluxonium to characterize the resonator and the nonlinearity imparted on it.

\section{Half-flux Characterization}
    
The most favorable operating point for the fluxonium as an ancillary qubit is at half-flux where its lifetime and coherence are maximized. 
Here, we found relaxation and (Hahn echo) coherence times of $T_1 = \qty{123}{\micro\second}$ and $T_{2}^E = \qty{90}{\micro\second}$ for the qubit (see \Cref{app:experimentsetup} for more information).
Our first objective is to characterize the qubit-resonator interaction at this operating point, and to demonstrate elementary control and readout of the resonator using the qubit (\cref{fig2:half_flux}).
To verify the coupling regime we measured the qubit spectrum after preparing a coherent state with amplitude $\alpha$ in the storage mode through a displacement pulse (\cref{fig2:half_flux}(a)).
The observation of clear number-split qubit resonances confirms the strong-dispersive coupling, with peak separation $\chi/2\pi = \qty{1.0}{\mega\hertz}$~\cite{schuster_resolving_2007}.
The strong dispersive regime allows Fock-state selective qubit rotations which is the basis for bosonic state control and readout.
Using this principle, we measured the resonator relaxation rate by monitoring the probability that the resonator is in the vacuum state $\ket{0}$ after a displacement. 
We extracted a single-photon lifetime of $\qty{12}{\micro\second}$, limited by the internal quality factor of the fabricated device (\cref{fig2:half_flux}(b)).

A central concern of our work is the extraction of the inherited self-Kerr nonlinearity of the storage mode.
Employing the toolkit provided by the dispersive coupling, we measured the magnitude of the Kerr effect with a `cavity Ramsey' interferometry experiment on the resonator using the sequence of resonator displacements and number-selective measurements shown in \Cref{fig2:half_flux}(c)~\cite{vlastakisDeterministicallyEncodingQuantum2013}.
The fringes observed confirm that the resonator mode remains phase coherent despite the short lifetime (Fig.~\ref{fig2:half_flux}(d)). 
The shift of the fringes at larger displacement amplitudes indicates a photon-number dependent detuning, i.e., the Kerr effect.
This detuning scales linearly with photon number, $|\alpha|^2$, from which we extract a self-Kerr coefficient of $K/2\pi = \SI{3.6}{\kilo\hertz}$.
To account for system losses, we fitted the Ramsey fringes at various displacements using a Lindblad master equation incorporating relevant loss mechanisms, yielding quantitative agreement with the experimental data (Fig.~\ref{fig2:half_flux}(e); see \cref{supplement_theory} and \cref{supplement_Kerr} for details).

\section{Flux dependence of $\chi$ and $K$}\label{sec_flux}

\begin{figure}
    \centering
    \includegraphics{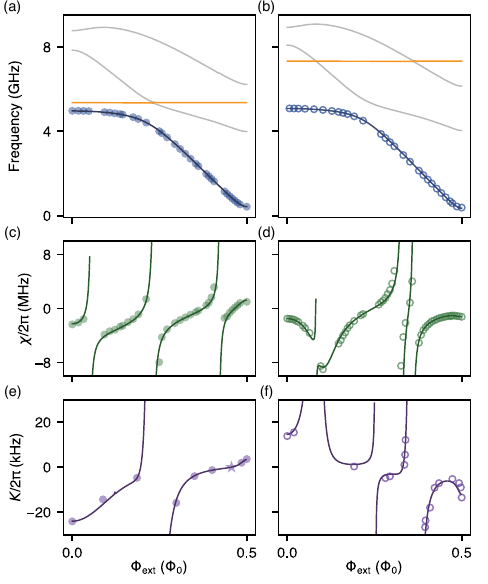}
    \caption{
    Hamiltonian parameters as function of external flux.
    (a),(b) Fluxonium spectra of device A (left; solid circles) and device B (right; open circles), fit to the $\ket{g}\rightarrow\ket{e}$ transition. 
    The higher transitions $\ket{g}\rightarrow\ket{f}$ and $\ket{g}\rightarrow\ket{h}$ are shown in gray.
    The storage resonator level (orange) crosses different higher fluxonium levels in each device. This results in distinct flux dependence of the dispersive shift $\chi$ shown in (c),(d), and the self-Kerr nonlinearity $K$ shown in (e),(f). 
    $K$ can change sign and cross zero at specific flux bias points (star; see Appendix \ref{supplement_cavity_ramsey} for raw data).
    Solid lines are expected parameters based on the fit to the fluxonium spectrum.
    All simulations are performed numerically and show good agreement with the measurements. 
    Error bars are smaller than the marker size (Appendix~\ref{error_analysis}).
    }
    \label{fig3:flux_characterization}
\end{figure}
    
\begin{figure*}[!htb]
    \centering
    \includegraphics{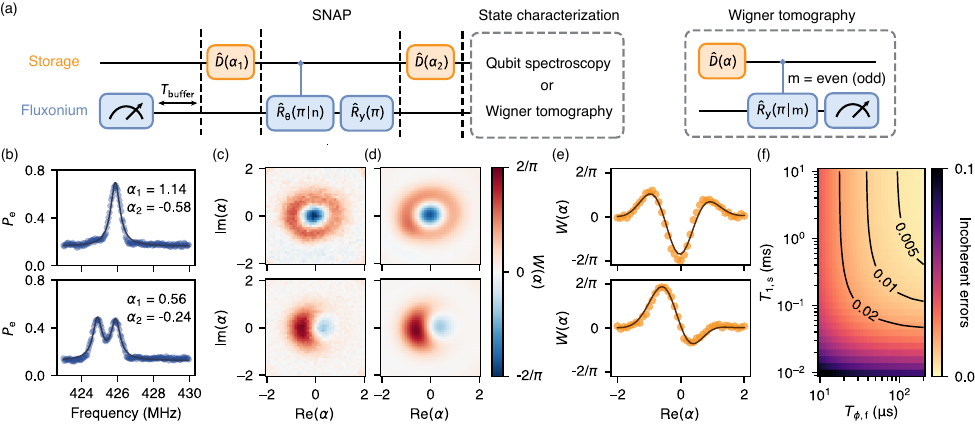}
    \caption{
    Bosonic control using the fluxonium. 
    (a) Pulse sequence for the preparation and characterization of Fock states in the storage resonator. 
    A SNAP gate is used to prepare specific Fock states, which are characterized using either qubit spectroscopy or Wigner tomography. 
    (b) Fluxonium spectroscopy with the storage resonator prepared in $\ket{1}$ (top) and $\frac{1}{\sqrt{2}}(\ket{0} - \ket{1})$ (bottom). 
    For the SNAP gate, we choose $\vec{\theta} = (\pi, 0, 0, \dots)$, with displacements indicated in the figure.
    (c) Measured Wigner functions of the prepared Fock states. 
    (d) Simulated Wigner functions using the master equation. Simulation includes both state preparation and Wigner measurement processes. 
    (e) Cuts of the measured and simulated Wigner functions along $\mathrm{Im}(\alpha)=0$. 
    (f) Simulated incoherent errors during the SNAP gate for $\vec{\theta}=(\pi, 0, 0, ...)$ and $\alpha_1=1$ as a function of storage resonator lifetime $T_{1,s}$ and fluxonium (pure) dephasing time $T_{\phi,f}$.
    }
    \label{fig4:SNAP}
\end{figure*}

Because the resonator's $\chi$ and $K$ depend heavily on the fluxonium level structure, an externally applied magnetic flux provides a control knob for \textit{in situ} adjustment of $\chi$ and $K$. 
For one, this control is useful for locating decoupled idling points \cite{atanasova_situ_2025}. 
In addition, mapping $\chi$ and $K$ across flux allows us to fully connect the circuit Hamiltonian (\cref{fluxonium_resonator_Hamiltonian}) to a flux-dependent cavity QED description and reinforce the predictability of interactions and nonlinearity.

We measured the fluxonium spectrum as a function of flux, and fitted the $\ket{g}\rightarrow\ket{e}$ transition (Fig.~\ref{fig3:flux_characterization}(a)) as well as the resonator spectrum (Appendix~\ref{Spectroscopic Characterization}) to \Cref{fluxonium_resonator_Hamiltonian}.
Using the extracted circuit parameters, we then numerically computed the expected flux dependence of $\chi$ and $K$.
The comparison with the measured $\chi$ and $K$ is shown in ~\Cref{fig3:flux_characterization}(c) and (e). 
The data show a markedly wide variation of $\chi$ and $K$ across flux, in excellent agreement with our numerical predictions. 
Notably, similar values of $\chi$ can correspond to drastically different values of $K$, highlighting the possibility of engineering distinct $K/\chi$ ratios for bosonic control.
Moreover, $K$ can change sign and cross zero at specific flux bias points, which could help improve control fidelity by suppressing Kerr-induced state evolution.
This tunability offered by the fluxonium thus allows dynamic reconfiguration of system properties for specific protocols.

To understand the dependence of $\chi$ and $K$ on resonator parameters, we repeated the same characterization with a second device (`device B').
It shares the same fluxonium design parameters as the initially presented device (`device A'), but features a different storage resonator frequency (Fig.~\ref{fig3:flux_characterization}(b); see Appendix~\ref{Spectroscopic Characterization} for system parameters).
This frequency difference causes the storage mode to intersect different higher-energy levels of the fluxonium, which leads to
device B's distinct flux dependence of $\chi$ and $K$ (Fig.~\ref{fig3:flux_characterization}(d) and (f)) from that observed in device A.
The pronounced difference in flux-dependent behavior between the two devices further illustrates the versatility of Hamiltonian engineering achievable with the fluxonium.

\section{Bosonic control}

A detailed understanding of the static system Hamiltonian allows us to predict coherent dynamics, and to design and implement bosonic control schemes. 
Here we demonstrate this capability as a proof of concept by preparing and characterizing storage resonator Fock states and their superpositions. 
We employed resonator displacement and the Selective Number-dependent Arbitrary Phase (SNAP) gate \cite{heeres_cavity_2015, krastanov_universal_2015, landgraf_fast_2024} to realize universal control over the storage mode (Fig.~\ref{fig4:SNAP}(a)). 
While other control schemes have been proposed~\cite{zheng_crossresonance_2025}, we adopted the SNAP protocol since it only relies on strong dispersive coupling to impart arbitrary phases, and allows for straightforward numerical optimization to suppress coherent errors.
We implemented the SNAP gate using optimized selective $\pi$ pulses followed by a fast $\pi$ pulse \cite{landgraf_fast_2024}. 
As a demonstration, we prepared both the single-photon Fock state $\ket{1}$ and the superposition state $(\ket{0} - \ket{1})/\sqrt{2}$. 

The prepared states were characterized using qubit spectroscopy and Wigner tomography. 
In Fig.~\ref{fig4:SNAP}(b), we show fluxonium qubit spectroscopy after state preparation. 
The measured spectra confirm that we prepared the intended Fock state components. 
We further leveraged the strong dispersive coupling to perform Wigner tomography, an essential ingredient for bosonic quantum information processing.
We note that the full experimental sequence lasted approximately $\SI{3.5}{\micro\second}$, which is a significant fraction ($>25\%$) of the storage mode relaxation time; the fidelity of both state preparation and tomography are thus significantly impacted by incoherent errors.
Nevertheless, we were able to prepare and characterize the Fock states, as evidenced by the experimentally measured Wigner functions  shown in \Cref{fig4:SNAP}(c). 
To account for imperfections, we simulated the Wigner function using the master equation, including both the SNAP gate and the Wigner measurement process (Fig.~\ref{fig4:SNAP}(d)).
Our simulations are in good quantitative agreement with the experimental data; a cut along $\mathrm{Im}(\alpha) = 0$ is shown in \Cref{fig4:SNAP}(e).
This agreement allows us to estimate a preparation fidelity of 79\% for $\ket{1}$ and 91\% for $(\ket{0} - \ket{1})/\sqrt{2}$ based on simulations.
The dominant sources of infidelity are qubit initialization inefficiency and $T_1$ decay of the storage mode (see \cref{supplement_SNAP} for a detailed breakdown of loss mechanisms).

While we have established the core capabilities needed for bosonic quantum information processing in our 2D fluxonium–resonator prototype, fidelities are constrained by the low cavity lifetime of our 2D resonator.
\Cref{fig4:SNAP}(f) shows the potential improvement in gate fidelity enabled by increased system lifetimes and coherence, such as those provided by a 3D architecture.
We simulated the SNAP gate infidelity due to incoherent errors as a function of the storage mode lifetime and qubit dephasing time, using our current system parameters and pulse durations (Appendix~\ref{supplement_SNAP}).
With a cavity lifetime of $\SI{1}{\milli\second}$ and a fluxonium dephasing time of $\SI{50}{\micro\second}$, the incoherent errors can already be suppressed to 1\%.
These results highlight the viability of universal, high-fidelity bosonic control in a fluxonium-resonator system using experimentally accessible parameters.

\section{Outlook}\label{sec:prospects}

We have experimentally demonstrated that the fluxonium is well-suited to act as an ancillary qubit for bosonic quantum information stored in superconducting microwave cavities.
We have shown strong dispersive qubit-resonator coupling with interactions and nonlinearities tailored by circuit design and tuned \emph{in situ}.
We have employed this coupling to demonstrate the manipulation and readout of the quantum state of the resonator, with fidelities limited mainly by the resonator lifetime of our planar prototype device.

\begin{figure}[!t]
    \centering
    \includegraphics{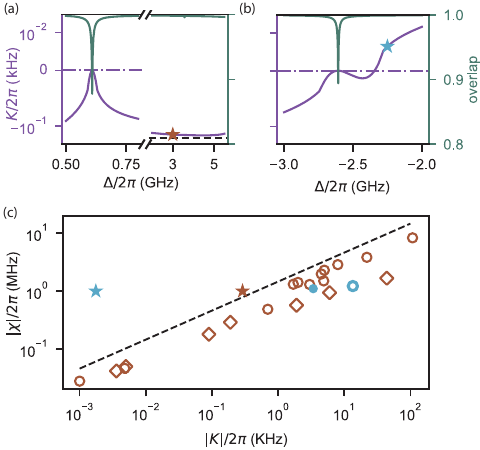}
    \caption{Relation of $\chi$ and $K$ for the fluxonium and transmon. 
    (a) $K$ (purple line) and overlap between bare and dressed states (green line) as a function of $\Delta$ for a (hypothetical, numerically optimized) transmon-resonator system with parameters $E_C/2\pi = \SI{530}{\mega\hertz}$ and $E_J/2\pi = \SI{26.5}{\giga\hertz}$ for fixed $|\chi|/2\pi = \SI{1}{\mega\hertz}$.
    Although $K$ reaches zero near $E_C \approx \Delta$, the qubit and resonator states become strongly hybridized at this point. The brown star shows example transmon parameters for the bound (black dashed line) on $\chi/K$ discussed in the main text.
    (b) The same quantities as in (a), shown for a fluxonium-resonator system with $E_C/2\pi = \SI{1.19}{\giga\hertz}$, $E_L/2\pi = \SI{556}{\mega\hertz}$, $E_J/2\pi = \SI{3.04}{\giga\hertz}$.  
    Blue star indicates a set of fluxonium parameters that beat the transmon bound, chosen with a detuning of \qty{100}{\mega\hertz} from the $K=0$ case, to illustrate that no unrealistic fine-tuning is required for small Kerr nonlinearity.
    (c) Our derived $\chi/K$ bound (dashed black line) plotted with values from the literature \cite{Pfaff2017,milul_superconducting_2023,heeres_cavity_2015,Axline2018,landgraf_fast_2024,Ofek2016,CampagneIbarcq2020,Sivak2023,Valadares2024,Cai2024,Yang2025,Gertler2021} for transmons (brown circles are measured data; brown diamonds have measured $\chi$ and simulated $K$), and devices A (blue circle) and B (open blue circle) of this work.}
    \label{fig_5}
\end{figure}

While our experiment has realized an interaction regime that is similar to typical transmon implementations, our modeling indicates that the fluxonium qubit can reach regimes that are not possible using the typically-employed transmon. 
For many bosonic control schemes we desire a large dispersive shift $\chi$ for fast operations but a small nonlinearity $K$.
As shown in \cref{sec_flux}, a fluxonium-resonator system allows for independent design of $\chi$ and $K$.  
This fact allows us to predict fluxonium parameters that can achieve a vanishing $K$ while still maintaining a large $\chi$; this parameter regime is extremely difficult to reach for the transmon.

The relationship between the dispersive shift and the Kerr nonlinearity for a transmon-resonator system is heavily constrained.
In the transmon-resonator system, $\chi$ and $K$ are both approximately proportional to the product of $E_C $ and the ratio of coupling strength and detuning, $g/\Delta$, to some power (see Appendix \ref{sup_transmon_limit} for details). 
The only way to achieve a high $\chi/K$ ratio given these constraints is to enter the regime $E_C \sim \Delta$ where the cavity and qubit become strongly hybridized, as shown in Fig.~\ref{fig_5}(a). This situation is typically not desirable if the aim is to protect cavity coherence.
Avoiding this strong hybridization while staying in the transmon limit allows us to estimate a lower bound on the Kerr nonlinearity, $|K| \gtrsim \chi^2/4E_C$. 
With reasonable bounds on transmon parameters we evaluate the bound numerically as $|K|/2\pi \gtrsim (\chi/2\pi)^2/(\qty{2.12}{GHz})$ (see \cref{sup_transmon_limit} for details). 
We plot this estimated bound on the $\chi/K$ ratio for the transmon as the dashed line in Fig.~\ref{fig_5}(c). A representative selection \cite{Pfaff2017,milul_superconducting_2023,heeres_cavity_2015,Axline2018,landgraf_fast_2024,Ofek2016,CampagneIbarcq2020,Sivak2023,Valadares2024,Cai2024,Yang2025,Gertler2021} of transmon $\chi$ and $K$ value pairs from the literature (see Table \ref{tab:transmon_literature}) are plotted in Fig. \ref{fig_5}(c), in agreement with this limit.

In contrast, the fluxonium can realize a wider range of $\chi$ and $K$, and we can find parameters that result in a vanishing $K$ with a large $\chi$ while maintaining low hybridization.
To illustrate this, we found an example circuit through numerical optimization that achieves this goal.
The simulated $K$ for the example fluxonium is plotted in \Cref{fig_5}(b) as a function of $\Delta$.  
Notably, when $K$ passes through zero, overlap between dressed and bare states (also plotted in \ref{fig_5}(b)) stays close to 1. 
To further illustrate that fluxonium can achieve a broader range of $\chi$ and $K$ values compared with the transmon, we contrast our example circuit (marked with a blue star) with the transmon case in~\Cref{fig_5}(c).
We see that our predicted large $\chi/K$ ratio fluxonium clearly falls outside of the transmon bound.

We conclude that the fluxonium is a beneficial pathway for bosonic quantum computation. 
In combination with long coherence times, its large anharmonicity and potential for negligible self-nonlinearity induced to the cavity make it a compelling choice for a high-performance control qubit with minimal detrimental side effects.
More broadly, our work demonstrates that the freedom of design available in superconducting artificial atoms provides unique opportunities for tailoring qubit-photon interactions that are not accessible in traditional cavity QED.

\section*{Acknowledgements}

We thank R.~Goncalves and K.~Chow for assistance with device fabrication, and S.~Chakram and S.~Shankar for valuable comments on the manuscript.
This research was carried out in part in the Materials Research Lab Central Facilities and the Holonyak Micro and Nanotechnology Lab, University of Illinois.
This work made use of a traveling-wave parametric amplifier provided by IBM.
We acknowledge support by the Air Force Office of Scientific Research under award number FA9550-21-1-032, and by the Army Research Office under award W911NF-23-1-0096. 
The views and conclusions contained in this document are those of the authors and should not be interpreted as representing the official policies, either expressed or implied, of the Army Research Office or the U.S. Government. 

\appendix

\section{Experimental setup}
\label{app:experimentsetup}

\subsection{Device fabrication}

The devices were fabricated with a Ta baselayer on a sapphire substrate. 
The Ta was patterned via photolithography and wet etching to form the ground plane, resonators, and qubit capacitor pads.
The qubits were made from $\mathrm{Al/AlO}_x/\mathrm{Al}$ Josephson junctions using the bridge-free fabrication process \cite{Lotkhov2011BridgeFree}. 
All measured devices were fabricated on the same chip and share a common transmission line.

\subsection{Cryogenic setup}

\begin{figure}[h!]
    \centering
    \includegraphics{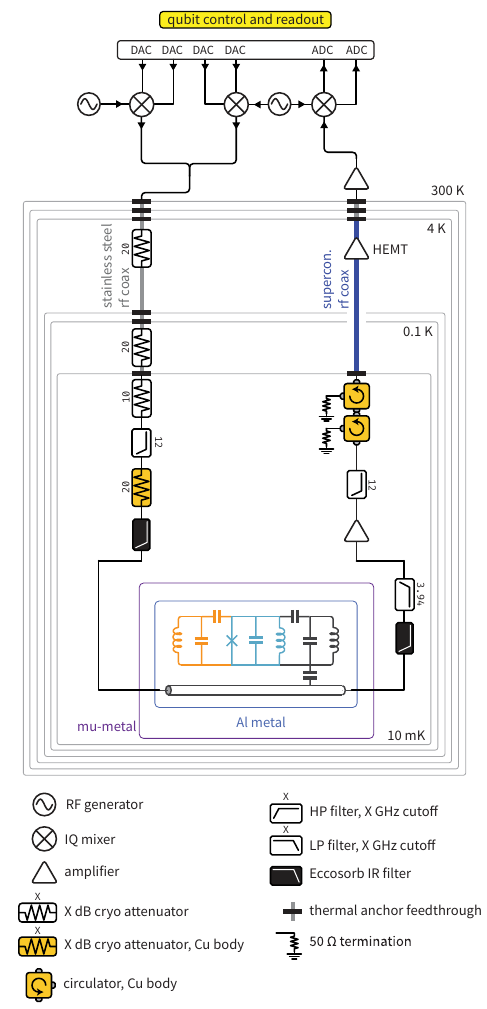}
    \caption{Measurement and cryogenic setup. 
    The readout tone and pump drives were synthesized using a \textit{Quantum Machines OPX+}, and up/downconverted via IQ mixing with external local oscillators (\textit{SignalCore SC5511A} and \textit{Rohde \& Schwarz SGS100A}).
    The tones were combined and sent through the input line in transmission. 
    A high-pass filter (\textit{VXHF-392+}) was used to filter out the qubit drive tone before amplification by a traveling-wave parametric amplifier.
    }
    \label{fig:supplement_fridge_wiring}
\end{figure}

The devices were cooled to \SI{10}{\milli\kelvin} and measured in an \textit{Oxford Instruments Triton 500} dilution refrigerator. 
A schematic of the measurement setup and device wiring is shown in ~\cref{fig:supplement_fridge_wiring}.
A global superconducting coil was used for flux biasing, with a \textit{Yokogawa GS200} serving as the current source.

\subsection{Spectroscopic Characterization}\label{Spectroscopic Characterization}

\begin{figure*}
    \centering
    \includegraphics{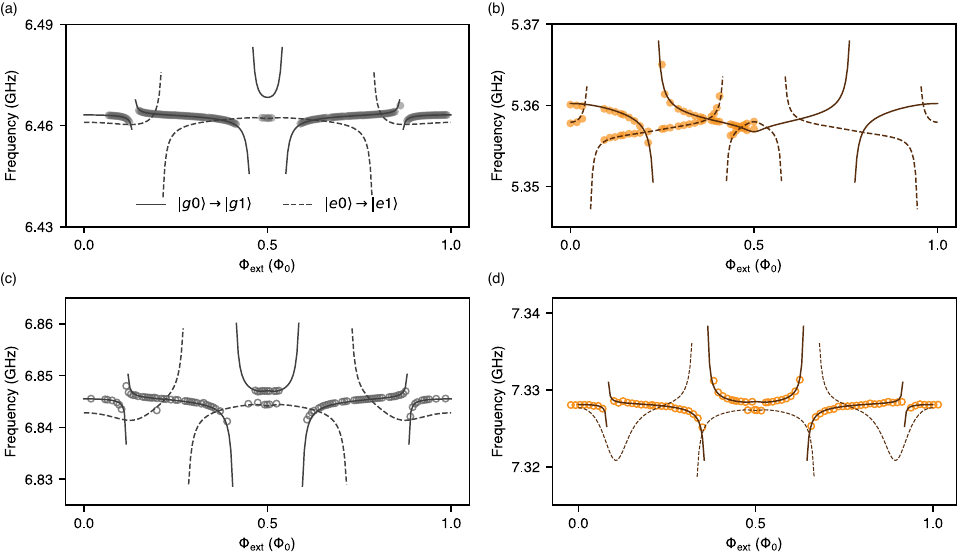}
    \caption{ 
    Resonator spectroscopy of the two devices.
    (a) Readout resonator spectroscopy and (b) Storage resonator spectroscopy of device A as a function of external flux. 
    Numerical simulation of $\ket{g0}\rightarrow\ket{g1}$ (solid line) and $\ket{e0}\rightarrow\ket{e1}$ show good agreement with the measurements. 
    (c), (d) Storage and readout resonator spectroscopy of device B.
    }
    \label{fig:supplement_system_characterization}
\end{figure*}

\begin{table}
    \centering
    \def\arraystretch{1.4} 
    \setlength{\tabcolsep}{12pt} 
    \begin{tabular}{c|c|c}
    \hline\hline
    Parameter& Device A& Device B\\
    \hline
    $E_C/2\pi$ & \SI{1.142}{\giga\hertz} & \SI{1.101}{\giga\hertz}\\
    $E_L/2\pi$ & \SI{0.559}{\giga\hertz} & \SI{0.564}{\giga\hertz} \\
    $E_J/2\pi$ & \SI{3.645}{\giga\hertz} & \SI{3.890}{\giga\hertz} \\
    $N_{JJ}$ & 116 & 116 \\
    $\tilde{\omega}/2\pi$ & \SI{5.3575}{\giga\hertz} & \SI{7.3269}{\giga\hertz}\\  
    $\tilde{\omega}_r/2\pi$ & \SI{6.4615}{\giga\hertz}& \SI{6.8435}{\giga\hertz}\\
    $g/2\pi$ & \SI{64}{\mega\hertz} & \SI{106}{\mega\hertz} \\  
    $g_{r}/2\pi$ & \SI{105}{\mega\hertz} & \SI{123}{\mega\hertz}\\ 
    $\kappa_r/2\pi$ & \SI{0.35}{\mega\hertz}& \SI{0.32}{\mega\hertz}\\
    $T_{1,s}$ & \SI{12.0 \pm 0.4}{\micro\second} & \SI{\sim 350}{\nano\second}\\
    \hline\hline
    Half flux& Device A& Device B\\
    \hline
    $\Omega/2\pi$ & \SI{424.88 \pm 0.01}{\mega\hertz} & \SI{358.72 \pm 0.01}{\mega\hertz}\\
    $\chi_{r}/2\pi$ & \SI{-6.38 \pm 0.04}{\mega\hertz} & \SI{2.85 \pm 0.01}{\mega\hertz}\\
    $\chi/2\pi$ & \SI{1.013 \pm 0.004}{\mega\hertz}& \SI{-1.22 \pm 0.01}{\mega\hertz}\\
    $K/2\pi$ & \SI{3.60\pm 0.04}{\kilo\hertz}& \SI{-13.5 \pm 0.4}{\kilo\hertz}\\    
    $T_{1,f}$ & \SI{123 \pm 4}{\micro\second} & \SI{85 \pm 6}{\micro\second} \\
    $T^R_{2,f}$ & \SI{16 \pm 1}{\micro\second} & \SI{17 \pm 4}{\micro\second} \\
    $T^E_{2,f}$ & \SI{90 \pm 4}{\micro\second} & \SI{62 \pm 4}{\micro\second} \\
    \hline\hline
    \end{tabular}
    \caption{
    Summary of measured and fitted system parameters for devices A and B.
    For device A, qubit $T_1$ and $T_2$ times were measured during the SNAP experiment and are used in all corresponding master equation simulations.
    For device B, $T_1$ and $T_2$ times were averaged from an 24-hour continuous time-domain measurements, as reported in Appendix~\ref{TLS}.}
    \label{tab:system_parameter}
\end{table}

Here we describe the full system Hamiltonian and present spectroscopic characterization of the readout and storage resonators.
The full system Hamiltonian is given by
\begin{multline}\label{full_Hamiltonian}
    \hat{H}_\text{full}/\hbar =  4E_C\hat{n}^2 + \frac{1}{2}E_L\hat{\phi}^2 - E_J\cos{(\phi-\phi_\mathrm{ext})} \\
     + \tilde{\omega}\hat{\tilde{a}}^\dagger\hat{\tilde{a}} + \tilde{\omega}_r\hat{\tilde{r}}^\dagger\hat{\tilde{r}} - ig\hat{n}(\hat{\tilde{a}}- \hat{\tilde{a}}^\dagger) - ig_{r}\hat{n}(\hat{\tilde{r}}- \hat{\tilde{r}}^\dagger).
\end{multline}
The operator $\hat{\tilde{r}}$ describes the readout resonator mode, and the subscript $r$ is used to distinguish quantities from those of the storage mode. 
By default, symbols without a subscript refer to the storage or qubit mode, consistent with the convention used in the main text.
In ~\cref{fig:supplement_system_characterization}, we show flux-dependent spectroscopy of the readout and storage resonators for both devices.
For the readout mode, its strong coupling to the transmission line allows direct characterization via single-tone spectroscopy (Fig.~\ref{fig:supplement_system_characterization}(a), (c)).
For the storage mode of device A, the resonance is not directly visible in the $S_{21}$ transmission due to its weak coupling. 
Instead, we probed the storage mode via the fluxonium qubit by applying a drive near the storage frequency, followed by a selective qubit $\pi$-pulse conditioned on the storage being in $\ket{0}$ (Fig.~\ref{fig:supplement_system_characterization}(b)).
Device B’s storage mode is coupled stronger and we directly measured its flux-dependent spectrum using single-tone spectroscopy (Fig.~\ref{fig:supplement_system_characterization}(d)).
From resonator spectroscopy, together with the fluxonium spectrum, and $\chi$, $K$ as shown in Fig.~\ref{fig3:flux_characterization}, we extracted the system parameters summarized in Table~\ref{tab:system_parameter}. 

\subsection{Error Analysis} \label{error_analysis}

Here, we briefly describe the fitting uncertainty associated with the system parameters.
In Table~\ref{tab:system_parameter}, the fluxonium circuit parameters, bare resonator frequencies, and their coupling strengths were tuned to simultaneously match spectroscopy data and measured $\chi$ and $K$.  
While we did not compute uncertainties for this procedure, obtaining a good fit required tuning each parameter to the precision reported.
Thus, the uncertainties in these parameters should be at least as small as the order of magnitude of the last digit reported.

For directly measured quantities, such as $\chi$ and $K$, we extracted the uncertainties from the covariance matrix of the fitting functions.
Specifically, the uncertainty in $\chi$ was obtained from a multi-Lorentzian fit to the spectrum, while the uncertainty in $K$ was extracted directly from a linear fit as a function of $\bar{n}$.
We note that the reported uncertainties do not account for temporal fluctuations, which are discussed separately in Appendix~\ref{TLS}.

The uncertainties in the qubit frequencies at half flux were not obtained from a covariance matrix, but were instead independently estimated from the measured dephasing rate, with $\delta\Omega/2\pi \sim 1/2\pi T_2^R$.

\section{Half flux characterization}

\subsection{Qubit initialization} \label{supplement_qubit_initialization}

\begin{figure}
    \centering
    \includegraphics{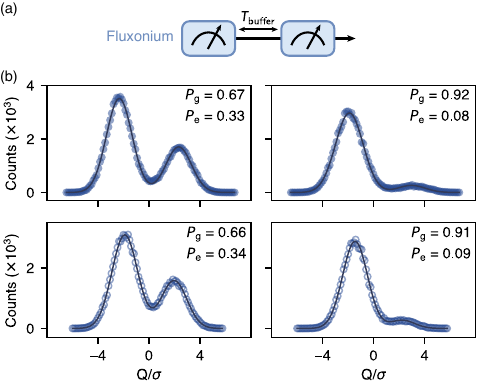}
    \caption{
    Cooling through measurement. 
    (a) Pulse sequence for measurement-based cooling. Two consecutive qubit measurements were performed, separated by a buffer time $T_\mathrm{buffer}$. The first measurement projected the fluxonium into its ground state, increasing the contrast of the second measurement.
    (b) Qubit population of device A (top) and device B (bottom), measured in the first measurement (left) and in the second measurement (right) after postselection.
    }
    \label{fig:supplement_readout_postselection}
\end{figure}

In this section, we describe the procedure used to prepare the qubit in its ground state at half flux and provide details on the initialization efficiency.
Qubit initialization is necessary due to the low transition frequency of the fluxonium at half flux.
We implemented qubit initialization using measurement-based cooling. 
Prior to each experiment, we applied a \SI{3}{\micro\second} readout pulse to perform single-shot measurement of the qubit state. 
We then post-selected data for which the qubit was in the ground state during this measurement pulse. 
To ensure the system returns to its computational subspace, we included a buffer time $T_\mathrm{buffer}$ of \SI{3}{\micro\second} ($\sim5\kappa_r$) after the readout to allow residual photons in the readout resonator to decay before starting the experiment.
In Fig.~\ref{fig:supplement_readout_postselection}, we show two consecutive single-shot readout measurements of the qubit, separated by a buffer time $T_\mathrm{buffer}$. 
The second measurement was post-selected based on the outcome of the first.
At thermal equilibrium, both devices had approximately 66\% qubit population in the ground state (Fig.~\ref{fig:supplement_readout_postselection}(b)).
After post-selection, we achieved greater than 90\% ground-state population.
For more complex bosonic control protocols where high-fidelity qubit initialization is critical, more effective active reset techniques are available for the fluxonium qubit \cite{vool2018driving, Zhang2021universal, Wang2024efficient, nie_parametrically_2024}.

\subsection{Readout background correction} \label{supplement_readout_background}

Here, we describe our general procedure used to correct for measurement background.
In the ideal case, the storage resonator is measured via the qubit with a constant readout background. 
Resonator cross talk and qubit decay after measurement-based cooling, however, cause a skewed background. 
For the reported experiments with background correction (indicated by an asterisk as $P_e^*$), we performed a control experiment where the qubit $\pi$ pulse was not played.
The time slot for the pulse was still preserved to account for any decay during this interval.
We subtracted the control data from the measurement to isolate the actual qubit signal.

\subsection{Device A displacement calibration} \label{supplement_displacement_calibration}

\begin{figure}
    \centering
    \includegraphics{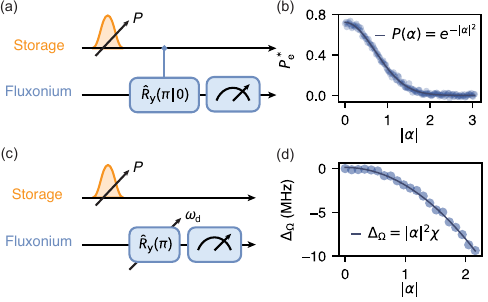}
    \caption{
    Displacement calibration of device A. 
    (a) Pulse sequence for displacement calibration at the flux sweet spots. 
    A Gaussian pulse of duration $\SI{80}{\nano\second}$ was
    used to populate the storage resonator.
    (b) Displacement calibration at half flux, using the pulse sequence shown in (a). 
    Background was subtracted using a control measurement.
    (c) Pulse sequence for displacement calibration away from the flux sweet spots. 
    Qubit spectroscopy was performed on the fluxonium to extract the induced AC Stark shift and calibrate the displacement.
    (d) Qubit frequency detuning as a function of displacement amplitude, measured using the pulse sequence shown in (c).
    }
    \label{fig:supplement_device2_alpha_calibration}
\end{figure}

Here, we describe the calibration procedures used to relate the drive amplitude to the displacement in the storage mode for device A. 
Depending on whether photon-number splitting is observable in the qubit spectrum, different calibration methods were employed.
At the flux sweet spots, where the fluxonium qubit is first-order insensitive to flux noise, the dephasing time reaches its maximum. 
Therefore, one can apply a selective $\pi$ pulse to the qubit to measure the storage mode’s overlap with $\ket{0}$, which is given by
\begin{equation}
    P(\alpha)=|\langle 0|\hat{D}_\alpha|0\rangle|^2=e^{-|\alpha|^2}.
\end{equation}
In Fig.~\ref{fig:supplement_device2_alpha_calibration}(a), we show the pulse sequence used to displace the storage mode to measure the probability of it being in the \(\ket{0}\) state.
A Gaussian pulse of duration \SI{80}{\nano\second} was applied to populate the storage mode, while the pump power was swept to calibrate the displacement amplitude.
The measured qubit population as function of displacement amplitude is shown in~\cref{fig:supplement_device2_alpha_calibration}(b).
Away from the flux sweet spots, where the fluxonium is sensitive to flux noise and the dispersive shift is smaller than the qubit linewidth, this method is no longer suitable for displacement calibration.
Alternatively, the displacement amplitude can be calibrated by measuring the AC Stark shift induced on the qubit by the photon population in the storage mode. This shift is given by
\begin{equation}
    \Omega(\alpha) = \Omega(0) + \chi|\alpha|^2.
\end{equation}
In Fig.~\ref{fig:supplement_device2_alpha_calibration}(c) and (d), we show the pulse sequence and the corresponding measured data used for the AC Stark shift calibration.
We note that the spectral width of the \SI{80}{\nano\second} storage pump pulse is significantly larger than the frequency shift of the storage resonator across an entire flux quantum. 
Consequently, the calibration performed at half-flux bias remains valid across the full flux range. Calibrations performed at different flux points using the two methods yield consistent results.

\subsection{Device B half-flux characterization} \label{supplement_device_B_characterization}

\begin{figure}
    \centering
    \includegraphics{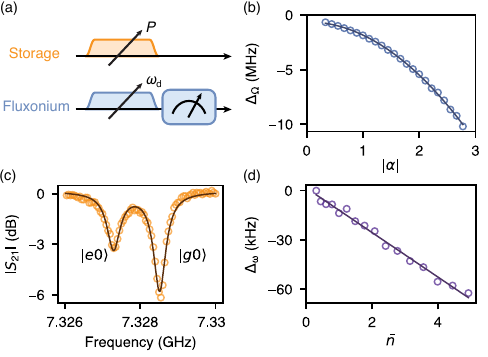}
    \caption{
    Half-flux characterization of device B. 
    (a) Pulse sequence for displacement calibration based on the AC Stark shift. Due to the short $T_1$ of the storage resonator, two overlapping $\SI{10}{\micro\second}$-long pulses were used for the resonator and qubit drives.
    (b) Displacement calibration at half flux, using the pulse sequence shown in (a) 
    (c) Storage resonator single-tone spectroscopy at half flux. Two peaks are visible, corresponding to the thermal population of the fluxonium qubit. The peaks are separated by a dispersive shift of $\chi/2\pi=\SI{-1.22}{MHz}$. 
    (d) Storage resonator frequency detuning as a function of average photon number. A linear fit yields a self-Kerr nonlinearity of $K/2\pi=\SI{-13.5}{\kilo\hertz}$. 
    }
    \label{fig:supplement_device1_characterization}
\end{figure}

In this section, we discuss the characterization of device B at half flux, which requires different methods from those used for device A.
Due to a nearby spurious mode, the storage mode of device B couples more strongly to the transmission line than intended, resulting in a reduced lifetime for the storage mode.
Consequently, displacement calibration via selective measurement is not feasible, as it relies on a sufficiently long storage mode lifetime.
Instead, we calibrated the displacement by measuring the AC Stark shift induced on the qubit. 
In Fig.~\ref{fig:supplement_device1_characterization}(a), we show the pulse sequence used for this measurement.
Unlike the pulse sequence shown in Fig.~\ref{fig:supplement_device2_alpha_calibration}(c), we used two overlapping \SI{10}{\micro\second} pulses to simultaneously populate the storage mode and perform qubit spectroscopy. 
The storage drive was applied on resonance with the resonator frequency corresponding to the qubit in the $\ket{g}$ state. 
We swept over a range of low powers for the storage drive to extract the AC Stark shift induced on the qubit.
Although not explicitly shown in the pulse sequence, we applied a cooling measurement prior to driving the storage mode on resonance with its $\ket{g}$ peak. 
Therefore, one can safely assume 
\begin{equation}
    \bar{n} = \bar{n}_g = |\alpha|^2  = |\alpha_g|^2, \bar{n}_e = 0,
\end{equation}
where $\bar{n}$ denotes the total mean photon number in the storage mode, while $\bar{n}_g\ (\bar{n}_e)$ refers to the mean photon number when the qubit is in the ground (excited) state.
    
In Fig.~\ref{fig:supplement_device1_characterization}(b), we show the measured AC Stark shift data, which exhibits a parabolic dependence, supporting our assumptions.
We directly measured the dispersive shift at half flux using single-tone spectroscopy, obtaining $\chi/2\pi = \SI{-1.22}{\mega\hertz}$ (Fig.~\ref{fig:supplement_device1_characterization}(c)). 
To extract the self-Kerr nonlinearity, we performed resonator single-tone spectroscopy as a function of pump power. 
Using the relation
\begin{equation}
 \omega_{\bar{n}+1}-\omega_{\bar{n}}=\bar{n}K, 
\end{equation}
we obtain $K/2\pi=\SI{-13.5}{\kilo\hertz}$.
We used the same procedures for extracting $\chi$ and $K$ across the entire flux range, as shown in Fig.~\ref{fig3:flux_characterization}(d) and (f) of the main text.

\subsection{System fluctuations due to TLS} \label{TLS}

\begin{figure}
    \centering
    \includegraphics{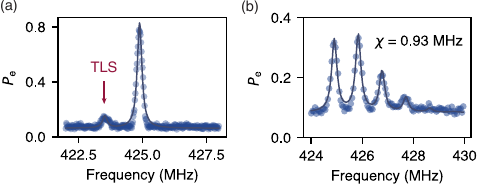}
    \caption{Evidence of TLS in Device A. 
    (a) Fluxonium two-tone spectroscopy at half flux.
    A small splitting approximately \SI{1.35}{\mega\hertz} away from the main fluxonium transition is observed, indicating coupling to a nearby TLS.
    (b) Photon-number splitting in the fluxonium spectrum, measured similarly to Fig.~\ref{fig2:half_flux}(b).
    However, the extracted dispersive shift $\chi$ differs due to fluctuations in the qubit–TLS coupling.
    }
    \label{fig:supplement_TLS_deviceA}
\end{figure} 

\begin{figure}
    \centering
    \includegraphics{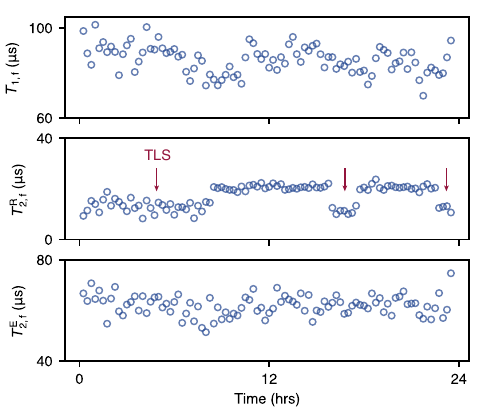}
    \caption{ 
    Time-domain measurements of device B over time. 
    The jump in device B's $T_\mathrm{2,R}$ data is attributed to fluctuations of a nearby two-level system.
    }
    \label{fig:supplement_time_domain_overtime_deviceB}
\end{figure}

During our measurements of both devices, we directly observed signatures of two-level systems (TLSs), including fluctuations in system parameters attributable to their presence.
For Device A at half flux, we observed a splitting in the fluxonium $\ket{g} \rightarrow \ket{e}$ transition due to coupling to a nearby TLS, as indicated by a small bump adjacent to the main transition peak in the two-tone spectroscopy shown in Fig.~\ref{fig:supplement_TLS_deviceA}(a).
Due to the fluctuating coupling to this TLS, we observed approximately 10\% variation in the qubit’s dispersive shift $\chi$ to the storage mode (Fig.~\ref{fig:supplement_TLS_deviceA}(b)). 
These changes occur on the timescale of several days, which is sufficiently slow to allow stable calibration during our measurements.
Similarly, for device B, although we did not directly observe a splitting in the qubit spectroscopy, we observed fluctuations in the qubit $T_{2}^R$ measurements over time (Fig.~\ref{fig:supplement_time_domain_overtime_deviceB}).

\section{Theoretical modeling}\label{supplement_theory}

\subsection{System Hamiltonian}\label{Appendix_sysmtem_hamiltonian}

To simulate the system spectrum, we numerically diagonalized the full Hamiltonian given in Eq.~(\ref{fluxonium_resonator_Hamiltonian}) using the scQubits \cite{scqubits_1,scqubits_2} Python package. 
From the spectrum, we extracted the qubit and resonator frequencies, as well as the dispersive shift $\chi$ and the self-Kerr nonlinearity $K$ of the storage mode.
At a specific flux bias point, we simulated the system using the effective Hamiltonian given in Eq.~(\ref{effective Hamiltonian}).

The qubit drive is modeled by
\begin{equation}\label{drive_hamiltonian}
    \hat{H}_\mathrm{drive}/\hbar = \sum_n\epsilon_n(t) e^{-i \omega_n t} \ket{e}\bra{g} + \text{h.c.},
\end{equation}
where $\epsilon_n(t)$ denotes the time-dependent amplitude of the $n$th microwave drive with pump frequency $\omega_n$.
For simplicity, we work in the rotating frame given by
\begin{equation}
\hat{H}_\mathrm{frame}/\hbar = \Omega\ket{e}\bra{e} + \omega \hat{a}^\dagger \hat{a}.  
\end{equation}
In this frame, the effective Hamiltonian and the drive Hamiltonian take the form
\begin{align}\label{RWA_hamiltonian}
    \hat{H}_\mathrm{eff, rot}/\hbar &=  \chi\hat{s}^\dagger\hat{s}\ket{e}\bra{e}+ \frac{K}{2}\hat{s}^\dagger\hat{s}^\dagger\hat{s}\hat{s}, \\ 
    \hat{H}_\mathrm{drive, rot}/\hbar &= \sum_n\epsilon_n(t) e^{-i \delta_n t} \ket{e}\bra{g} + \text{h.c.},
\end{align}
where $\delta_n = \omega_n - \Omega$ is the detuning between the $n$th drive frequency and the fluxonium transition frequency. 
A microwave drive on the storage resonator can be included analogously by replacing $\ket{e}\bra{g}$ with the annihilation operator $\hat{a}$. 
The total Hamiltonian $\hat{H}_\mathrm{eff, rot} + \hat{H}_\mathrm{drive, rot}$ provides a sufficiently accurate model for our system in the rotating frame.

\subsection{Master equation simulation} \label{supplement_master_simualtion}

To incorporate decoherence and energy relaxation, we simulated the system dynamics by numerically solving the Lindblad master equation,
\begin{equation}\label{master_equation}
    \frac{\partial\hat{\rho}}{\partial t}=-i[\hat{H}_\text{eff,rot}+\hat{H}_\text{drive,rot}, \hat{\rho}] + \sum_{k}[L_k\rho L_k^\dagger-\frac{1}{2}\{L_k^\dagger L_k,\hat{\rho}\}],
\end{equation}
where $\hat{\rho}$ is the system density operator and $L_k$ are the collapse operators describing dissipation. 
We included the dominant loss mechanisms using the collapse operators: storage-mode energy decay $\sqrt{\kappa}\hat{a}$, fluxonium relaxation $\sqrt{\Gamma_\downarrow}\ket{g}\bra{e}$, fluxonium excitation $\sqrt{\Gamma_\uparrow}\ket{e}\bra{g}$, and pure dephasing $\sqrt{\frac{\Gamma_\phi}{2}}( \ket{e}\bra{e} - \ket{g}\bra{g})$. 
In addition to incoherent loss, we incorporated qubit initialization error by setting the initial density matrix to a mixed state. 
Simulations were carried out using the Python package QuTiP \cite{qutip3}.

\subsection{Derivation of Transmon $K$ limit}
\label{sup_transmon_limit}

\begin{table}[h]
    \centering
    \def\arraystretch{1.5} 
    \setlength{\tabcolsep}{5pt} 
    \begin{tabular}{c|c|c|c}
    \hline\hline
    $|K|/2\pi \:\SI{}{\kilo\hertz}$& $|\chi|/2\pi\: \SI{}{\mega\hertz}$&$K$ type& Reference\\
    \hline
    22 $\pm$ 2& 3.825 $\pm$ 0.001 & data & \cite{Pfaff2017}\\
    $3.6\times 10^{-3}$ & $4.2\times 10^{-2}$ &simulation&\cite{milul_superconducting_2023}\\
    107.9 $\pm$ 0.5&8.2813 $\pm$ 0.001 & data & \cite{heeres_cavity_2015}\\
    8&2.86&data&\cite{Axline2018}\\
    5&2.29&data&\cite{Axline2018}\\
    0.699(6)&0.4861(3)&data&\cite{landgraf_fast_2024}\\
    4.5&1.97&data&\cite{Ofek2016}\\
    $1\times 10^{-3}$&$2.8\times10^{-2}$&data&\cite{CampagneIbarcq2020}\\
    $4.8\times 10^{-3}$&$4.65\times10^{-2}$&data&\cite{Sivak2023}\\
    44&1.67&simulation&\cite{Valadares2024}\\
    6&0.94&simulation&\cite{Valadares2024}\\
    1.9&0.57&simulation&\cite{Valadares2024}\\
    0.19&0.29&simulation&\cite{Valadares2024}\\
    $9\times10^{-2}$&0.18&simulation&\cite{Valadares2024}\\
    $5\times10^{-3}$&$5\times10^{-2}$&simulation&\cite{Valadares2024}\\
    3&1.3&data&\cite{Cai2024}\\
    2&1.411&data&\cite{Cai2024}\\
    4.9 $\pm$ 0.1&1.499 $\pm$ 0.003&data&\cite{Yang2025}\\
    1.7&1.313&data&\cite{Gertler2021}\\
    
    \hline\hline
    \end{tabular}
    \caption{Literature values for transmon $\chi$ and $K$.  Each row corresponds to a unique device, except for the six values from \cite{Valadares2024}.  Those values were measured with a SQUID transmon, with each row corresponding to a different flux point.  Some of the $K$ values are measured, while some were simulated by the respective authors.  The distinction is made in column 3.  Uncertainties are given as originally reported.}
    \label{tab:transmon_literature}
\end{table}

As derived in \cite{blais_qed}, $\chi$ and $K$ for a transmon dispersively coupled to a cavity are given by 
\begin{gather}
    \label{supeq_blais_kerr}
    {K} \approx - E_C \left( \frac{g}{\Delta}\right)^4,\\
    \label{supeq_blais_chi}
    {\chi} \approx -\frac{2g^2 E_C}{\Delta(\Delta - E_C)}.
\end{gather}
Here, we are setting $\hbar = 1$, so all quantities in \cref{supeq_blais_kerr,supeq_blais_chi} are in units of angular frequency.
The detuning $\Delta$ is defined with the sign convention $\Delta = \tilde{\omega}_q - \tilde{\omega}$ \cite{blais_qed}, where $\tilde{\omega}_q$ and $\tilde{\omega}$ denote the bare frequencies of the transmon and cavity, respectively.

We want to find the smallest possible $K$ for a given $\chi$.  
If we impose a lower limit $|\chi| \ge \chi_0$, then 
\begin{equation}
    g^2 \gtrsim \frac{\chi_0 |\Delta (\Delta - E_C)|}{2 E_C}.
\end{equation}  
Plugging this into the expression for $K$ yields
\begin{equation}
\label{supeq_transmon_limit1}
    \left|{K}\right| \approx E_C \left( \frac{g}{\Delta}\right)^4 \gtrsim  \frac{\chi_0(\Delta - E_C)^2}{4E_C\Delta^2}.
\end{equation}
Initially, it appears that for constant $\chi_0$ we could have $|K|\rightarrow0$ by taking $\Delta \rightarrow E_C$.  
However, while that parameter regime can be useful in some situations \cite{amazon_bosonic}, we aim to keep the storage mode as linear as possible and demand minimal hybridization between the cavity and qubit modes.
To measure the hybridization, we computed the overlap between the bare and dressed cavity states, where an overlap less than unity indicates finite qubit-cavity hybridization.
To compute the overlap, we numerically diagonalized the transmon-storage system, approximated by 
\begin{multline}
\label{supeq_transmon_hamiltonian}
    H_{\textrm{approx}}/\hbar = 4 E_C \hat n^2 -\frac{1}{2} E_J \sum_{n = -15}^{15}(\ket{n}\bra{n+1} +  \textrm{h.c.})\\
    +\tilde{\omega} \hat{\tilde{a}}^\dagger\hat{\tilde{a}} - i g_n  \hat n(\hat{\tilde{a}}- \hat{\tilde{a}}^\dagger) + \textrm{h.c.},
\end{multline}
with scQubits \cite{scqubits_1,scqubits_2}.  
Note that in \cref{supeq_transmon_hamiltonian}, we have truncated the transmon Hilbert space in the charge basis.  Further, though not shown explicitly in \cref{supeq_transmon_hamiltonian}, we have truncated the storage Hilbert space dimension to 5 Fock states.
Then we averaged the bare $\rightarrow$ dressed overlap for the states $\ket{g;0},\;\ket{e;0}$, $\;\ket{g;1},\;\ket{e;1}$.
We find that the overlap drops dramatically when $\Delta \sim E_C$, exactly where $|K| \rightarrow0$ (see Fig. \ref{fig_5} (a)).

Hence, for low hybridization, we must move to the regime $|\Delta| \gg E_C$.  In this regime, we have $(\Delta - E_C)^2 \approx \Delta^2$.  This simplifies eq. \ref{supeq_transmon_limit1} to
\begin{equation}\label{supeq_intermediate_transmon_limit}
    \left|{K}\right|  \gtrsim  \frac{\chi_0^2}{4E_C}
\end{equation}
In principle, we can reduce this lower bound by increasing $E_C$.
However, $E_C$ is bounded by practical limitations.
In particular, we don't want the transmon frequency $\tilde{\omega}_q \simeq \sqrt{8E_CE_J} - E_C$ to be too large, and want to stay in the transmon regime $E_J \ge 50E_C$ \cite{koch_transmon_regime,schreier_transmon_regime}.  
We require that the qubit frequency be less than $\SI{10}{\giga\hertz}$, which leads to $E_C/2\pi \le \SI{10}{\giga\hertz} /19 \approx \SI{530}{\mega\hertz}$.
Plugging this into \cref{supeq_intermediate_transmon_limit}, we find the bound described in the main text:
\begin{equation}
\label{supeq_transmon_limit3}
    \left|{K}\right|  \gtrsim  \frac{\chi_0^2}{2\pi \cdot\SI{2120}{\mega\hertz}}
\end{equation}

\subsection{Higher order nonlinearities}
\begin{figure}
    \centering
    \includegraphics{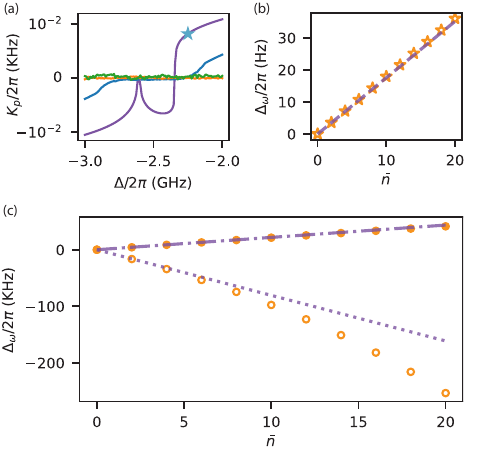}
    \caption{Higher order Kerr in the fluxonium devices.
    (a) Simulated $p$-th order shift $K_p$ as a function of detuning for the same fluxonium as presented in Fig. \ref{fig_5} for $p=$ 2 (purple), 3 (blue), 4 (orange), and 5 (green).  At each $\Delta$, $g$ is tuned such that $|\chi| = \SI{1}{\mega\hertz}$.  The blue star at $\Delta/2\pi = \SI{-2.25}{\mega\hertz}$ is the same example point as in Fig. \ref{fig_5}.
    (b) At the blue star point of sub-figure a, the simulated deviation $\Delta_\omega$ (orange stars) of the cavity frequency is plotted for different photon numbers.  
    The deviation only due to $K_2/2\pi = \SI{1.76}{\hertz}$ is plotted for comparison (dashed purple line).
    (c) The simulated deviation $\Delta_\omega$ for devices A and B (closed and open circles, respectively).  The deviations only due to $K_2$ are plotted as dash-dotted and dotted lines, respectively.
    }
    \label{fig:supplement_higher_kerr}
\end{figure}
Our effective Hamiltonian (\cref{effective Hamiltonian}) is truncated to the leading-order non-linearity $K\hat{a}^\dagger\hat{a}^\dagger\hat{a}\hat{a}/2$.  
However, the full non-linearity of the dressed cavity mode generically includes higher-order terms.  
The terms that are relevant when the qubit is in the ground state can be written as
\begin{equation}
    \sum_{p = 2}^\infty \frac{K_p}{p!} \hat a^{\dagger p} \hat a^p.
\end{equation}
Note that $K_2 \equiv K$.
The terms in the sum are normal-ordered so that the $p-$th term only contributes to the energy once the cavity has at least $p$ photons.

Because the bosonic control experiments in this work only need a few cavity photons, the effects of higher $K_p$ are not seen. But later experiments to create bosonic codes will need larger photon numbers.  
Thus, here we present simulated higher-order $K_p$ for the low-$K$ fluxonium proposed in section \ref{sec:prospects} (see Fiq. \ref{fig_5}) to verify its viability for general bosonic encoding.

To compute $K_p$, we first found the dressed energies with scQubits \cite{scqubits_1,scqubits_2}.  
If $E_{g,n}$ is the energy when the fluxonium is in the ground state and the cavity has $n$ photons, then the coefficients $K_p$ are given by
\begin{equation}
    K_p = \sum_{j=0}^p (-1)^{(j+p)}{p\choose j} E_{g,j}
\end{equation}
The $K_p$ are plotted as a function of detuning in \cref{fig:supplement_higher_kerr}(a) up to $p=5$ ($p=2$ is shown also in \cref{fig_5}).  
We find that $K_p$ decreases with $p$ in the simulated range of $\Delta$ and starting with $p\ge 4$, the value of $K_p$ reaches the level of numerical noise in our simulation.
 
As an additional illustration of the non-linearity, we computed the deviation of the cavity frequency away from its single-photon value up to 20 photons (\cref{fig:supplement_higher_kerr}(b)).  
More precisely, $\Delta_{\omega}(n) = \omega(n) - \omega(0)$, with $\omega(n) = E_{g,n+1} - E_{g,n}$.  
The actual $\Delta_{\omega}$ is compared to the deviation accounted for by taking only $K_2$.
We computed the same quantities for devices A and B (\cref{fig:supplement_higher_kerr}(c)).
Up to 20 photons, the deviation is well described by just $K_2$ for both device A and our proposed device.
For device B, however, higher-order corrections become significant after approximately 10 photons.

\section{$K$ measurement}\label{supplement_Kerr}
\subsection{Cavity Ramsey simulation} \label{supplement_cavity_ramsey}

\begin{figure}
    \centering
    \includegraphics{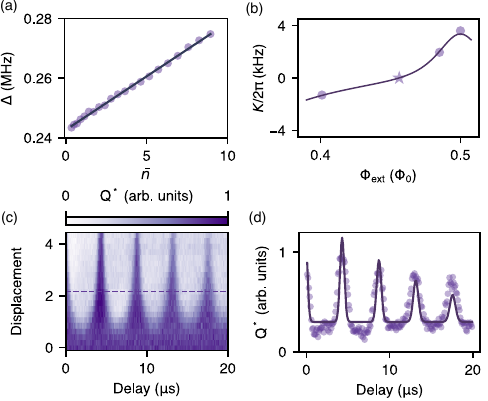}
    \caption{
    (a) Detuning extracted from the Ramsey data shown in Fig.~\ref{fig2:half_flux}(d). The detuning exhibits a linear dependence on the average photon number. A linear fit yields a self-Kerr nonlinearity of $K/2\pi = \SI{3.6}{\kilo\hertz}$.
    (b) Zoom-in of the data from Fig.~\ref{fig3:flux_characterization}(e) near the Kerr zero-crossing point. The data shows a clear sign change in the extracted Kerr, indicating a transition through $K/2\pi = \SI{0}{\kilo\hertz}$.
    (c) Cavity Ramsey raw data as a function of displacement taken at the flux bias indicated by the star marker shown in (b). 
    No appreciable change in detuning is observed.
    (d) Data taken at displacement amplitude $|\alpha| = 2.2$, from which we extract a detuning of $\Delta/2\pi \approx \SI{0.227}{\mega\hertz}$.
    The resonance observed in the data is not as sharply peaked as in simulation, due to imperfect selectivity in the measurement arising from reduced qubit coherence away from the sweet spot.
    Based on the detuning and the swept displacement range, we estimate the minimum Kerr that can be reliably resolved in this experiment to be $K/2\pi\approx\SI{300}{\hertz}$.
    }
    \label{fig:deviceA_cavity_ramsey}
\end{figure}

Under the pulse sequence shown in Fig.~\ref{fig2:half_flux}(c), and not considering dissipation, the storage resonator evolves under the unitary transformation
\begin{equation}
\hat{U}(t) = \exp\left[-i\left(\Delta\hat{a}^\dagger \hat{a} + \frac{K}{2} \hat{a}^\dagger \hat{a}^\dagger \hat{a} \hat{a} \right)t\right],
\end{equation}
where $\Delta$ denotes the detuning between the storage-mode frequency and the pump frequency. 
In this case, the Ramsey signal is given by the return probability of the coherent state under the evolution,
\begin{equation}
    P(t) = \left| \left\langle \alpha| \hat{U}(t) | \alpha \right\rangle \right|^2.
\end{equation}
For short evolution times, the phase accumulation of a coherent state under this unitary can be approximated as linear in time,
\begin{equation}
    \hat{U}(t)\ket{\alpha}\approx\ket{\alpha e^{-i\phi(t)}},\phi(t) = (\Delta + K|\alpha|^2)t.
\end{equation}
Thus, the effective detuning experienced by the coherent state grows linearly with the mean photon number $\bar{n}$, with a slope given by $K$, as shown in Fig.~\ref{fig:deviceA_cavity_ramsey}(a) for half flux. 

Due to Kerr-evolution of the coherent state, the Ramsey signal undergoes Gaussian decay at short evolution times.
However, the finite lifetime of the storage resonator leads to further incoherent decay. 
To capture both effects, we modeled the cavity Ramsey signal using the master equation detailed in \cref{supplement_master_simualtion}.

We note that away from the flux sweet spot, the reduced qubit coherence limits the ability to perform photon-number-selective measurements.
As a result, the measured cavity Ramsey data exhibit reduced contrast, leading to broadened fringes and deviations from the exact predictions of the master equation simulation (\cref{fig:deviceA_cavity_ramsey}(d)).
However, extracting the Kerr coefficient from the measured detuning remains possible, as the frequency shift is still reliably observed despite the reduced measurement selectivity.
Accordingly, all Kerr values shown in Fig.~\ref{fig3:flux_characterization}(e) were extracted from the measured detuning rather than from full master equation fits. 
At flux sweet spots, where selective measurement was feasible, we also verified the extracted Kerr values against full master equation simulations for consistency.

For bosonic encoding, self-Kerr nonlinearity is generally undesirable, as it induces photon-number-dependent dephasing. 
For fluxonium, it is possible to flux-tune the system to operating points where $K$ is small.
In \Cref{fig:deviceA_cavity_ramsey}(b), we show a zoom-in of the same data presented in \Cref{fig3:flux_characterization}(e) near half flux.
Both the data and simulation clearly show the Kerr coefficient transitioning from positive to negative values, indicating a zero-crossing around $\Phi_\mathrm{ext}\approx0.456\,\Phi_0$.
In \Cref{fig:deviceA_cavity_ramsey}(c), we show the cavity Ramsey data as a function of displacement, taken at $\Phi_\mathrm{ext} \approx 0.4563\,\Phi_0$.
At this flux point, as single-shot readout was not feasible, we plot the measured quadrature signal instead of the actual qubit population.
As evidenced by the data, there is no visible change in detuning, which remains approximately constant at $\Delta/2\pi \approx \SI{0.227}{\mega\hertz}$ (\cref{fig:deviceA_cavity_ramsey}(d)).
To resolve $K$, we estimate that the shift in the position of the final Ramsey fringe at the largest displacement must be at least \SI{400}{\nano\second}.
Achieving such a shift would require a Kerr of at least $K/2\pi \gtrsim \SI{300}{\hertz}$, thereby placing an upper bound on $K$ at this flux point.
Improved resolution of smaller Kerr would require a cavity with higher coherence. 

\subsection{Q function characterization of $K$}

\begin{figure}
    \centering
    \includegraphics{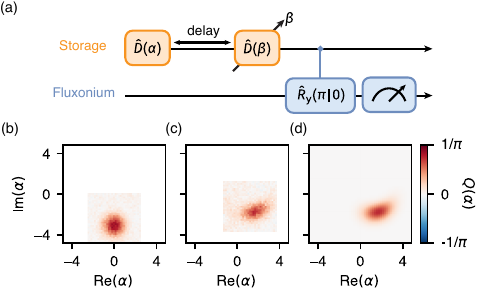}
    \caption{
    Q function under Kerr evolution.
    (a) Pulse sequence used to characterize Kerr by measuring Q function.
    A delay time was inserted between state preparation and Q function measurement, during which the coherent state evolved under the Kerr effect. 
    (b) The Q function of a coherent state $|\alpha|=3$ with zero delay time.
    Only regional data were acquired to save measurement time.
    (c) The Q function of a coherent state that evolves under Kerr interaction for \SI{5}{\micro\second}.
    Due to the Kerr effect, the initially circular distribution becomes elongated in phase space.
    (d) Simulation of the dynamics shown in (c) using the measured system parameters.
    The simulation agrees well with the experimental data.}
    \label{fig:supplement_Q_function_Kerr}
\end{figure}

An alternative method to extract $K$ is through the time-evolution of the Husimi Q-function.
By measuring the overlap between a displaced resonator state and $\ket{0}$, we can effectively construct the Q function, defined by 
\begin{equation}
    Q(\alpha)=\frac{1}{\pi}\left<\alpha\left|\hat{\rho} \right|\alpha\right>
    =\frac{1}{\pi}\left<0\left|\hat{D}^\dagger_\alpha\hat{\rho} \hat{D}_\alpha\right|0\right>.
\end{equation}
Measuring the Q function enables us to identify the Kerr nonlinearity by observing the evolution of the quantum state in phase space (Fig.~\ref{fig:supplement_Q_function_Kerr}(a)), where the Q function is given by
\begin{align}
    Q(\alpha, t)=\frac{1}{\pi}\left<\alpha\left|\hat{U}^\dagger(t)\hat{\rho} \hat{U}(t)\right|\alpha\right>, \\  
    \hat{U}(t) = \exp\left[-i\left( \frac{K}{2} \hat{a}^\dagger\hat{a}^\dagger \hat{a} \hat{a} \right)t\right].
\end{align}
In Fig.~\ref{fig:supplement_Q_function_Kerr}(b), we prepared a coherent state with amplitude $|\alpha| = 3$ and immediately measured its Q function.
As expected for a coherent state, the resulting Q function exhibits a Gaussian-shaped peak centered at $|\alpha| = 3$ in phase space.
However, when we waited for \SI{5}{\micro\second} before measuring the Q function, the state evolved under the Kerr interaction, leading to a shearing of the Q function distribution, as shown in Fig.~\ref{fig:supplement_Q_function_Kerr}(c).
Photon loss during the evolution caused the Q function peak to shift inward from $|\alpha| = 3$.
We simulated the evolution using a master equation with the system parameters listed in Table~\ref{tab:system_parameter}, which successfully reproduces the data (Fig.~\ref{fig:supplement_Q_function_Kerr}(d)).
The consistency between simulation and measurement provides additional validation of the Kerr nonlinearity extracted from the cavity Ramsey data.

\section{Wigner tomography and SNAP}\label{supplement_Q_function}

\subsection{Power Rabi calibration}

\begin{figure}
    \centering
    \includegraphics{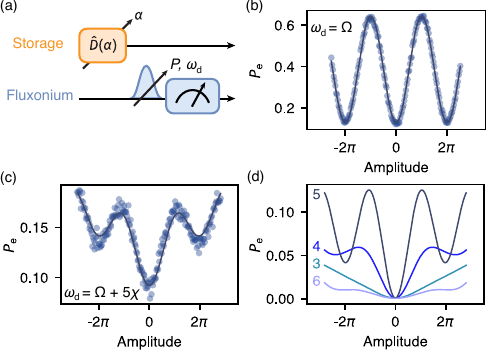}
    \caption{
    Photon-number selective power Rabi calibration. 
    (a) Pulse sequence for the Rabi calibration. A Gaussian pulse of duration $\SI{1.6}{\micro\second}$ was used for the qubit $\pi$ pulse. 
    (b) Power Rabi calibrated at the $\ket{0}$ peak. The data was fit using a cosine function
    (c) Power Rabi calibrated at the $\ket{5}$ peak. Due to a combination of photon decay, qubit dephasing, and imperfect selectivity of the drive pulse, the Rabi oscillation curves upward. The data was fit using a master equation simulation.
    (d) Simulation of the time evolution of each Fock state under the drive conditions used in (c).
    }
    \label{fig:supplement_wigner_power_rabi_calibration}
\end{figure}

This section outlines the power Rabi calibration procedure used for the Wigner tomography measurement.
We performed Wigner tomography of the storage resonator by applying parity-selective measurements, where we drove the fluxonium at frequencies corresponding to even and odd photon number peaks, up to the 10th peak (Fig~\ref{fig4:SNAP}(a)). 
The difference between the measured qubit excitation probabilities after a prior resonator displacement yields the Wigner function.
The drives can be described by 
\begin{equation}
    \hat{H}_\mathrm{drive,rot}/\hbar 
    = \sum_{\substack{n = \text{even,} \\ \text{odd}}} 
    \frac{\pi}{2T} e^{-in\chi t} \ket{e}\bra{g} + \text{h.c.}
\end{equation}
where $T$ is the duration of the Rabi pulse.
We therefore need to perform a Rabi amplitude calibration at each photon-number-resolved peak by preparing a coherent state in the storage resonator and driving the fluxonium at the corresponding transition frequency (Fig.~\ref{fig:supplement_wigner_power_rabi_calibration}(a)). 
For the Wigner measurements, we used \SI{1.6}{\micro\second} multiplexed Gaussian pulses to selectively address each peak. 
In Fig.~\ref{fig:supplement_wigner_power_rabi_calibration}(b), we show the Rabi calibration performed at the 0$^\text{th}$ peak, which exhibits a well-defined Rabi oscillation.
However, for higher photon-number peaks, the calibration signal deviates from ideal Rabi oscillations and instead displays an overall rising trend with oscillatory features (Fig.~\ref{fig:supplement_wigner_power_rabi_calibration}(c)).
We fitted the measured data using the master equation and extracted the qubit population associated with the neighboring photon-number states (Fig.~\ref{fig:supplement_wigner_power_rabi_calibration}(d)).
The significant population in the $\ket{3}$ and $\ket{4}$ states during the pulse arises from the fact that the storage resonator undergoes substantial photon loss due to its limited relaxation time.
The residual oscillations observed in the $\ket{4}$ and $\ket{6}$ populations suggest imperfect selectivity of the Rabi pulse, leading to weak off-resonant excitation of neighboring photon-number states.
However, we could not use a longer Rabi pulse to improve selectivity due to the strong damping of the storage resonator.
Consequently, the Wigner function we reconstruct deviates from the actual quantum state of the resonator. 
However, this deviation is well captured by our master equation simulation, as evidenced by the agreement between our simulation and experimental data in Fig.~\ref{fig4:SNAP}(c) and (d).

\subsection{SNAP simulation and fidelity} \label{supplement_SNAP}

\begin{figure}
    \centering
    \includegraphics{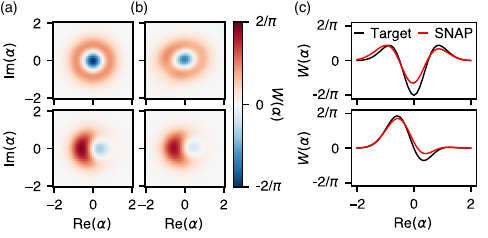}
    \caption{
    Wigner function comparison.
    (a) Target Wigner functions for state $\ket{1}$ (top) and $\frac{1}{\sqrt{2}}(\ket{0}-\ket{1})$ (bottom).  
    (b) Simulated Wigner functions produced by the SNAP gate, assuming ideal Wigner measurement.
    (c) Cuts of the target and simulated Wigner functions along $\mathrm{Im}(\alpha)=0$. 
    }
    \label{fig:supplement_wigner_comparison}
\end{figure}

\begin{table}[t]
    \centering
    \def\arraystretch{1.4} 
    \setlength{\tabcolsep}{12pt} 
    \begin{tabular}{c|c|c}
    \hline\hline
    Loss mechanism & $\ket{1}$& $\ket{0} - \ket{1}$\\
    \hline
    Intrinsic limit & $\sim\!1.9\%$ & $\sim\!0.05\%$ \\
    Qubit initialization & $\sim\!7.2\%$ & $\sim\!4.7\%$ \\
    Storage $T_1$ decay & $\sim\!6.6\%$ & $\sim\!1.6\%$ \\
    Qubit depolarization & $\sim\!0.5\%$ & $\sim\!0.3\%$ \\
    Qubit dephasing & $\sim\!1.7\%$ & $\sim\!1.4\%$ \\
    Coherent errors & $\sim\!2.9\%$ & $\sim\!0.7\%$ \\
    \hline\hline
    \end{tabular}
    \caption{
    Breakdown of the loss contributions for the initialization of $\ket{1}$ and $\ket{0} - \ket{1})/\sqrt{2}$ using SNAP.}
    \label{tab:SNAP_loss_breakdown}
\end{table}

Since the state preparation fidelity cannot be directly obtained from the overlap of the reconstructed Wigner function, we simulated the full SNAP process to evaluate the fidelity. 
The SNAP process was implemented by applying photon-number-selective $\pi$ pulses on the fluxonium, followed by a fast unconditional $\pi$ pulse. 
We used \SI{1.6}{\micro\second} multiplexed Gaussian pulses for the selective pulses and a short \SI{40}{\nano\second} Gaussian pulse for the fast unconditional $\pi$ pulse.
The selective pulses were numerically optimized to suppress coherent errors, following the algorithm described in Ref.~\cite{landgraf_fast_2024}.
The drive pulses are described by 
\begin{gather}
    \hat{H}_\mathrm{slow}/\hbar = \frac{\pi}{2T_\mathrm{slow}} \sum_{n} e^{i(\omega_n t + \alpha_n - \Delta\omega_n T_\mathrm{slow}/2)} \ket{e}\bra{g} + \text{h.c.}, \\
    \hat{H}_\mathrm{fast}/\hbar = \frac{\pi}{2T_\mathrm{fast}} \ket{e}\bra{g} + \text{h.c.}
\end{gather}
where $T_\mathrm{slow}$ and $T_\mathrm{fast}$ denote the durations of the slow and fast Rabi pulses, respectively.
$\lambda_n,\omega_n$ and $\alpha_n$ are the optimized Fock-state-dependent amplitudes, frequencies, and phases of the selective drives, with $\Delta\omega_n=\omega_n+n\chi$.
In Fig.~\ref{fig:supplement_wigner_comparison}, we show the Wigner functions of the ideal target state and the output state resulting from the simulated SNAP process, along with a center cut at $\mathrm{Im}(\alpha) = 0$.
From the overlap of the Wigner functions, we extract a state preparation fidelity of 79\% for the Fock state $\ket{1}$ and 91\% for the superposition state $(\ket{0} - \ket{1})/\sqrt{2}$.
The lower preparation fidelity of the Fock state $\ket{1}$ arises from the larger displacement required to initialize the state, which results in increased photon loss during the SNAP sequence.

A more detailed breakdown of the loss mechanisms during the SNAP sequence is provided in Table~\ref{tab:SNAP_loss_breakdown}.
The intrinsic limit refers to the minimum infidelity achievable with our chosen displacement and SNAP parameters.
To compute the contribution of each loss channel, we simulated the master equation while adding one loss channel at a time.
The resulting contributions were then weighted and summed to account for the loss in addition to the intrinsic limit, using the following equation
\begin{equation}
    \epsilon_{i,\mathrm{weighted}} = \epsilon_i\cdot\frac{\epsilon_\mathrm{total}-\epsilon_\mathrm{intrinsic}}{\sum_i (\epsilon_i)}.
\end{equation}
Here, \(\epsilon_{i,\mathrm{weighted}}\) is the final weighted error reported for loss channel \(i\) in Table~\ref{tab:SNAP_loss_breakdown}.
\(\epsilon_i\) denotes the unweighted error contribution from a specific loss channel, \(\epsilon_\mathrm{total}\) is the total simulated preparation error, and \(\epsilon_\mathrm{intrinsic}\) is the intrinsic error associated with the finite fidelity of the chosen SNAP parameters.
This rescaling accounts for the fact that the different loss channels are not strictly independent and ensures that the weighted sum of individual contributions matches the total observed additional error beyond the intrinsic limit.
The coherent errors in the table are primarily attributed to phase accumulation during the nonselective $\pi$ pulse, which arises because the pulse is not resonant with all spectral peaks.
However, the coherent errors could potentially be corrected with improved optimization algorithms or shorter unselective $\pi$ pulse.

Based on Table~\ref{tab:SNAP_loss_breakdown}, aside from qubit initialization, the main sources of incoherent loss are storage $T_1$ decay and qubit $T_2$ dephasing during the selective $\pi$ pulse.
Errors from qubit initialization could be addressed with improved active reset methods, whereas mitigating incoherent loss requires better-cohering qubit and resonator systems.
To estimate the extent to which improved system coherence suppresses incoherent errors, we simulated the incoherent loss as a function of storage $T_1$ and qubit $T_2$, as shown in Fig.~\ref{fig4:SNAP}(f).
Since the unselective pulse is so short that it incurs essentially no incoherent loss, we only simulated the loss during the selective $\pi$ pulse.
To simplify the simulation, we used representative parameters consistent with the experiment, choosing $\vec{\theta} = (\pi, 0, 0, ...)$ and $\alpha = 1$.
We also used pulse parameters identical to those in the experiment.
We ran the master equation simulation while sweeping over the coherence times, and subtract out coherent errors to obtain the incoherent loss contribution.

\vspace{5mm}


\begin{thebibliography}{62}%
\makeatletter
\providecommand \@ifxundefined [1]{%
 \@ifx{#1\undefined}
}%
\providecommand \@ifnum [1]{%
 \ifnum #1\expandafter \@firstoftwo
 \else \expandafter \@secondoftwo
 \fi
}%
\providecommand \@ifx [1]{%
 \ifx #1\expandafter \@firstoftwo
 \else \expandafter \@secondoftwo
 \fi
}%
\providecommand \natexlab [1]{#1}%
\providecommand \enquote  [1]{``#1''}%
\providecommand \bibnamefont  [1]{#1}%
\providecommand \bibfnamefont [1]{#1}%
\providecommand \citenamefont [1]{#1}%
\providecommand \href@noop [0]{\@secondoftwo}%
\providecommand \href [0]{\begingroup \@sanitize@url \@href}%
\providecommand \@href[1]{\@@startlink{#1}\@@href}%
\providecommand \@@href[1]{\endgroup#1\@@endlink}%
\providecommand \@sanitize@url [0]{\catcode `\\12\catcode `\$12\catcode `\&12\catcode `\#12\catcode `\^12\catcode `\_12\catcode `\%12\relax}%
\providecommand \@@startlink[1]{}%
\providecommand \@@endlink[0]{}%
\providecommand \url  [0]{\begingroup\@sanitize@url \@url }%
\providecommand \@url [1]{\endgroup\@href {#1}{\urlprefix }}%
\providecommand \urlprefix  [0]{URL }%
\providecommand \Eprint [0]{\href }%
\providecommand \doibase [0]{https://doi.org/}%
\providecommand \selectlanguage [0]{\@gobble}%
\providecommand \bibinfo  [0]{\@secondoftwo}%
\providecommand \bibfield  [0]{\@secondoftwo}%
\providecommand \translation [1]{[#1]}%
\providecommand \BibitemOpen [0]{}%
\providecommand \bibitemStop [0]{}%
\providecommand \bibitemNoStop [0]{.\EOS\space}%
\providecommand \EOS [0]{\spacefactor3000\relax}%
\providecommand \BibitemShut  [1]{\csname bibitem#1\endcsname}%
\let\auto@bib@innerbib\@empty
\bibitem [{\citenamefont {Vitali}\ \emph {et~al.}(1998)\citenamefont {Vitali}, \citenamefont {Tombesi},\ and\ \citenamefont {Milburn}}]{vitali_quantumstate_1998}%
  \BibitemOpen
  \bibfield  {author} {\bibinfo {author} {\bibfnamefont {D.}~\bibnamefont {Vitali}}, \bibinfo {author} {\bibfnamefont {P.}~\bibnamefont {Tombesi}},\ and\ \bibinfo {author} {\bibfnamefont {G.~J.}\ \bibnamefont {Milburn}},\ }\bibfield  {title} {\bibinfo {title} {Quantum-state protection in cavities},\ }\href {https://doi.org/10.1103/PhysRevA.57.4930} {\bibfield  {journal} {\bibinfo  {journal} {Physical Review A}\ }\textbf {\bibinfo {volume} {57}},\ \bibinfo {pages} {4930} (\bibinfo {year} {1998})}\BibitemShut {NoStop}%
\bibitem [{\citenamefont {Gottesman}\ \emph {et~al.}(2001)\citenamefont {Gottesman}, \citenamefont {Kitaev},\ and\ \citenamefont {Preskill}}]{gottesman_encoding_2001}%
  \BibitemOpen
  \bibfield  {author} {\bibinfo {author} {\bibfnamefont {D.}~\bibnamefont {Gottesman}}, \bibinfo {author} {\bibfnamefont {A.}~\bibnamefont {Kitaev}},\ and\ \bibinfo {author} {\bibfnamefont {J.}~\bibnamefont {Preskill}},\ }\bibfield  {title} {\bibinfo {title} {Encoding a qubit in an oscillator},\ }\href {https://doi.org/10.1103/PhysRevA.64.012310} {\bibfield  {journal} {\bibinfo  {journal} {Physical Review A}\ }\textbf {\bibinfo {volume} {64}},\ \bibinfo {pages} {012310} (\bibinfo {year} {2001})}\BibitemShut {NoStop}%
\bibitem [{\citenamefont {Zippilli}\ \emph {et~al.}(2003)\citenamefont {Zippilli}, \citenamefont {Vitali}, \citenamefont {Tombesi},\ and\ \citenamefont {Raimond}}]{zippilli_scheme_2003}%
  \BibitemOpen
  \bibfield  {author} {\bibinfo {author} {\bibfnamefont {S.}~\bibnamefont {Zippilli}}, \bibinfo {author} {\bibfnamefont {D.}~\bibnamefont {Vitali}}, \bibinfo {author} {\bibfnamefont {P.}~\bibnamefont {Tombesi}},\ and\ \bibinfo {author} {\bibfnamefont {J.-M.}\ \bibnamefont {Raimond}},\ }\bibfield  {title} {\bibinfo {title} {Scheme for decoherence control in microwave cavities},\ }\href {https://doi.org/10.1103/PhysRevA.67.052101} {\bibfield  {journal} {\bibinfo  {journal} {Physical Review A}\ }\textbf {\bibinfo {volume} {67}},\ \bibinfo {pages} {052101} (\bibinfo {year} {2003})}\BibitemShut {NoStop}%
\bibitem [{\citenamefont {Michael}\ \emph {et~al.}(2016)\citenamefont {Michael}, \citenamefont {Silveri}, \citenamefont {Brierley}, \citenamefont {Albert}, \citenamefont {Salmilehto}, \citenamefont {Jiang},\ and\ \citenamefont {Girvin}}]{michael_new_2016}%
  \BibitemOpen
  \bibfield  {author} {\bibinfo {author} {\bibfnamefont {M.~H.}\ \bibnamefont {Michael}}, \bibinfo {author} {\bibfnamefont {M.}~\bibnamefont {Silveri}}, \bibinfo {author} {\bibfnamefont {R.~T.}\ \bibnamefont {Brierley}}, \bibinfo {author} {\bibfnamefont {V.~V.}\ \bibnamefont {Albert}}, \bibinfo {author} {\bibfnamefont {J.}~\bibnamefont {Salmilehto}}, \bibinfo {author} {\bibfnamefont {L.}~\bibnamefont {Jiang}},\ and\ \bibinfo {author} {\bibfnamefont {S.~M.}\ \bibnamefont {Girvin}},\ }\bibfield  {title} {\bibinfo {title} {New {{Class}} of {{Quantum Error-Correcting Codes}} for a {{Bosonic Mode}}},\ }\href {https://doi.org/10.1103/PhysRevX.6.031006} {\bibfield  {journal} {\bibinfo  {journal} {Physical Review X}\ }\textbf {\bibinfo {volume} {6}},\ \bibinfo {pages} {031006} (\bibinfo {year} {2016})}\BibitemShut {NoStop}%
\bibitem [{\citenamefont {Cai}\ \emph {et~al.}(2021)\citenamefont {Cai}, \citenamefont {Ma}, \citenamefont {Wang}, \citenamefont {Zou},\ and\ \citenamefont {Sun}}]{cai_bosonic_2021}%
  \BibitemOpen
  \bibfield  {author} {\bibinfo {author} {\bibfnamefont {W.}~\bibnamefont {Cai}}, \bibinfo {author} {\bibfnamefont {Y.}~\bibnamefont {Ma}}, \bibinfo {author} {\bibfnamefont {W.}~\bibnamefont {Wang}}, \bibinfo {author} {\bibfnamefont {C.-L.}\ \bibnamefont {Zou}},\ and\ \bibinfo {author} {\bibfnamefont {L.}~\bibnamefont {Sun}},\ }\bibfield  {title} {\bibinfo {title} {Bosonic quantum error correction codes in superconducting quantum circuits},\ }\href {https://doi.org/10.1016/j.fmre.2020.12.006} {\bibfield  {journal} {\bibinfo  {journal} {Fundamental Research}\ }\textbf {\bibinfo {volume} {1}},\ \bibinfo {pages} {50} (\bibinfo {year} {2021})}\BibitemShut {NoStop}%
\bibitem [{\citenamefont {Ma}\ \emph {et~al.}(2021)\citenamefont {Ma}, \citenamefont {Puri}, \citenamefont {Schoelkopf}, \citenamefont {Devoret}, \citenamefont {Girvin},\ and\ \citenamefont {Jiang}}]{ma_quantum_2021}%
  \BibitemOpen
  \bibfield  {author} {\bibinfo {author} {\bibfnamefont {W.-L.}\ \bibnamefont {Ma}}, \bibinfo {author} {\bibfnamefont {S.}~\bibnamefont {Puri}}, \bibinfo {author} {\bibfnamefont {R.~J.}\ \bibnamefont {Schoelkopf}}, \bibinfo {author} {\bibfnamefont {M.~H.}\ \bibnamefont {Devoret}}, \bibinfo {author} {\bibfnamefont {S.}~\bibnamefont {Girvin}},\ and\ \bibinfo {author} {\bibfnamefont {L.}~\bibnamefont {Jiang}},\ }\bibfield  {title} {\bibinfo {title} {Quantum control of bosonic modes with superconducting circuits},\ }\href {https://doi.org/10.1016/j.scib.2021.05.024} {\bibfield  {journal} {\bibinfo  {journal} {Science Bulletin}\ }\textbf {\bibinfo {volume} {66}},\ \bibinfo {pages} {1789} (\bibinfo {year} {2021})}\BibitemShut {NoStop}%
\bibitem [{\citenamefont {Reagor}\ \emph {et~al.}(2016)\citenamefont {Reagor}, \citenamefont {Pfaff}, \citenamefont {Axline}, \citenamefont {Heeres}, \citenamefont {Ofek}, \citenamefont {Sliwa}, \citenamefont {Holland}, \citenamefont {Wang}, \citenamefont {Blumoff}, \citenamefont {Chou}, \citenamefont {Hatridge}, \citenamefont {Frunzio}, \citenamefont {Devoret}, \citenamefont {Jiang},\ and\ \citenamefont {Schoelkopf}}]{reagor_quantum_2016}%
  \BibitemOpen
  \bibfield  {author} {\bibinfo {author} {\bibfnamefont {M.}~\bibnamefont {Reagor}}, \bibinfo {author} {\bibfnamefont {W.}~\bibnamefont {Pfaff}}, \bibinfo {author} {\bibfnamefont {C.}~\bibnamefont {Axline}}, \bibinfo {author} {\bibfnamefont {R.~W.}\ \bibnamefont {Heeres}}, \bibinfo {author} {\bibfnamefont {N.}~\bibnamefont {Ofek}}, \bibinfo {author} {\bibfnamefont {K.}~\bibnamefont {Sliwa}}, \bibinfo {author} {\bibfnamefont {E.}~\bibnamefont {Holland}}, \bibinfo {author} {\bibfnamefont {C.}~\bibnamefont {Wang}}, \bibinfo {author} {\bibfnamefont {J.}~\bibnamefont {Blumoff}}, \bibinfo {author} {\bibfnamefont {K.}~\bibnamefont {Chou}}, \bibinfo {author} {\bibfnamefont {M.~J.}\ \bibnamefont {Hatridge}}, \bibinfo {author} {\bibfnamefont {L.}~\bibnamefont {Frunzio}}, \bibinfo {author} {\bibfnamefont {M.~H.}\ \bibnamefont {Devoret}}, \bibinfo {author} {\bibfnamefont {L.}~\bibnamefont {Jiang}},\ and\ \bibinfo {author} {\bibfnamefont {R.~J.}\ \bibnamefont {Schoelkopf}},\ }\bibfield  {title} {\bibinfo {title} {Quantum memory with millisecond coherence in circuit {{QED}}},\ }\href {https://doi.org/10.1103/PhysRevB.94.014506} {\bibfield  {journal} {\bibinfo  {journal} {Physical Review B}\ }\textbf {\bibinfo {volume} {94}},\ \bibinfo {pages} {014506} (\bibinfo {year} {2016})}\BibitemShut {NoStop}%
\bibitem [{\citenamefont {Chakram}\ \emph {et~al.}(2021)\citenamefont {Chakram}, \citenamefont {Oriani}, \citenamefont {Naik}, \citenamefont {Dixit}, \citenamefont {He}, \citenamefont {Agrawal}, \citenamefont {Kwon},\ and\ \citenamefont {Schuster}}]{chakram_seamless_2021}%
  \BibitemOpen
  \bibfield  {author} {\bibinfo {author} {\bibfnamefont {S.}~\bibnamefont {Chakram}}, \bibinfo {author} {\bibfnamefont {A.~E.}\ \bibnamefont {Oriani}}, \bibinfo {author} {\bibfnamefont {R.~K.}\ \bibnamefont {Naik}}, \bibinfo {author} {\bibfnamefont {A.~V.}\ \bibnamefont {Dixit}}, \bibinfo {author} {\bibfnamefont {K.}~\bibnamefont {He}}, \bibinfo {author} {\bibfnamefont {A.}~\bibnamefont {Agrawal}}, \bibinfo {author} {\bibfnamefont {H.}~\bibnamefont {Kwon}},\ and\ \bibinfo {author} {\bibfnamefont {D.~I.}\ \bibnamefont {Schuster}},\ }\bibfield  {title} {\bibinfo {title} {Seamless {{High- Q Microwave Cavities}} for {{Multimode Circuit Quantum Electrodynamics}}},\ }\href {https://doi.org/10.1103/PhysRevLett.127.107701} {\bibfield  {journal} {\bibinfo  {journal} {Physical Review Letters}\ }\textbf {\bibinfo {volume} {127}},\ \bibinfo {pages} {107701} (\bibinfo {year} {2021})}\BibitemShut {NoStop}%
\bibitem [{\citenamefont {Milul}\ \emph {et~al.}(2023)\citenamefont {Milul}, \citenamefont {Guttel}, \citenamefont {Goldblatt}, \citenamefont {Hazanov}, \citenamefont {Joshi}, \citenamefont {Chausovsky}, \citenamefont {Kahn}, \citenamefont {{\c C}ifty{\"u}rek}, \citenamefont {Lafont},\ and\ \citenamefont {Rosenblum}}]{milul_superconducting_2023}%
  \BibitemOpen
  \bibfield  {author} {\bibinfo {author} {\bibfnamefont {O.}~\bibnamefont {Milul}}, \bibinfo {author} {\bibfnamefont {B.}~\bibnamefont {Guttel}}, \bibinfo {author} {\bibfnamefont {U.}~\bibnamefont {Goldblatt}}, \bibinfo {author} {\bibfnamefont {S.}~\bibnamefont {Hazanov}}, \bibinfo {author} {\bibfnamefont {L.~M.}\ \bibnamefont {Joshi}}, \bibinfo {author} {\bibfnamefont {D.}~\bibnamefont {Chausovsky}}, \bibinfo {author} {\bibfnamefont {N.}~\bibnamefont {Kahn}}, \bibinfo {author} {\bibfnamefont {E.}~\bibnamefont {{\c C}ifty{\"u}rek}}, \bibinfo {author} {\bibfnamefont {F.}~\bibnamefont {Lafont}},\ and\ \bibinfo {author} {\bibfnamefont {S.}~\bibnamefont {Rosenblum}},\ }\bibfield  {title} {\bibinfo {title} {Superconducting {{Cavity Qubit}} with {{Tens}} of {{Milliseconds Single-Photon Coherence Time}}},\ }\href {https://doi.org/10.1103/PRXQuantum.4.030336} {\bibfield  {journal} {\bibinfo  {journal} {PRX Quantum}\ }\textbf {\bibinfo {volume} {4}},\ \bibinfo {pages} {030336} (\bibinfo {year} {2023})}\BibitemShut {NoStop}%
\bibitem [{\citenamefont {Reagor}\ \emph {et~al.}(2013)\citenamefont {Reagor}, \citenamefont {Paik}, \citenamefont {Catelani}, \citenamefont {Sun}, \citenamefont {Axline}, \citenamefont {Holland}, \citenamefont {Pop}, \citenamefont {Masluk}, \citenamefont {Brecht}, \citenamefont {Frunzio}, \citenamefont {Devoret}, \citenamefont {Glazman},\ and\ \citenamefont {Schoelkopf}}]{reagorReaching10Ms2013}%
  \BibitemOpen
  \bibfield  {author} {\bibinfo {author} {\bibfnamefont {M.}~\bibnamefont {Reagor}}, \bibinfo {author} {\bibfnamefont {H.}~\bibnamefont {Paik}}, \bibinfo {author} {\bibfnamefont {G.}~\bibnamefont {Catelani}}, \bibinfo {author} {\bibfnamefont {L.}~\bibnamefont {Sun}}, \bibinfo {author} {\bibfnamefont {C.}~\bibnamefont {Axline}}, \bibinfo {author} {\bibfnamefont {E.}~\bibnamefont {Holland}}, \bibinfo {author} {\bibfnamefont {I.~M.}\ \bibnamefont {Pop}}, \bibinfo {author} {\bibfnamefont {N.~A.}\ \bibnamefont {Masluk}}, \bibinfo {author} {\bibfnamefont {T.}~\bibnamefont {Brecht}}, \bibinfo {author} {\bibfnamefont {L.}~\bibnamefont {Frunzio}}, \bibinfo {author} {\bibfnamefont {M.~H.}\ \bibnamefont {Devoret}}, \bibinfo {author} {\bibfnamefont {L.}~\bibnamefont {Glazman}},\ and\ \bibinfo {author} {\bibfnamefont {R.~J.}\ \bibnamefont {Schoelkopf}},\ }\bibfield  {title} {\bibinfo {title} {Reaching 10 ms single photon lifetimes for superconducting aluminum cavities},\ }\href {https://doi.org/10.1063/1.4807015} {\bibfield  {journal} {\bibinfo  {journal} {Applied Physics Letters}\ }\textbf {\bibinfo {volume} {102}},\ \bibinfo {pages} {192604} (\bibinfo {year} {2013})}\BibitemShut {NoStop}%
\bibitem [{\citenamefont {Rosenblum}\ \emph {et~al.}(2018)\citenamefont {Rosenblum}, \citenamefont {Reinhold}, \citenamefont {Mirrahimi}, \citenamefont {Jiang}, \citenamefont {Frunzio},\ and\ \citenamefont {Schoelkopf}}]{rosenblum_faulttolerant_2018}%
  \BibitemOpen
  \bibfield  {author} {\bibinfo {author} {\bibfnamefont {S.}~\bibnamefont {Rosenblum}}, \bibinfo {author} {\bibfnamefont {P.}~\bibnamefont {Reinhold}}, \bibinfo {author} {\bibfnamefont {M.}~\bibnamefont {Mirrahimi}}, \bibinfo {author} {\bibfnamefont {L.}~\bibnamefont {Jiang}}, \bibinfo {author} {\bibfnamefont {L.}~\bibnamefont {Frunzio}},\ and\ \bibinfo {author} {\bibfnamefont {R.~J.}\ \bibnamefont {Schoelkopf}},\ }\bibfield  {title} {\bibinfo {title} {Fault-tolerant detection of a quantum error},\ }\href {https://doi.org/10.1126/science.aat3996} {\bibfield  {journal} {\bibinfo  {journal} {Science}\ }\textbf {\bibinfo {volume} {361}},\ \bibinfo {pages} {266} (\bibinfo {year} {2018})}\BibitemShut {NoStop}%
\bibitem [{\citenamefont {Huang}\ \emph {et~al.}(2025)\citenamefont {Huang}, \citenamefont {DiNapoli}, \citenamefont {Rockwood}, \citenamefont {Yuan}, \citenamefont {Narasimhan}, \citenamefont {Gupta}, \citenamefont {Bal}, \citenamefont {Crisa}, \citenamefont {Garattoni}, \citenamefont {Lu}, \citenamefont {Jiang},\ and\ \citenamefont {Chakram}}]{huang_fast_2025}%
  \BibitemOpen
  \bibfield  {author} {\bibinfo {author} {\bibfnamefont {J.}~\bibnamefont {Huang}}, \bibinfo {author} {\bibfnamefont {T.~J.}\ \bibnamefont {DiNapoli}}, \bibinfo {author} {\bibfnamefont {G.}~\bibnamefont {Rockwood}}, \bibinfo {author} {\bibfnamefont {M.}~\bibnamefont {Yuan}}, \bibinfo {author} {\bibfnamefont {P.}~\bibnamefont {Narasimhan}}, \bibinfo {author} {\bibfnamefont {E.}~\bibnamefont {Gupta}}, \bibinfo {author} {\bibfnamefont {M.}~\bibnamefont {Bal}}, \bibinfo {author} {\bibfnamefont {F.}~\bibnamefont {Crisa}}, \bibinfo {author} {\bibfnamefont {S.}~\bibnamefont {Garattoni}}, \bibinfo {author} {\bibfnamefont {Y.}~\bibnamefont {Lu}}, \bibinfo {author} {\bibfnamefont {L.}~\bibnamefont {Jiang}},\ and\ \bibinfo {author} {\bibfnamefont {S.}~\bibnamefont {Chakram}},\ }\href@noop {} {\bibinfo {title} {Fast {{Sideband Control}} of a {{Weakly Coupled Multimode Bosonic Memory}}}} (\bibinfo {year} {2025}),\ \Eprint {https://arxiv.org/abs/2503.10623} {arXiv:2503.10623 [quant-ph]} \BibitemShut {NoStop}%
\bibitem [{\citenamefont {Law}\ and\ \citenamefont {Eberly}(1996)}]{lawArbitraryControlQuantum1996}%
  \BibitemOpen
  \bibfield  {author} {\bibinfo {author} {\bibfnamefont {C.~K.}\ \bibnamefont {Law}}\ and\ \bibinfo {author} {\bibfnamefont {J.~H.}\ \bibnamefont {Eberly}},\ }\bibfield  {title} {\bibinfo {title} {Arbitrary {{Control}} of a {{Quantum Electromagnetic Field}}},\ }\href {https://doi.org/10.1103/PhysRevLett.76.1055} {\bibfield  {journal} {\bibinfo  {journal} {Physical Review Letters}\ }\textbf {\bibinfo {volume} {76}},\ \bibinfo {pages} {1055} (\bibinfo {year} {1996})}\BibitemShut {NoStop}%
\bibitem [{\citenamefont {Lloyd}\ and\ \citenamefont {Braunstein}(1999)}]{lloyd_quantum_1999}%
  \BibitemOpen
  \bibfield  {author} {\bibinfo {author} {\bibfnamefont {S.}~\bibnamefont {Lloyd}}\ and\ \bibinfo {author} {\bibfnamefont {S.~L.}\ \bibnamefont {Braunstein}},\ }\bibfield  {title} {\bibinfo {title} {Quantum {{Computation}} over {{Continuous Variables}}},\ }\href {https://doi.org/10.1103/PhysRevLett.82.1784} {\bibfield  {journal} {\bibinfo  {journal} {Physical Review Letters}\ }\textbf {\bibinfo {volume} {82}},\ \bibinfo {pages} {1784} (\bibinfo {year} {1999})}\BibitemShut {NoStop}%
\bibitem [{\citenamefont {Hofheinz}\ \emph {et~al.}(2009)\citenamefont {Hofheinz}, \citenamefont {Wang}, \citenamefont {Ansmann}, \citenamefont {Bialczak}, \citenamefont {Lucero}, \citenamefont {Neeley}, \citenamefont {O'Connell}, \citenamefont {Sank}, \citenamefont {Wenner}, \citenamefont {Martinis},\ and\ \citenamefont {Cleland}}]{hofheinz_synthesizing_2009}%
  \BibitemOpen
  \bibfield  {author} {\bibinfo {author} {\bibfnamefont {M.}~\bibnamefont {Hofheinz}}, \bibinfo {author} {\bibfnamefont {H.}~\bibnamefont {Wang}}, \bibinfo {author} {\bibfnamefont {M.}~\bibnamefont {Ansmann}}, \bibinfo {author} {\bibfnamefont {R.~C.}\ \bibnamefont {Bialczak}}, \bibinfo {author} {\bibfnamefont {E.}~\bibnamefont {Lucero}}, \bibinfo {author} {\bibfnamefont {M.}~\bibnamefont {Neeley}}, \bibinfo {author} {\bibfnamefont {A.~D.}\ \bibnamefont {O'Connell}}, \bibinfo {author} {\bibfnamefont {D.}~\bibnamefont {Sank}}, \bibinfo {author} {\bibfnamefont {J.}~\bibnamefont {Wenner}}, \bibinfo {author} {\bibfnamefont {J.~M.}\ \bibnamefont {Martinis}},\ and\ \bibinfo {author} {\bibfnamefont {A.~N.}\ \bibnamefont {Cleland}},\ }\bibfield  {title} {\bibinfo {title} {Synthesizing arbitrary quantum states in a superconducting resonator},\ }\href {https://doi.org/10.1038/nature08005} {\bibfield  {journal} {\bibinfo  {journal} {Nature}\ }\textbf {\bibinfo {volume} {459}},\ \bibinfo {pages} {546} (\bibinfo {year} {2009})}\BibitemShut {NoStop}%
\bibitem [{\citenamefont {Krastanov}\ \emph {et~al.}(2015)\citenamefont {Krastanov}, \citenamefont {Albert}, \citenamefont {Shen}, \citenamefont {Zou}, \citenamefont {Heeres}, \citenamefont {Vlastakis}, \citenamefont {Schoelkopf},\ and\ \citenamefont {Jiang}}]{krastanov_universal_2015}%
  \BibitemOpen
  \bibfield  {author} {\bibinfo {author} {\bibfnamefont {S.}~\bibnamefont {Krastanov}}, \bibinfo {author} {\bibfnamefont {V.~V.}\ \bibnamefont {Albert}}, \bibinfo {author} {\bibfnamefont {C.}~\bibnamefont {Shen}}, \bibinfo {author} {\bibfnamefont {C.-L.}\ \bibnamefont {Zou}}, \bibinfo {author} {\bibfnamefont {R.~W.}\ \bibnamefont {Heeres}}, \bibinfo {author} {\bibfnamefont {B.}~\bibnamefont {Vlastakis}}, \bibinfo {author} {\bibfnamefont {R.~J.}\ \bibnamefont {Schoelkopf}},\ and\ \bibinfo {author} {\bibfnamefont {L.}~\bibnamefont {Jiang}},\ }\bibfield  {title} {\bibinfo {title} {Universal control of an oscillator with dispersive coupling to a qubit},\ }\href {https://doi.org/10.1103/PhysRevA.92.040303} {\bibfield  {journal} {\bibinfo  {journal} {Physical Review A}\ }\textbf {\bibinfo {volume} {92}},\ \bibinfo {pages} {040303} (\bibinfo {year} {2015})}\BibitemShut {NoStop}%
\bibitem [{\citenamefont {Lutterbach}\ and\ \citenamefont {Davidovich}(1997)}]{lutterbach_method_1997}%
  \BibitemOpen
  \bibfield  {author} {\bibinfo {author} {\bibfnamefont {L.~G.}\ \bibnamefont {Lutterbach}}\ and\ \bibinfo {author} {\bibfnamefont {L.}~\bibnamefont {Davidovich}},\ }\bibfield  {title} {\bibinfo {title} {Method for {{Direct Measurement}} of the {{Wigner Function}} in {{Cavity QED}} and {{Ion Traps}}},\ }\href {https://doi.org/10.1103/PhysRevLett.78.2547} {\bibfield  {journal} {\bibinfo  {journal} {Physical Review Letters}\ }\textbf {\bibinfo {volume} {78}},\ \bibinfo {pages} {2547} (\bibinfo {year} {1997})}\BibitemShut {NoStop}%
\bibitem [{\citenamefont {Bertet}\ \emph {et~al.}(2002)\citenamefont {Bertet}, \citenamefont {Auffeves}, \citenamefont {Maioli}, \citenamefont {Osnaghi}, \citenamefont {Meunier}, \citenamefont {Brune}, \citenamefont {Raimond},\ and\ \citenamefont {Haroche}}]{bertet_direct_2002}%
  \BibitemOpen
  \bibfield  {author} {\bibinfo {author} {\bibfnamefont {P.}~\bibnamefont {Bertet}}, \bibinfo {author} {\bibfnamefont {A.}~\bibnamefont {Auffeves}}, \bibinfo {author} {\bibfnamefont {P.}~\bibnamefont {Maioli}}, \bibinfo {author} {\bibfnamefont {S.}~\bibnamefont {Osnaghi}}, \bibinfo {author} {\bibfnamefont {T.}~\bibnamefont {Meunier}}, \bibinfo {author} {\bibfnamefont {M.}~\bibnamefont {Brune}}, \bibinfo {author} {\bibfnamefont {J.~M.}\ \bibnamefont {Raimond}},\ and\ \bibinfo {author} {\bibfnamefont {S.}~\bibnamefont {Haroche}},\ }\bibfield  {title} {\bibinfo {title} {Direct {{Measurement}} of the {{Wigner Function}} of a {{One-Photon Fock State}} in a {{Cavity}}},\ }\href {https://doi.org/10.1103/PhysRevLett.89.200402} {\bibfield  {journal} {\bibinfo  {journal} {Physical Review Letters}\ }\textbf {\bibinfo {volume} {89}},\ \bibinfo {pages} {200402} (\bibinfo {year} {2002})}\BibitemShut {NoStop}%
\bibitem [{\citenamefont {Vlastakis}\ \emph {et~al.}(2013)\citenamefont {Vlastakis}, \citenamefont {Kirchmair}, \citenamefont {Leghtas}, \citenamefont {Nigg}, \citenamefont {Frunzio}, \citenamefont {Girvin}, \citenamefont {Mirrahimi}, \citenamefont {Devoret},\ and\ \citenamefont {Schoelkopf}}]{vlastakisDeterministicallyEncodingQuantum2013}%
  \BibitemOpen
  \bibfield  {author} {\bibinfo {author} {\bibfnamefont {B.}~\bibnamefont {Vlastakis}}, \bibinfo {author} {\bibfnamefont {G.}~\bibnamefont {Kirchmair}}, \bibinfo {author} {\bibfnamefont {Z.}~\bibnamefont {Leghtas}}, \bibinfo {author} {\bibfnamefont {S.~E.}\ \bibnamefont {Nigg}}, \bibinfo {author} {\bibfnamefont {L.}~\bibnamefont {Frunzio}}, \bibinfo {author} {\bibfnamefont {S.~M.}\ \bibnamefont {Girvin}}, \bibinfo {author} {\bibfnamefont {M.}~\bibnamefont {Mirrahimi}}, \bibinfo {author} {\bibfnamefont {M.~H.}\ \bibnamefont {Devoret}},\ and\ \bibinfo {author} {\bibfnamefont {R.~J.}\ \bibnamefont {Schoelkopf}},\ }\bibfield  {title} {\bibinfo {title} {Deterministically {{Encoding Quantum Information Using}} 100-{{Photon Schrodinger Cat States}}},\ }\href {https://doi.org/10.1126/science.1243289} {\bibfield  {journal} {\bibinfo  {journal} {Science}\ }\textbf {\bibinfo {volume} {342}},\ \bibinfo {pages} {607} (\bibinfo {year} {2013})}\BibitemShut {NoStop}%
\bibitem [{\citenamefont {Heeres}\ \emph {et~al.}(2017)\citenamefont {Heeres}, \citenamefont {Reinhold}, \citenamefont {Ofek}, \citenamefont {Frunzio}, \citenamefont {Jiang}, \citenamefont {Devoret},\ and\ \citenamefont {Schoelkopf}}]{heeres_implementing_2017}%
  \BibitemOpen
  \bibfield  {author} {\bibinfo {author} {\bibfnamefont {R.~W.}\ \bibnamefont {Heeres}}, \bibinfo {author} {\bibfnamefont {P.}~\bibnamefont {Reinhold}}, \bibinfo {author} {\bibfnamefont {N.}~\bibnamefont {Ofek}}, \bibinfo {author} {\bibfnamefont {L.}~\bibnamefont {Frunzio}}, \bibinfo {author} {\bibfnamefont {L.}~\bibnamefont {Jiang}}, \bibinfo {author} {\bibfnamefont {M.~H.}\ \bibnamefont {Devoret}},\ and\ \bibinfo {author} {\bibfnamefont {R.~J.}\ \bibnamefont {Schoelkopf}},\ }\bibfield  {title} {\bibinfo {title} {Implementing a universal gate set on a logical qubit encoded in an oscillator},\ }\href {https://doi.org/10.1038/s41467-017-00045-1} {\bibfield  {journal} {\bibinfo  {journal} {Nature Communications}\ }\textbf {\bibinfo {volume} {8}},\ \bibinfo {pages} {94} (\bibinfo {year} {2017})}\BibitemShut {NoStop}%
\bibitem [{\citenamefont {Chakram}\ \emph {et~al.}(2022)\citenamefont {Chakram}, \citenamefont {He}, \citenamefont {Dixit}, \citenamefont {Oriani}, \citenamefont {Naik}, \citenamefont {Leung}, \citenamefont {Kwon}, \citenamefont {Ma}, \citenamefont {Jiang},\ and\ \citenamefont {Schuster}}]{chakram_multimode_2022}%
  \BibitemOpen
  \bibfield  {author} {\bibinfo {author} {\bibfnamefont {S.}~\bibnamefont {Chakram}}, \bibinfo {author} {\bibfnamefont {K.}~\bibnamefont {He}}, \bibinfo {author} {\bibfnamefont {A.~V.}\ \bibnamefont {Dixit}}, \bibinfo {author} {\bibfnamefont {A.~E.}\ \bibnamefont {Oriani}}, \bibinfo {author} {\bibfnamefont {R.~K.}\ \bibnamefont {Naik}}, \bibinfo {author} {\bibfnamefont {N.}~\bibnamefont {Leung}}, \bibinfo {author} {\bibfnamefont {H.}~\bibnamefont {Kwon}}, \bibinfo {author} {\bibfnamefont {W.-L.}\ \bibnamefont {Ma}}, \bibinfo {author} {\bibfnamefont {L.}~\bibnamefont {Jiang}},\ and\ \bibinfo {author} {\bibfnamefont {D.~I.}\ \bibnamefont {Schuster}},\ }\bibfield  {title} {\bibinfo {title} {Multimode photon blockade},\ }\href {https://doi.org/10.1038/s41567-022-01630-y} {\bibfield  {journal} {\bibinfo  {journal} {Nature Physics}\ }\textbf {\bibinfo {volume} {18}},\ \bibinfo {pages} {879} (\bibinfo {year} {2022})}\BibitemShut {NoStop}%
\bibitem [{\citenamefont {Kirchmair}\ \emph {et~al.}(2013)\citenamefont {Kirchmair}, \citenamefont {Vlastakis}, \citenamefont {Leghtas}, \citenamefont {Nigg}, \citenamefont {Paik}, \citenamefont {Ginossar}, \citenamefont {Mirrahimi}, \citenamefont {Frunzio}, \citenamefont {Girvin},\ and\ \citenamefont {Schoelkopf}}]{kirchmairObservationQuantumState2013}%
  \BibitemOpen
  \bibfield  {author} {\bibinfo {author} {\bibfnamefont {G.}~\bibnamefont {Kirchmair}}, \bibinfo {author} {\bibfnamefont {B.}~\bibnamefont {Vlastakis}}, \bibinfo {author} {\bibfnamefont {Z.}~\bibnamefont {Leghtas}}, \bibinfo {author} {\bibfnamefont {S.~E.}\ \bibnamefont {Nigg}}, \bibinfo {author} {\bibfnamefont {H.}~\bibnamefont {Paik}}, \bibinfo {author} {\bibfnamefont {E.}~\bibnamefont {Ginossar}}, \bibinfo {author} {\bibfnamefont {M.}~\bibnamefont {Mirrahimi}}, \bibinfo {author} {\bibfnamefont {L.}~\bibnamefont {Frunzio}}, \bibinfo {author} {\bibfnamefont {S.~M.}\ \bibnamefont {Girvin}},\ and\ \bibinfo {author} {\bibfnamefont {R.~J.}\ \bibnamefont {Schoelkopf}},\ }\bibfield  {title} {\bibinfo {title} {Observation of quantum state collapse and revival due to the single-photon {{Kerr}} effect},\ }\href {https://doi.org/10.1038/nature11902} {\bibfield  {journal} {\bibinfo  {journal} {Nature}\ }\textbf {\bibinfo {volume} {495}},\ \bibinfo {pages} {205} (\bibinfo {year} {2013})}\BibitemShut {NoStop}%
\bibitem [{\citenamefont {Ofek}\ \emph {et~al.}(2016)\citenamefont {Ofek}, \citenamefont {Petrenko}, \citenamefont {Heeres}, \citenamefont {Reinhold}, \citenamefont {Leghtas}, \citenamefont {Vlastakis}, \citenamefont {Liu}, \citenamefont {Frunzio}, \citenamefont {Girvin}, \citenamefont {Jiang}, \citenamefont {Mirrahimi}, \citenamefont {Devoret},\ and\ \citenamefont {Schoelkopf}}]{Ofek2016}%
  \BibitemOpen
  \bibfield  {author} {\bibinfo {author} {\bibfnamefont {N.}~\bibnamefont {Ofek}}, \bibinfo {author} {\bibfnamefont {A.}~\bibnamefont {Petrenko}}, \bibinfo {author} {\bibfnamefont {R.}~\bibnamefont {Heeres}}, \bibinfo {author} {\bibfnamefont {P.}~\bibnamefont {Reinhold}}, \bibinfo {author} {\bibfnamefont {Z.}~\bibnamefont {Leghtas}}, \bibinfo {author} {\bibfnamefont {B.}~\bibnamefont {Vlastakis}}, \bibinfo {author} {\bibfnamefont {Y.}~\bibnamefont {Liu}}, \bibinfo {author} {\bibfnamefont {L.}~\bibnamefont {Frunzio}}, \bibinfo {author} {\bibfnamefont {S.~M.}\ \bibnamefont {Girvin}}, \bibinfo {author} {\bibfnamefont {L.}~\bibnamefont {Jiang}}, \bibinfo {author} {\bibfnamefont {M.}~\bibnamefont {Mirrahimi}}, \bibinfo {author} {\bibfnamefont {M.~H.}\ \bibnamefont {Devoret}},\ and\ \bibinfo {author} {\bibfnamefont {R.~J.}\ \bibnamefont {Schoelkopf}},\ }\bibfield  {title} {\bibinfo {title} {Extending the lifetime of a quantum bit with error correction in superconducting circuits},\ }\href {https://doi.org/10.1038/nature18949} {\bibfield  {journal} {\bibinfo  {journal} {Nature}\ }\textbf {\bibinfo {volume} {536}},\ \bibinfo {pages} {441} (\bibinfo {year} {2016})}\BibitemShut {NoStop}%
\bibitem [{\citenamefont {Gertler}\ \emph {et~al.}(2021)\citenamefont {Gertler}, \citenamefont {Baker}, \citenamefont {Li}, \citenamefont {Shirol}, \citenamefont {Koch},\ and\ \citenamefont {Wang}}]{Gertler2021}%
  \BibitemOpen
  \bibfield  {author} {\bibinfo {author} {\bibfnamefont {J.~M.}\ \bibnamefont {Gertler}}, \bibinfo {author} {\bibfnamefont {B.}~\bibnamefont {Baker}}, \bibinfo {author} {\bibfnamefont {J.}~\bibnamefont {Li}}, \bibinfo {author} {\bibfnamefont {S.}~\bibnamefont {Shirol}}, \bibinfo {author} {\bibfnamefont {J.}~\bibnamefont {Koch}},\ and\ \bibinfo {author} {\bibfnamefont {C.}~\bibnamefont {Wang}},\ }\bibfield  {title} {\bibinfo {title} {Protecting a bosonic qubit with autonomous quantum error correction},\ }\href {https://doi.org/10.1038/s41586-021-03257-0} {\bibfield  {journal} {\bibinfo  {journal} {Nature}\ }\textbf {\bibinfo {volume} {590}},\ \bibinfo {pages} {243} (\bibinfo {year} {2021})}\BibitemShut {NoStop}%
\bibitem [{\citenamefont {Sivak}\ \emph {et~al.}(2023)\citenamefont {Sivak}, \citenamefont {Eickbusch}, \citenamefont {Royer}, \citenamefont {Singh}, \citenamefont {Tsioutsios}, \citenamefont {Ganjam}, \citenamefont {Miano}, \citenamefont {Brock}, \citenamefont {Ding}, \citenamefont {Frunzio}, \citenamefont {Girvin}, \citenamefont {Schoelkopf},\ and\ \citenamefont {Devoret}}]{Sivak2023}%
  \BibitemOpen
  \bibfield  {author} {\bibinfo {author} {\bibfnamefont {V.~V.}\ \bibnamefont {Sivak}}, \bibinfo {author} {\bibfnamefont {A.}~\bibnamefont {Eickbusch}}, \bibinfo {author} {\bibfnamefont {B.}~\bibnamefont {Royer}}, \bibinfo {author} {\bibfnamefont {S.}~\bibnamefont {Singh}}, \bibinfo {author} {\bibfnamefont {I.}~\bibnamefont {Tsioutsios}}, \bibinfo {author} {\bibfnamefont {S.}~\bibnamefont {Ganjam}}, \bibinfo {author} {\bibfnamefont {A.}~\bibnamefont {Miano}}, \bibinfo {author} {\bibfnamefont {B.~L.}\ \bibnamefont {Brock}}, \bibinfo {author} {\bibfnamefont {A.~Z.}\ \bibnamefont {Ding}}, \bibinfo {author} {\bibfnamefont {L.}~\bibnamefont {Frunzio}}, \bibinfo {author} {\bibfnamefont {S.~M.}\ \bibnamefont {Girvin}}, \bibinfo {author} {\bibfnamefont {R.~J.}\ \bibnamefont {Schoelkopf}},\ and\ \bibinfo {author} {\bibfnamefont {M.~H.}\ \bibnamefont {Devoret}},\ }\bibfield  {title} {\bibinfo {title} {Real-time quantum error correction beyond break-even},\ }\href {https://doi.org/10.1038/s41586-023-05782-6} {\bibfield  {journal} {\bibinfo  {journal} {Nature}\ }\textbf {\bibinfo {volume} {616}},\ \bibinfo {pages} {50} (\bibinfo {year} {2023})}\BibitemShut {NoStop}%
\bibitem [{\citenamefont {Ni}\ \emph {et~al.}(2023)\citenamefont {Ni}, \citenamefont {Li}, \citenamefont {Deng}, \citenamefont {Cai}, \citenamefont {Zhang}, \citenamefont {Wang}, \citenamefont {Yang}, \citenamefont {Yu}, \citenamefont {Yan}, \citenamefont {Liu}, \citenamefont {Zou}, \citenamefont {Sun}, \citenamefont {Zheng}, \citenamefont {Xu},\ and\ \citenamefont {Yu}}]{ni_beating_2023}%
  \BibitemOpen
  \bibfield  {author} {\bibinfo {author} {\bibfnamefont {Z.}~\bibnamefont {Ni}}, \bibinfo {author} {\bibfnamefont {S.}~\bibnamefont {Li}}, \bibinfo {author} {\bibfnamefont {X.}~\bibnamefont {Deng}}, \bibinfo {author} {\bibfnamefont {Y.}~\bibnamefont {Cai}}, \bibinfo {author} {\bibfnamefont {L.}~\bibnamefont {Zhang}}, \bibinfo {author} {\bibfnamefont {W.}~\bibnamefont {Wang}}, \bibinfo {author} {\bibfnamefont {Z.-B.}\ \bibnamefont {Yang}}, \bibinfo {author} {\bibfnamefont {H.}~\bibnamefont {Yu}}, \bibinfo {author} {\bibfnamefont {F.}~\bibnamefont {Yan}}, \bibinfo {author} {\bibfnamefont {S.}~\bibnamefont {Liu}}, \bibinfo {author} {\bibfnamefont {C.-L.}\ \bibnamefont {Zou}}, \bibinfo {author} {\bibfnamefont {L.}~\bibnamefont {Sun}}, \bibinfo {author} {\bibfnamefont {S.-B.}\ \bibnamefont {Zheng}}, \bibinfo {author} {\bibfnamefont {Y.}~\bibnamefont {Xu}},\ and\ \bibinfo {author} {\bibfnamefont {D.}~\bibnamefont {Yu}},\ }\bibfield  {title} {\bibinfo {title} {Beating the break-even point with a discrete-variable-encoded logical qubit},\ }\href {https://doi.org/10.1038/s41586-023-05784-4} {\bibfield  {journal} {\bibinfo  {journal} {Nature}\ }\textbf {\bibinfo {volume} {616}},\ \bibinfo {pages} {56} (\bibinfo {year} {2023})}\BibitemShut {NoStop}%
\bibitem [{\citenamefont {Lescanne}\ \emph {et~al.}(2020)\citenamefont {Lescanne}, \citenamefont {Villiers}, \citenamefont {Peronnin}, \citenamefont {Sarlette}, \citenamefont {Delbecq}, \citenamefont {Huard}, \citenamefont {Kontos}, \citenamefont {Mirrahimi},\ and\ \citenamefont {Leghtas}}]{lescanne_exponential_2020}%
  \BibitemOpen
  \bibfield  {author} {\bibinfo {author} {\bibfnamefont {R.}~\bibnamefont {Lescanne}}, \bibinfo {author} {\bibfnamefont {M.}~\bibnamefont {Villiers}}, \bibinfo {author} {\bibfnamefont {T.}~\bibnamefont {Peronnin}}, \bibinfo {author} {\bibfnamefont {A.}~\bibnamefont {Sarlette}}, \bibinfo {author} {\bibfnamefont {M.}~\bibnamefont {Delbecq}}, \bibinfo {author} {\bibfnamefont {B.}~\bibnamefont {Huard}}, \bibinfo {author} {\bibfnamefont {T.}~\bibnamefont {Kontos}}, \bibinfo {author} {\bibfnamefont {M.}~\bibnamefont {Mirrahimi}},\ and\ \bibinfo {author} {\bibfnamefont {Z.}~\bibnamefont {Leghtas}},\ }\bibfield  {title} {\bibinfo {title} {Exponential suppression of bit-flips in a qubit encoded in an oscillator},\ }\href {https://doi.org/10.1038/s41567-020-0824-x} {\bibfield  {journal} {\bibinfo  {journal} {Nature Physics}\ }\textbf {\bibinfo {volume} {16}},\ \bibinfo {pages} {509} (\bibinfo {year} {2020})}\BibitemShut {NoStop}%
\bibitem [{\citenamefont {Berdou}\ \emph {et~al.}(2023)\citenamefont {Berdou}, \citenamefont {Murani}, \citenamefont {R{\'e}glade}, \citenamefont {Smith}, \citenamefont {Villiers}, \citenamefont {Palomo}, \citenamefont {Rosticher}, \citenamefont {Denis}, \citenamefont {Morfin}, \citenamefont {Delbecq}, \citenamefont {Kontos}, \citenamefont {Pankratova}, \citenamefont {Rautschke}, \citenamefont {Peronnin}, \citenamefont {Sellem}, \citenamefont {Rouchon}, \citenamefont {Sarlette}, \citenamefont {Mirrahimi}, \citenamefont {{Campagne-Ibarcq}}, \citenamefont {Jezouin}, \citenamefont {Lescanne},\ and\ \citenamefont {Leghtas}}]{berdou_one_2023}%
  \BibitemOpen
  \bibfield  {author} {\bibinfo {author} {\bibfnamefont {C.}~\bibnamefont {Berdou}}, \bibinfo {author} {\bibfnamefont {A.}~\bibnamefont {Murani}}, \bibinfo {author} {\bibfnamefont {U.}~\bibnamefont {R{\'e}glade}}, \bibinfo {author} {\bibfnamefont {W.}~\bibnamefont {Smith}}, \bibinfo {author} {\bibfnamefont {M.}~\bibnamefont {Villiers}}, \bibinfo {author} {\bibfnamefont {J.}~\bibnamefont {Palomo}}, \bibinfo {author} {\bibfnamefont {M.}~\bibnamefont {Rosticher}}, \bibinfo {author} {\bibfnamefont {A.}~\bibnamefont {Denis}}, \bibinfo {author} {\bibfnamefont {P.}~\bibnamefont {Morfin}}, \bibinfo {author} {\bibfnamefont {M.}~\bibnamefont {Delbecq}}, \bibinfo {author} {\bibfnamefont {T.}~\bibnamefont {Kontos}}, \bibinfo {author} {\bibfnamefont {N.}~\bibnamefont {Pankratova}}, \bibinfo {author} {\bibfnamefont {F.}~\bibnamefont {Rautschke}}, \bibinfo {author} {\bibfnamefont {T.}~\bibnamefont {Peronnin}}, \bibinfo {author} {\bibfnamefont {L.-A.}\ \bibnamefont {Sellem}}, \bibinfo {author} {\bibfnamefont {P.}~\bibnamefont {Rouchon}}, \bibinfo {author} {\bibfnamefont {A.}~\bibnamefont {Sarlette}}, \bibinfo {author} {\bibfnamefont {M.}~\bibnamefont {Mirrahimi}}, \bibinfo {author} {\bibfnamefont {P.}~\bibnamefont {{Campagne-Ibarcq}}}, \bibinfo {author} {\bibfnamefont {S.}~\bibnamefont {Jezouin}}, \bibinfo {author} {\bibfnamefont {R.}~\bibnamefont {Lescanne}},\ and\ \bibinfo {author} {\bibfnamefont {Z.}~\bibnamefont {Leghtas}},\ }\bibfield  {title} {\bibinfo {title} {One {{Hundred Second Bit-Flip Time}} in a {{Two-Photon Dissipative Oscillator}}},\ }\href {https://doi.org/10.1103/PRXQuantum.4.020350} {\bibfield  {journal} {\bibinfo  {journal} {PRX Quantum}\ }\textbf {\bibinfo {volume} {4}},\ \bibinfo {pages} {020350} (\bibinfo {year} {2023})}\BibitemShut {NoStop}%
\bibitem [{\citenamefont {Eickbusch}\ \emph {et~al.}(2022)\citenamefont {Eickbusch}, \citenamefont {Sivak}, \citenamefont {Ding}, \citenamefont {Elder}, \citenamefont {Jha}, \citenamefont {Venkatraman}, \citenamefont {Royer}, \citenamefont {Girvin}, \citenamefont {Schoelkopf},\ and\ \citenamefont {Devoret}}]{eickbusch_fast_2022}%
  \BibitemOpen
  \bibfield  {author} {\bibinfo {author} {\bibfnamefont {A.}~\bibnamefont {Eickbusch}}, \bibinfo {author} {\bibfnamefont {V.}~\bibnamefont {Sivak}}, \bibinfo {author} {\bibfnamefont {A.~Z.}\ \bibnamefont {Ding}}, \bibinfo {author} {\bibfnamefont {S.~S.}\ \bibnamefont {Elder}}, \bibinfo {author} {\bibfnamefont {S.~R.}\ \bibnamefont {Jha}}, \bibinfo {author} {\bibfnamefont {J.}~\bibnamefont {Venkatraman}}, \bibinfo {author} {\bibfnamefont {B.}~\bibnamefont {Royer}}, \bibinfo {author} {\bibfnamefont {S.~M.}\ \bibnamefont {Girvin}}, \bibinfo {author} {\bibfnamefont {R.~J.}\ \bibnamefont {Schoelkopf}},\ and\ \bibinfo {author} {\bibfnamefont {M.~H.}\ \bibnamefont {Devoret}},\ }\bibfield  {title} {\bibinfo {title} {Fast universal control of an oscillator with weak dispersive coupling to a qubit},\ }\href {https://doi.org/10.1038/s41567-022-01776-9} {\bibfield  {journal} {\bibinfo  {journal} {Nature Physics}\ }\textbf {\bibinfo {volume} {18}},\ \bibinfo {pages} {1464} (\bibinfo {year} {2022})}\BibitemShut {NoStop}%
\bibitem [{\citenamefont {Putterman}\ \emph {et~al.}(2025)\citenamefont {Putterman}, \citenamefont {Noh}, \citenamefont {Hann}, \citenamefont {MacCabe}, \citenamefont {Aghaeimeibodi}, \citenamefont {Patel}, \citenamefont {Lee}, \citenamefont {Jones}, \citenamefont {Moradinejad}, \citenamefont {Rodriguez}, \citenamefont {Mahuli}, \citenamefont {Rose}, \citenamefont {Owens}, \citenamefont {Levine}, \citenamefont {Rosenfeld}, \citenamefont {Reinhold}, \citenamefont {Moncelsi}, \citenamefont {Alcid}, \citenamefont {Alidoust}, \citenamefont {{Arrangoiz-Arriola}}, \citenamefont {Barnett}, \citenamefont {Bienias}, \citenamefont {Carson}, \citenamefont {Chen}, \citenamefont {Chen}, \citenamefont {Chinkezian}, \citenamefont {Chisholm}, \citenamefont {Chou}, \citenamefont {Clerk}, \citenamefont {Clifford}, \citenamefont {Cosmic}, \citenamefont {Curiel}, \citenamefont {Davis}, \citenamefont {DeLorenzo}, \citenamefont {D'Ewart}, \citenamefont {Diky}, \citenamefont {D'Souza}, \citenamefont {Dumitrescu}, \citenamefont {Eisenmann}, \citenamefont {Elkhouly}, \citenamefont {Evenbly}, \citenamefont {Fang}, \citenamefont {Fang}, \citenamefont {Fling}, \citenamefont {Fon}, \citenamefont {Garcia}, \citenamefont {Gorshkov}, \citenamefont {Grant}, \citenamefont {Gray}, \citenamefont {Grimberg}, \citenamefont {Grimsmo}, \citenamefont {Haim}, \citenamefont {Hand}, \citenamefont {He}, \citenamefont {Hernandez}, \citenamefont {Hover}, \citenamefont {Hung}, \citenamefont {Hunt}, \citenamefont {Iverson}, \citenamefont {Jarrige}, \citenamefont {Jaskula}, \citenamefont {Jiang}, \citenamefont {Kalaee}, \citenamefont {Karabalin}, \citenamefont {Karalekas}, \citenamefont {Keller}, \citenamefont {Khalajhedayati}, \citenamefont {Kubica}, \citenamefont {Lee}, \citenamefont {Leroux}, \citenamefont {Lieu}, \citenamefont {Ly}, \citenamefont {Madrigal}, \citenamefont {Marcaud}, \citenamefont {McCabe}, \citenamefont {Miles}, \citenamefont {Milsted}, \citenamefont {Minguzzi}, \citenamefont {Mishra}, \citenamefont {Mukherjee}, \citenamefont {Naghiloo}, \citenamefont {Oblepias}, \citenamefont {Ortuno}, \citenamefont {Pagdilao}, \citenamefont {Pancotti}, \citenamefont {Panduro}, \citenamefont {Paquette}, \citenamefont {Park}, \citenamefont {Peairs}, \citenamefont {Perello}, \citenamefont {Peterson}, \citenamefont {Ponte}, \citenamefont {Preskill}, \citenamefont {Qiao}, \citenamefont {Refael}, \citenamefont {Resnick}, \citenamefont {Retzker}, \citenamefont {Reyna}, \citenamefont {Runyan}, \citenamefont {Ryan}, \citenamefont {Sahmoud}, \citenamefont {Sanchez}, \citenamefont {Sanil}, \citenamefont {Sankar}, \citenamefont {Sato}, \citenamefont {Scaffidi}, \citenamefont {Siavoshi}, \citenamefont {Sivarajah}, \citenamefont {Skogland}, \citenamefont {Su}, \citenamefont {Swenson}, \citenamefont {Teo}, \citenamefont {Tomada}, \citenamefont {Torlai}, \citenamefont {Wollack}, \citenamefont {Ye}, \citenamefont {Zerrudo}, \citenamefont {Zhang}, \citenamefont {Brand{\~a}o}, \citenamefont {Matheny},\ and\ \citenamefont {Painter}}]{amazon_bosonic}%
  \BibitemOpen
  \bibfield  {author} {\bibinfo {author} {\bibfnamefont {H.}~\bibnamefont {Putterman}}, \bibinfo {author} {\bibfnamefont {K.}~\bibnamefont {Noh}}, \bibinfo {author} {\bibfnamefont {C.~T.}\ \bibnamefont {Hann}}, \bibinfo {author} {\bibfnamefont {G.~S.}\ \bibnamefont {MacCabe}}, \bibinfo {author} {\bibfnamefont {S.}~\bibnamefont {Aghaeimeibodi}}, \bibinfo {author} {\bibfnamefont {R.~N.}\ \bibnamefont {Patel}}, \bibinfo {author} {\bibfnamefont {M.}~\bibnamefont {Lee}}, \bibinfo {author} {\bibfnamefont {W.~M.}\ \bibnamefont {Jones}}, \bibinfo {author} {\bibfnamefont {H.}~\bibnamefont {Moradinejad}}, \bibinfo {author} {\bibfnamefont {R.}~\bibnamefont {Rodriguez}}, \bibinfo {author} {\bibfnamefont {N.}~\bibnamefont {Mahuli}}, \bibinfo {author} {\bibfnamefont {J.}~\bibnamefont {Rose}}, \bibinfo {author} {\bibfnamefont {J.~C.}\ \bibnamefont {Owens}}, \bibinfo {author} {\bibfnamefont {H.}~\bibnamefont {Levine}}, \bibinfo {author} {\bibfnamefont {E.}~\bibnamefont {Rosenfeld}}, \bibinfo {author} {\bibfnamefont {P.}~\bibnamefont {Reinhold}}, \bibinfo {author} {\bibfnamefont {L.}~\bibnamefont {Moncelsi}}, \bibinfo {author} {\bibfnamefont {J.~A.}\ \bibnamefont {Alcid}}, \bibinfo {author} {\bibfnamefont {N.}~\bibnamefont {Alidoust}}, \bibinfo {author} {\bibfnamefont {P.}~\bibnamefont {{Arrangoiz-Arriola}}}, \bibinfo {author} {\bibfnamefont {J.}~\bibnamefont {Barnett}}, \bibinfo {author} {\bibfnamefont {P.}~\bibnamefont {Bienias}}, \bibinfo {author} {\bibfnamefont {H.~A.}\ \bibnamefont {Carson}}, \bibinfo {author} {\bibfnamefont {C.}~\bibnamefont {Chen}}, \bibinfo {author} {\bibfnamefont {L.}~\bibnamefont {Chen}}, \bibinfo {author} {\bibfnamefont {H.}~\bibnamefont {Chinkezian}}, \bibinfo {author} {\bibfnamefont {E.~M.}\ \bibnamefont {Chisholm}}, \bibinfo {author} {\bibfnamefont {M.-H.}\ \bibnamefont {Chou}}, \bibinfo {author} {\bibfnamefont {A.}~\bibnamefont {Clerk}}, \bibinfo {author} {\bibfnamefont {A.}~\bibnamefont {Clifford}}, \bibinfo {author} {\bibfnamefont {R.}~\bibnamefont {Cosmic}}, \bibinfo {author} {\bibfnamefont {A.~V.}\ \bibnamefont {Curiel}}, \bibinfo {author} {\bibfnamefont {E.}~\bibnamefont {Davis}}, \bibinfo {author} {\bibfnamefont {L.}~\bibnamefont {DeLorenzo}}, \bibinfo {author} {\bibfnamefont {J.~M.}\ \bibnamefont {D'Ewart}}, \bibinfo {author} {\bibfnamefont {A.}~\bibnamefont {Diky}}, \bibinfo {author} {\bibfnamefont {N.}~\bibnamefont {D'Souza}}, \bibinfo {author} {\bibfnamefont {P.~T.}\ \bibnamefont {Dumitrescu}}, \bibinfo {author} {\bibfnamefont {S.}~\bibnamefont {Eisenmann}}, \bibinfo {author} {\bibfnamefont {E.}~\bibnamefont {Elkhouly}}, \bibinfo {author} {\bibfnamefont {G.}~\bibnamefont {Evenbly}}, \bibinfo {author} {\bibfnamefont {M.~T.}\ \bibnamefont {Fang}}, \bibinfo {author} {\bibfnamefont {Y.}~\bibnamefont {Fang}}, \bibinfo {author} {\bibfnamefont {M.~J.}\ \bibnamefont {Fling}}, \bibinfo {author} {\bibfnamefont {W.}~\bibnamefont {Fon}}, \bibinfo {author} {\bibfnamefont {G.}~\bibnamefont {Garcia}}, \bibinfo {author} {\bibfnamefont {A.~V.}\ \bibnamefont {Gorshkov}}, \bibinfo {author} {\bibfnamefont {J.~A.}\ \bibnamefont {Grant}}, \bibinfo {author} {\bibfnamefont {M.~J.}\ \bibnamefont {Gray}}, \bibinfo {author} {\bibfnamefont {S.}~\bibnamefont {Grimberg}}, \bibinfo {author} {\bibfnamefont {A.~L.}\ \bibnamefont {Grimsmo}}, \bibinfo {author} {\bibfnamefont {A.}~\bibnamefont {Haim}}, \bibinfo {author} {\bibfnamefont {J.}~\bibnamefont {Hand}}, \bibinfo {author} {\bibfnamefont {Y.}~\bibnamefont {He}}, \bibinfo {author} {\bibfnamefont {M.}~\bibnamefont {Hernandez}}, \bibinfo {author} {\bibfnamefont {D.}~\bibnamefont {Hover}}, \bibinfo {author} {\bibfnamefont {J.~S.~C.}\ \bibnamefont {Hung}}, \bibinfo {author} {\bibfnamefont {M.}~\bibnamefont {Hunt}}, \bibinfo {author} {\bibfnamefont {J.}~\bibnamefont {Iverson}}, \bibinfo {author} {\bibfnamefont {I.}~\bibnamefont {Jarrige}}, \bibinfo {author} {\bibfnamefont {J.-C.}\ \bibnamefont {Jaskula}}, \bibinfo {author} {\bibfnamefont {L.}~\bibnamefont {Jiang}}, \bibinfo {author} {\bibfnamefont {M.}~\bibnamefont {Kalaee}}, \bibinfo {author} {\bibfnamefont {R.}~\bibnamefont {Karabalin}}, \bibinfo {author} {\bibfnamefont {P.~J.}\ \bibnamefont {Karalekas}}, \bibinfo {author} {\bibfnamefont {A.~J.}\ \bibnamefont {Keller}}, \bibinfo {author} {\bibfnamefont {A.}~\bibnamefont {Khalajhedayati}}, \bibinfo {author} {\bibfnamefont {A.}~\bibnamefont {Kubica}}, \bibinfo {author} {\bibfnamefont {H.}~\bibnamefont {Lee}}, \bibinfo {author} {\bibfnamefont {C.}~\bibnamefont {Leroux}}, \bibinfo {author} {\bibfnamefont {S.}~\bibnamefont {Lieu}}, \bibinfo {author} {\bibfnamefont {V.}~\bibnamefont {Ly}}, \bibinfo {author} {\bibfnamefont {K.~V.}\ \bibnamefont {Madrigal}}, \bibinfo {author} {\bibfnamefont {G.}~\bibnamefont {Marcaud}}, \bibinfo {author} {\bibfnamefont {G.}~\bibnamefont {McCabe}}, \bibinfo {author} {\bibfnamefont {C.}~\bibnamefont {Miles}}, \bibinfo {author} {\bibfnamefont {A.}~\bibnamefont {Milsted}}, \bibinfo {author} {\bibfnamefont {J.}~\bibnamefont {Minguzzi}}, \bibinfo {author} {\bibfnamefont {A.}~\bibnamefont {Mishra}}, \bibinfo {author} {\bibfnamefont {B.}~\bibnamefont {Mukherjee}}, \bibinfo {author} {\bibfnamefont {M.}~\bibnamefont {Naghiloo}}, \bibinfo {author} {\bibfnamefont {E.}~\bibnamefont {Oblepias}}, \bibinfo {author} {\bibfnamefont {G.}~\bibnamefont {Ortuno}}, \bibinfo {author} {\bibfnamefont {J.}~\bibnamefont {Pagdilao}}, \bibinfo {author} {\bibfnamefont {N.}~\bibnamefont {Pancotti}}, \bibinfo {author} {\bibfnamefont {A.}~\bibnamefont {Panduro}}, \bibinfo {author} {\bibfnamefont {{\relax JP}.}~\bibnamefont {Paquette}}, \bibinfo {author} {\bibfnamefont {M.}~\bibnamefont {Park}}, \bibinfo {author} {\bibfnamefont {G.~A.}\ \bibnamefont {Peairs}}, \bibinfo {author} {\bibfnamefont {D.}~\bibnamefont {Perello}}, \bibinfo {author} {\bibfnamefont {E.~C.}\ \bibnamefont {Peterson}}, \bibinfo {author} {\bibfnamefont {S.}~\bibnamefont {Ponte}}, \bibinfo {author} {\bibfnamefont {J.}~\bibnamefont {Preskill}}, \bibinfo {author} {\bibfnamefont {J.}~\bibnamefont {Qiao}}, \bibinfo {author} {\bibfnamefont {G.}~\bibnamefont {Refael}}, \bibinfo {author} {\bibfnamefont {R.}~\bibnamefont {Resnick}}, \bibinfo {author} {\bibfnamefont {A.}~\bibnamefont {Retzker}}, \bibinfo {author} {\bibfnamefont {O.~A.}\ \bibnamefont {Reyna}}, \bibinfo {author} {\bibfnamefont {M.}~\bibnamefont {Runyan}}, \bibinfo {author} {\bibfnamefont {C.~A.}\ \bibnamefont {Ryan}}, \bibinfo {author} {\bibfnamefont {A.}~\bibnamefont {Sahmoud}}, \bibinfo {author} {\bibfnamefont {E.}~\bibnamefont {Sanchez}}, \bibinfo {author} {\bibfnamefont {R.}~\bibnamefont {Sanil}}, \bibinfo {author} {\bibfnamefont {K.}~\bibnamefont {Sankar}}, \bibinfo {author} {\bibfnamefont {Y.}~\bibnamefont {Sato}}, \bibinfo {author} {\bibfnamefont {T.}~\bibnamefont {Scaffidi}}, \bibinfo {author} {\bibfnamefont {S.}~\bibnamefont {Siavoshi}}, \bibinfo {author} {\bibfnamefont {P.}~\bibnamefont {Sivarajah}}, \bibinfo {author} {\bibfnamefont {T.}~\bibnamefont {Skogland}}, \bibinfo {author} {\bibfnamefont {C.-J.}\ \bibnamefont {Su}}, \bibinfo {author} {\bibfnamefont {L.~J.}\ \bibnamefont {Swenson}}, \bibinfo {author} {\bibfnamefont {S.~M.}\ \bibnamefont {Teo}}, \bibinfo {author} {\bibfnamefont {A.}~\bibnamefont {Tomada}}, \bibinfo {author} {\bibfnamefont {G.}~\bibnamefont {Torlai}}, \bibinfo {author} {\bibfnamefont {E.~A.}\ \bibnamefont {Wollack}}, \bibinfo {author} {\bibfnamefont {Y.}~\bibnamefont {Ye}}, \bibinfo {author} {\bibfnamefont {J.~A.}\ \bibnamefont {Zerrudo}}, \bibinfo {author} {\bibfnamefont {K.}~\bibnamefont {Zhang}}, \bibinfo {author} {\bibfnamefont {F.~G. S.~L.}\ \bibnamefont {Brand{\~a}o}}, \bibinfo {author} {\bibfnamefont {M.~H.}\ \bibnamefont {Matheny}},\ and\ \bibinfo {author} {\bibfnamefont {O.}~\bibnamefont {Painter}},\ }\bibfield  {title} {\bibinfo {title} {Hardware-efficient quantum error correction via concatenated bosonic qubits},\ }\href {https://doi.org/10.1038/s41586-025-08642-7} {\bibfield  {journal} {\bibinfo  {journal} {Nature}\ }\textbf {\bibinfo {volume} {638}},\ \bibinfo {pages} {927} (\bibinfo {year} {2025})}\BibitemShut {NoStop}%
\bibitem [{\citenamefont {Vrajitoarea}\ \emph {et~al.}(2020)\citenamefont {Vrajitoarea}, \citenamefont {Huang}, \citenamefont {Groszkowski}, \citenamefont {Koch},\ and\ \citenamefont {Houck}}]{vrajitoareaQuantumControlOscillator2020}%
  \BibitemOpen
  \bibfield  {author} {\bibinfo {author} {\bibfnamefont {A.}~\bibnamefont {Vrajitoarea}}, \bibinfo {author} {\bibfnamefont {Z.}~\bibnamefont {Huang}}, \bibinfo {author} {\bibfnamefont {P.}~\bibnamefont {Groszkowski}}, \bibinfo {author} {\bibfnamefont {J.}~\bibnamefont {Koch}},\ and\ \bibinfo {author} {\bibfnamefont {A.~A.}\ \bibnamefont {Houck}},\ }\bibfield  {title} {\bibinfo {title} {Quantum control of an oscillator using a stimulated {{Josephson}} nonlinearity},\ }\href {https://doi.org/10.1038/s41567-019-0703-5} {\bibfield  {journal} {\bibinfo  {journal} {Nature Physics}\ }\textbf {\bibinfo {volume} {16}},\ \bibinfo {pages} {211} (\bibinfo {year} {2020})}\BibitemShut {NoStop}%
\bibitem [{\citenamefont {Grimm}\ \emph {et~al.}(2020)\citenamefont {Grimm}, \citenamefont {Frattini}, \citenamefont {Puri}, \citenamefont {Mundhada}, \citenamefont {Touzard}, \citenamefont {Mirrahimi}, \citenamefont {Girvin}, \citenamefont {Shankar},\ and\ \citenamefont {Devoret}}]{grimm_stabilization_2020}%
  \BibitemOpen
  \bibfield  {author} {\bibinfo {author} {\bibfnamefont {A.}~\bibnamefont {Grimm}}, \bibinfo {author} {\bibfnamefont {N.~E.}\ \bibnamefont {Frattini}}, \bibinfo {author} {\bibfnamefont {S.}~\bibnamefont {Puri}}, \bibinfo {author} {\bibfnamefont {S.~O.}\ \bibnamefont {Mundhada}}, \bibinfo {author} {\bibfnamefont {S.}~\bibnamefont {Touzard}}, \bibinfo {author} {\bibfnamefont {M.}~\bibnamefont {Mirrahimi}}, \bibinfo {author} {\bibfnamefont {S.~M.}\ \bibnamefont {Girvin}}, \bibinfo {author} {\bibfnamefont {S.}~\bibnamefont {Shankar}},\ and\ \bibinfo {author} {\bibfnamefont {M.~H.}\ \bibnamefont {Devoret}},\ }\bibfield  {title} {\bibinfo {title} {Stabilization and operation of a {{Kerr-cat}} qubit},\ }\href {https://doi.org/10.1038/s41586-020-2587-z} {\bibfield  {journal} {\bibinfo  {journal} {Nature}\ }\textbf {\bibinfo {volume} {584}},\ \bibinfo {pages} {205} (\bibinfo {year} {2020})}\BibitemShut {NoStop}%
\bibitem [{\citenamefont {Ding}\ \emph {et~al.}(2024)\citenamefont {Ding}, \citenamefont {Brock}, \citenamefont {Eickbusch}, \citenamefont {Koottandavida}, \citenamefont {Frattini}, \citenamefont {Cortinas}, \citenamefont {Joshi}, \citenamefont {de~Graaf}, \citenamefont {Chapman}, \citenamefont {Ganjam}, \citenamefont {Frunzio}, \citenamefont {Schoelkopf},\ and\ \citenamefont {Devoret}}]{ding_quantum_2024}%
  \BibitemOpen
  \bibfield  {author} {\bibinfo {author} {\bibfnamefont {A.~Z.}\ \bibnamefont {Ding}}, \bibinfo {author} {\bibfnamefont {B.~L.}\ \bibnamefont {Brock}}, \bibinfo {author} {\bibfnamefont {A.}~\bibnamefont {Eickbusch}}, \bibinfo {author} {\bibfnamefont {A.}~\bibnamefont {Koottandavida}}, \bibinfo {author} {\bibfnamefont {N.~E.}\ \bibnamefont {Frattini}}, \bibinfo {author} {\bibfnamefont {R.~G.}\ \bibnamefont {Cortinas}}, \bibinfo {author} {\bibfnamefont {V.~R.}\ \bibnamefont {Joshi}}, \bibinfo {author} {\bibfnamefont {S.~J.}\ \bibnamefont {de~Graaf}}, \bibinfo {author} {\bibfnamefont {B.~J.}\ \bibnamefont {Chapman}}, \bibinfo {author} {\bibfnamefont {S.}~\bibnamefont {Ganjam}}, \bibinfo {author} {\bibfnamefont {L.}~\bibnamefont {Frunzio}}, \bibinfo {author} {\bibfnamefont {R.~J.}\ \bibnamefont {Schoelkopf}},\ and\ \bibinfo {author} {\bibfnamefont {M.~H.}\ \bibnamefont {Devoret}},\ }\href {https://doi.org/10.48550/arXiv.2407.10940} {\bibinfo {title} {Quantum {{Control}} of an {{Oscillator}} with a {{Kerr-cat Qubit}}}} (\bibinfo {year} {2024}),\ \Eprint {https://arxiv.org/abs/2407.10940} {arXiv:2407.10940 [quant-ph]} \BibitemShut {NoStop}%
\bibitem [{\citenamefont {R{\'e}glade}\ \emph {et~al.}(2024)\citenamefont {R{\'e}glade}, \citenamefont {Bocquet}, \citenamefont {Gautier}, \citenamefont {Cohen}, \citenamefont {Marquet}, \citenamefont {Albertinale}, \citenamefont {Pankratova}, \citenamefont {Hall{\'e}n}, \citenamefont {Rautschke}, \citenamefont {Sellem}, \citenamefont {Rouchon}, \citenamefont {Sarlette}, \citenamefont {Mirrahimi}, \citenamefont {{Campagne-Ibarcq}}, \citenamefont {Lescanne}, \citenamefont {Jezouin},\ and\ \citenamefont {Leghtas}}]{réglade_quantum_2024}%
  \BibitemOpen
  \bibfield  {author} {\bibinfo {author} {\bibfnamefont {U.}~\bibnamefont {R{\'e}glade}}, \bibinfo {author} {\bibfnamefont {A.}~\bibnamefont {Bocquet}}, \bibinfo {author} {\bibfnamefont {R.}~\bibnamefont {Gautier}}, \bibinfo {author} {\bibfnamefont {J.}~\bibnamefont {Cohen}}, \bibinfo {author} {\bibfnamefont {A.}~\bibnamefont {Marquet}}, \bibinfo {author} {\bibfnamefont {E.}~\bibnamefont {Albertinale}}, \bibinfo {author} {\bibfnamefont {N.}~\bibnamefont {Pankratova}}, \bibinfo {author} {\bibfnamefont {M.}~\bibnamefont {Hall{\'e}n}}, \bibinfo {author} {\bibfnamefont {F.}~\bibnamefont {Rautschke}}, \bibinfo {author} {\bibfnamefont {L.-A.}\ \bibnamefont {Sellem}}, \bibinfo {author} {\bibfnamefont {P.}~\bibnamefont {Rouchon}}, \bibinfo {author} {\bibfnamefont {A.}~\bibnamefont {Sarlette}}, \bibinfo {author} {\bibfnamefont {M.}~\bibnamefont {Mirrahimi}}, \bibinfo {author} {\bibfnamefont {P.}~\bibnamefont {{Campagne-Ibarcq}}}, \bibinfo {author} {\bibfnamefont {R.}~\bibnamefont {Lescanne}}, \bibinfo {author} {\bibfnamefont {S.}~\bibnamefont {Jezouin}},\ and\ \bibinfo {author} {\bibfnamefont {Z.}~\bibnamefont {Leghtas}},\ }\bibfield  {title} {\bibinfo {title} {Quantum control of a cat qubit with bit-flip times exceeding ten seconds},\ }\href {https://doi.org/10.1038/s41586-024-07294-3} {\bibfield  {journal} {\bibinfo  {journal} {Nature}\ }\textbf {\bibinfo {volume} {629}},\ \bibinfo {pages} {778} (\bibinfo {year} {2024})}\BibitemShut {NoStop}%
\bibitem [{\citenamefont {Manucharyan}\ \emph {et~al.}(2009)\citenamefont {Manucharyan}, \citenamefont {Koch}, \citenamefont {Glazman},\ and\ \citenamefont {Devoret}}]{manucharyan_fluxonium_2009}%
  \BibitemOpen
  \bibfield  {author} {\bibinfo {author} {\bibfnamefont {V.~E.}\ \bibnamefont {Manucharyan}}, \bibinfo {author} {\bibfnamefont {J.}~\bibnamefont {Koch}}, \bibinfo {author} {\bibfnamefont {L.~I.}\ \bibnamefont {Glazman}},\ and\ \bibinfo {author} {\bibfnamefont {M.~H.}\ \bibnamefont {Devoret}},\ }\bibfield  {title} {\bibinfo {title} {Fluxonium: {{Single Cooper-Pair Circuit Free}} of {{Charge Offsets}}},\ }\href {https://doi.org/10.1126/science.1175552} {\bibfield  {journal} {\bibinfo  {journal} {Science}\ }\textbf {\bibinfo {volume} {326}},\ \bibinfo {pages} {113} (\bibinfo {year} {2009})}\BibitemShut {NoStop}%
\bibitem [{\citenamefont {Somoroff}\ \emph {et~al.}(2023)\citenamefont {Somoroff}, \citenamefont {Ficheux}, \citenamefont {Mencia}, \citenamefont {Xiong}, \citenamefont {Kuzmin},\ and\ \citenamefont {Manucharyan}}]{somoroff_millisecond_2023a}%
  \BibitemOpen
  \bibfield  {author} {\bibinfo {author} {\bibfnamefont {A.}~\bibnamefont {Somoroff}}, \bibinfo {author} {\bibfnamefont {Q.}~\bibnamefont {Ficheux}}, \bibinfo {author} {\bibfnamefont {R.~A.}\ \bibnamefont {Mencia}}, \bibinfo {author} {\bibfnamefont {H.}~\bibnamefont {Xiong}}, \bibinfo {author} {\bibfnamefont {R.}~\bibnamefont {Kuzmin}},\ and\ \bibinfo {author} {\bibfnamefont {V.~E.}\ \bibnamefont {Manucharyan}},\ }\bibfield  {title} {\bibinfo {title} {Millisecond {{Coherence}} in a {{Superconducting Qubit}}},\ }\href {https://doi.org/10.1103/PhysRevLett.130.267001} {\bibfield  {journal} {\bibinfo  {journal} {Physical Review Letters}\ }\textbf {\bibinfo {volume} {130}},\ \bibinfo {pages} {267001} (\bibinfo {year} {2023})}\BibitemShut {NoStop}%
\bibitem [{\citenamefont {Ding}\ \emph {et~al.}(2023)\citenamefont {Ding}, \citenamefont {Hays}, \citenamefont {Sung}, \citenamefont {Kannan}, \citenamefont {An}, \citenamefont {Di~Paolo}, \citenamefont {Karamlou}, \citenamefont {Hazard}, \citenamefont {Azar}, \citenamefont {Kim}, \citenamefont {Niedzielski}, \citenamefont {Melville}, \citenamefont {Schwartz}, \citenamefont {Yoder}, \citenamefont {Orlando}, \citenamefont {Gustavsson}, \citenamefont {Grover}, \citenamefont {Serniak},\ and\ \citenamefont {Oliver}}]{leon_high_2023}%
  \BibitemOpen
  \bibfield  {author} {\bibinfo {author} {\bibfnamefont {L.}~\bibnamefont {Ding}}, \bibinfo {author} {\bibfnamefont {M.}~\bibnamefont {Hays}}, \bibinfo {author} {\bibfnamefont {Y.}~\bibnamefont {Sung}}, \bibinfo {author} {\bibfnamefont {B.}~\bibnamefont {Kannan}}, \bibinfo {author} {\bibfnamefont {J.}~\bibnamefont {An}}, \bibinfo {author} {\bibfnamefont {A.}~\bibnamefont {Di~Paolo}}, \bibinfo {author} {\bibfnamefont {A.~H.}\ \bibnamefont {Karamlou}}, \bibinfo {author} {\bibfnamefont {T.~M.}\ \bibnamefont {Hazard}}, \bibinfo {author} {\bibfnamefont {K.}~\bibnamefont {Azar}}, \bibinfo {author} {\bibfnamefont {D.~K.}\ \bibnamefont {Kim}}, \bibinfo {author} {\bibfnamefont {B.~M.}\ \bibnamefont {Niedzielski}}, \bibinfo {author} {\bibfnamefont {A.}~\bibnamefont {Melville}}, \bibinfo {author} {\bibfnamefont {M.~E.}\ \bibnamefont {Schwartz}}, \bibinfo {author} {\bibfnamefont {J.~L.}\ \bibnamefont {Yoder}}, \bibinfo {author} {\bibfnamefont {T.~P.}\ \bibnamefont {Orlando}}, \bibinfo {author} {\bibfnamefont {S.}~\bibnamefont {Gustavsson}}, \bibinfo {author} {\bibfnamefont {J.~A.}\ \bibnamefont {Grover}}, \bibinfo {author} {\bibfnamefont {K.}~\bibnamefont {Serniak}},\ and\ \bibinfo {author} {\bibfnamefont {W.~D.}\ \bibnamefont {Oliver}},\ }\bibfield  {title} {\bibinfo {title} {High-fidelity, frequency-flexible two-qubit fluxonium gates with a transmon coupler},\ }\href {https://doi.org/10.1103/PhysRevX.13.031035} {\bibfield  {journal} {\bibinfo  {journal} {Physical Review X}\ }\textbf {\bibinfo {volume} {13}},\ \bibinfo {pages} {031035} (\bibinfo {year} {2023})}\BibitemShut {NoStop}%
\bibitem [{\citenamefont {Wang}\ \emph {et~al.}(2025)\citenamefont {Wang}, \citenamefont {Lu}, \citenamefont {Zhan}, \citenamefont {Ma}, \citenamefont {Wu}, \citenamefont {Sun}, \citenamefont {Deng}, \citenamefont {Bai}, \citenamefont {Bao}, \citenamefont {Chang}, \citenamefont {Gao}, \citenamefont {Gao}, \citenamefont {Gong}, \citenamefont {Hu}, \citenamefont {Hu}, \citenamefont {Ji}, \citenamefont {Ma}, \citenamefont {Mao}, \citenamefont {Song}, \citenamefont {Tang}, \citenamefont {Wang}, \citenamefont {Wang}, \citenamefont {Wang}, \citenamefont {Xia}, \citenamefont {Xu}, \citenamefont {Zhan}, \citenamefont {Zhang}, \citenamefont {Zhou}, \citenamefont {Zhu}, \citenamefont {Zhu}, \citenamefont {Zhu}, \citenamefont {Zhu}, \citenamefont {Shi}, \citenamefont {Zhao},\ and\ \citenamefont {Deng}}]{wang_high_2025}%
  \BibitemOpen
  \bibfield  {author} {\bibinfo {author} {\bibfnamefont {F.}~\bibnamefont {Wang}}, \bibinfo {author} {\bibfnamefont {K.}~\bibnamefont {Lu}}, \bibinfo {author} {\bibfnamefont {H.}~\bibnamefont {Zhan}}, \bibinfo {author} {\bibfnamefont {L.}~\bibnamefont {Ma}}, \bibinfo {author} {\bibfnamefont {F.}~\bibnamefont {Wu}}, \bibinfo {author} {\bibfnamefont {H.}~\bibnamefont {Sun}}, \bibinfo {author} {\bibfnamefont {H.}~\bibnamefont {Deng}}, \bibinfo {author} {\bibfnamefont {Y.}~\bibnamefont {Bai}}, \bibinfo {author} {\bibfnamefont {F.}~\bibnamefont {Bao}}, \bibinfo {author} {\bibfnamefont {X.}~\bibnamefont {Chang}}, \bibinfo {author} {\bibfnamefont {R.}~\bibnamefont {Gao}}, \bibinfo {author} {\bibfnamefont {X.}~\bibnamefont {Gao}}, \bibinfo {author} {\bibfnamefont {G.}~\bibnamefont {Gong}}, \bibinfo {author} {\bibfnamefont {L.}~\bibnamefont {Hu}}, \bibinfo {author} {\bibfnamefont {R.}~\bibnamefont {Hu}}, \bibinfo {author} {\bibfnamefont {H.}~\bibnamefont {Ji}}, \bibinfo {author} {\bibfnamefont {X.}~\bibnamefont {Ma}}, \bibinfo {author} {\bibfnamefont {L.}~\bibnamefont {Mao}}, \bibinfo {author} {\bibfnamefont {Z.}~\bibnamefont {Song}}, \bibinfo {author} {\bibfnamefont {C.}~\bibnamefont {Tang}}, \bibinfo {author} {\bibfnamefont {H.}~\bibnamefont {Wang}}, \bibinfo {author} {\bibfnamefont {T.}~\bibnamefont {Wang}}, \bibinfo {author} {\bibfnamefont {Z.}~\bibnamefont {Wang}}, \bibinfo {author} {\bibfnamefont {T.}~\bibnamefont {Xia}}, \bibinfo {author} {\bibfnamefont {H.}~\bibnamefont {Xu}}, \bibinfo {author} {\bibfnamefont {Z.}~\bibnamefont {Zhan}}, \bibinfo {author} {\bibfnamefont {G.}~\bibnamefont {Zhang}}, \bibinfo {author} {\bibfnamefont {T.}~\bibnamefont {Zhou}}, \bibinfo {author} {\bibfnamefont {M.}~\bibnamefont {Zhu}}, \bibinfo {author} {\bibfnamefont {Q.}~\bibnamefont {Zhu}}, \bibinfo {author} {\bibfnamefont {S.}~\bibnamefont {Zhu}}, \bibinfo {author} {\bibfnamefont {X.}~\bibnamefont {Zhu}}, \bibinfo {author} {\bibfnamefont {Y.}~\bibnamefont {Shi}}, \bibinfo {author} {\bibfnamefont {H.-H.}\ \bibnamefont {Zhao}},\ and\ \bibinfo {author} {\bibfnamefont {C.}~\bibnamefont {Deng}},\ }\bibfield  {title} {\bibinfo {title} {High coherence fluxonium manufactured with a wafer-scale uniformity process},\ }\href {https://doi.org/10.1103/PhysRevApplied.23.044064} {\bibfield  {journal} {\bibinfo  {journal} {Physical Review Applied}\ }\textbf {\bibinfo {volume} {23}},\ \bibinfo {pages} {044064} (\bibinfo {year} {2025})}\BibitemShut {NoStop}%
\bibitem [{\citenamefont {Atanasova}\ \emph {et~al.}(2025)\citenamefont {Atanasova}, \citenamefont {Yang}, \citenamefont {{H{\"o}nigl-Decrinis}}, \citenamefont {Gusenkova}, \citenamefont {Pop},\ and\ \citenamefont {Kirchmair}}]{atanasova_situ_2025}%
  \BibitemOpen
  \bibfield  {author} {\bibinfo {author} {\bibfnamefont {D.}~\bibnamefont {Atanasova}}, \bibinfo {author} {\bibfnamefont {I.}~\bibnamefont {Yang}}, \bibinfo {author} {\bibfnamefont {T.}~\bibnamefont {{H{\"o}nigl-Decrinis}}}, \bibinfo {author} {\bibfnamefont {D.}~\bibnamefont {Gusenkova}}, \bibinfo {author} {\bibfnamefont {I.}~\bibnamefont {Pop}},\ and\ \bibinfo {author} {\bibfnamefont {G.}~\bibnamefont {Kirchmair}},\ }\bibfield  {title} {\bibinfo {title} {{\emph{In }}{{{\emph{Situ}}}} {{Tunable Interaction}} with an {{Invertible Sign}} between a {{Fluxonium}} and a {{Post Cavity}}},\ }\href {https://doi.org/10.1103/PRXQuantum.6.020318} {\bibfield  {journal} {\bibinfo  {journal} {PRX Quantum}\ }\textbf {\bibinfo {volume} {6}},\ \bibinfo {pages} {020318} (\bibinfo {year} {2025})}\BibitemShut {NoStop}%
\bibitem [{\citenamefont {Zhu}\ \emph {et~al.}(2013)\citenamefont {Zhu}, \citenamefont {Ferguson}, \citenamefont {Manucharyan},\ and\ \citenamefont {Koch}}]{zhu_circuit_2013}%
  \BibitemOpen
  \bibfield  {author} {\bibinfo {author} {\bibfnamefont {G.}~\bibnamefont {Zhu}}, \bibinfo {author} {\bibfnamefont {D.~G.}\ \bibnamefont {Ferguson}}, \bibinfo {author} {\bibfnamefont {V.~E.}\ \bibnamefont {Manucharyan}},\ and\ \bibinfo {author} {\bibfnamefont {J.}~\bibnamefont {Koch}},\ }\bibfield  {title} {\bibinfo {title} {Circuit {{QED}} with fluxonium qubits: {{Theory}} of the dispersive regime},\ }\href {https://doi.org/10.1103/PhysRevB.87.024510} {\bibfield  {journal} {\bibinfo  {journal} {Physical Review B}\ }\textbf {\bibinfo {volume} {87}},\ \bibinfo {pages} {024510} (\bibinfo {year} {2013})}\BibitemShut {NoStop}%
\bibitem [{\citenamefont {Smith}\ \emph {et~al.}(2016)\citenamefont {Smith}, \citenamefont {Kou}, \citenamefont {Vool}, \citenamefont {Pop}, \citenamefont {Frunzio}, \citenamefont {Schoelkopf},\ and\ \citenamefont {Devoret}}]{smith_quantization_2016}%
  \BibitemOpen
  \bibfield  {author} {\bibinfo {author} {\bibfnamefont {W.~C.}\ \bibnamefont {Smith}}, \bibinfo {author} {\bibfnamefont {A.}~\bibnamefont {Kou}}, \bibinfo {author} {\bibfnamefont {U.}~\bibnamefont {Vool}}, \bibinfo {author} {\bibfnamefont {I.~M.}\ \bibnamefont {Pop}}, \bibinfo {author} {\bibfnamefont {L.}~\bibnamefont {Frunzio}}, \bibinfo {author} {\bibfnamefont {R.~J.}\ \bibnamefont {Schoelkopf}},\ and\ \bibinfo {author} {\bibfnamefont {M.~H.}\ \bibnamefont {Devoret}},\ }\bibfield  {title} {\bibinfo {title} {Quantization of inductively shunted superconducting circuits},\ }\href {https://doi.org/10.1103/PhysRevB.94.144507} {\bibfield  {journal} {\bibinfo  {journal} {Physical Review B}\ }\textbf {\bibinfo {volume} {94}},\ \bibinfo {pages} {144507} (\bibinfo {year} {2016})}\BibitemShut {NoStop}%
\bibitem [{\citenamefont {Schuster}\ \emph {et~al.}(2007)\citenamefont {Schuster}, \citenamefont {Houck}, \citenamefont {Schreier}, \citenamefont {Wallraff}, \citenamefont {Gambetta}, \citenamefont {Blais}, \citenamefont {Frunzio}, \citenamefont {Majer}, \citenamefont {Johnson}, \citenamefont {Devoret}, \citenamefont {Girvin},\ and\ \citenamefont {Schoelkopf}}]{schuster_resolving_2007}%
  \BibitemOpen
  \bibfield  {author} {\bibinfo {author} {\bibfnamefont {D.~I.}\ \bibnamefont {Schuster}}, \bibinfo {author} {\bibfnamefont {A.~A.}\ \bibnamefont {Houck}}, \bibinfo {author} {\bibfnamefont {J.~A.}\ \bibnamefont {Schreier}}, \bibinfo {author} {\bibfnamefont {A.}~\bibnamefont {Wallraff}}, \bibinfo {author} {\bibfnamefont {J.~M.}\ \bibnamefont {Gambetta}}, \bibinfo {author} {\bibfnamefont {A.}~\bibnamefont {Blais}}, \bibinfo {author} {\bibfnamefont {L.}~\bibnamefont {Frunzio}}, \bibinfo {author} {\bibfnamefont {J.}~\bibnamefont {Majer}}, \bibinfo {author} {\bibfnamefont {B.}~\bibnamefont {Johnson}}, \bibinfo {author} {\bibfnamefont {M.~H.}\ \bibnamefont {Devoret}}, \bibinfo {author} {\bibfnamefont {S.~M.}\ \bibnamefont {Girvin}},\ and\ \bibinfo {author} {\bibfnamefont {R.~J.}\ \bibnamefont {Schoelkopf}},\ }\bibfield  {title} {\bibinfo {title} {Resolving photon number states in a superconducting circuit},\ }\href {https://doi.org/10.1038/nature05461} {\bibfield  {journal} {\bibinfo  {journal} {Nature}\ }\textbf {\bibinfo {volume} {445}},\ \bibinfo {pages} {515} (\bibinfo {year} {2007})}\BibitemShut {NoStop}%
\bibitem [{\citenamefont {Heeres}\ \emph {et~al.}(2015)\citenamefont {Heeres}, \citenamefont {Vlastakis}, \citenamefont {Holland}, \citenamefont {Krastanov}, \citenamefont {Albert}, \citenamefont {Frunzio}, \citenamefont {Jiang},\ and\ \citenamefont {Schoelkopf}}]{heeres_cavity_2015}%
  \BibitemOpen
  \bibfield  {author} {\bibinfo {author} {\bibfnamefont {R.~W.}\ \bibnamefont {Heeres}}, \bibinfo {author} {\bibfnamefont {B.}~\bibnamefont {Vlastakis}}, \bibinfo {author} {\bibfnamefont {E.}~\bibnamefont {Holland}}, \bibinfo {author} {\bibfnamefont {S.}~\bibnamefont {Krastanov}}, \bibinfo {author} {\bibfnamefont {V.~V.}\ \bibnamefont {Albert}}, \bibinfo {author} {\bibfnamefont {L.}~\bibnamefont {Frunzio}}, \bibinfo {author} {\bibfnamefont {L.}~\bibnamefont {Jiang}},\ and\ \bibinfo {author} {\bibfnamefont {R.~J.}\ \bibnamefont {Schoelkopf}},\ }\bibfield  {title} {\bibinfo {title} {Cavity state manipulation using photon-number selective phase gates},\ }\href {https://doi.org/10.1103/PhysRevLett.115.137002} {\bibfield  {journal} {\bibinfo  {journal} {Physical Review Letters}\ }\textbf {\bibinfo {volume} {115}},\ \bibinfo {pages} {137002} (\bibinfo {year} {2015})}\BibitemShut {NoStop}%
\bibitem [{\citenamefont {Landgraf}\ \emph {et~al.}(2024)\citenamefont {Landgraf}, \citenamefont {Fl{\"u}hmann}, \citenamefont {F{\"o}sel}, \citenamefont {Marquardt},\ and\ \citenamefont {Schoelkopf}}]{landgraf_fast_2024}%
  \BibitemOpen
  \bibfield  {author} {\bibinfo {author} {\bibfnamefont {J.}~\bibnamefont {Landgraf}}, \bibinfo {author} {\bibfnamefont {C.}~\bibnamefont {Fl{\"u}hmann}}, \bibinfo {author} {\bibfnamefont {T.}~\bibnamefont {F{\"o}sel}}, \bibinfo {author} {\bibfnamefont {F.}~\bibnamefont {Marquardt}},\ and\ \bibinfo {author} {\bibfnamefont {R.~J.}\ \bibnamefont {Schoelkopf}},\ }\bibfield  {title} {\bibinfo {title} {Fast quantum control of cavities using an improved protocol without coherent errors},\ }\href {https://doi.org/10.1103/PhysRevLett.133.260802} {\bibfield  {journal} {\bibinfo  {journal} {Physical Review Letters}\ }\textbf {\bibinfo {volume} {133}},\ \bibinfo {pages} {260802} (\bibinfo {year} {2024})}\BibitemShut {NoStop}%
\bibitem [{\citenamefont {Zheng}\ \emph {et~al.}(2025)\citenamefont {Zheng}, \citenamefont {Lieu}, \citenamefont {Rosenfeld}, \citenamefont {Noh},\ and\ \citenamefont {Hann}}]{zheng_crossresonance_2025}%
  \BibitemOpen
  \bibfield  {author} {\bibinfo {author} {\bibfnamefont {G.}~\bibnamefont {Zheng}}, \bibinfo {author} {\bibfnamefont {S.}~\bibnamefont {Lieu}}, \bibinfo {author} {\bibfnamefont {E.~L.}\ \bibnamefont {Rosenfeld}}, \bibinfo {author} {\bibfnamefont {K.}~\bibnamefont {Noh}},\ and\ \bibinfo {author} {\bibfnamefont {C.~T.}\ \bibnamefont {Hann}},\ }\bibfield  {title} {\bibinfo {title} {Cross-resonance control of an oscillator with an auxiliary fluxonium qubit},\ }\href {https://doi.org/10.1103/PhysRevApplied.23.024067} {\bibfield  {journal} {\bibinfo  {journal} {Physical Review Applied}\ }\textbf {\bibinfo {volume} {23}},\ \bibinfo {pages} {024067} (\bibinfo {year} {2025})}\BibitemShut {NoStop}%
\bibitem [{\citenamefont {Pfaff}\ \emph {et~al.}(2017)\citenamefont {Pfaff}, \citenamefont {Axline}, \citenamefont {Burkhart}, \citenamefont {Vool}, \citenamefont {Reinhold}, \citenamefont {Frunzio}, \citenamefont {Jiang}, \citenamefont {Devoret},\ and\ \citenamefont {Schoelkopf}}]{Pfaff2017}%
  \BibitemOpen
  \bibfield  {author} {\bibinfo {author} {\bibfnamefont {W.}~\bibnamefont {Pfaff}}, \bibinfo {author} {\bibfnamefont {C.~J.}\ \bibnamefont {Axline}}, \bibinfo {author} {\bibfnamefont {L.~D.}\ \bibnamefont {Burkhart}}, \bibinfo {author} {\bibfnamefont {U.}~\bibnamefont {Vool}}, \bibinfo {author} {\bibfnamefont {P.}~\bibnamefont {Reinhold}}, \bibinfo {author} {\bibfnamefont {L.}~\bibnamefont {Frunzio}}, \bibinfo {author} {\bibfnamefont {L.}~\bibnamefont {Jiang}}, \bibinfo {author} {\bibfnamefont {M.~H.}\ \bibnamefont {Devoret}},\ and\ \bibinfo {author} {\bibfnamefont {R.~J.}\ \bibnamefont {Schoelkopf}},\ }\bibfield  {title} {\bibinfo {title} {Controlled release of multiphoton quantum states from a microwave cavity memory},\ }\href {https://doi.org/10.1038/nphys4143} {\bibfield  {journal} {\bibinfo  {journal} {Nature Physics}\ }\textbf {\bibinfo {volume} {13}},\ \bibinfo {pages} {882} (\bibinfo {year} {2017})}\BibitemShut {NoStop}%
\bibitem [{\citenamefont {Axline}\ \emph {et~al.}(2018)\citenamefont {Axline}, \citenamefont {Burkhart}, \citenamefont {Pfaff}, \citenamefont {Zhang}, \citenamefont {Chou}, \citenamefont {{Campagne-Ibarcq}}, \citenamefont {Reinhold}, \citenamefont {Frunzio}, \citenamefont {Girvin}, \citenamefont {Jiang}, \citenamefont {Devoret},\ and\ \citenamefont {Schoelkopf}}]{Axline2018}%
  \BibitemOpen
  \bibfield  {author} {\bibinfo {author} {\bibfnamefont {C.~J.}\ \bibnamefont {Axline}}, \bibinfo {author} {\bibfnamefont {L.~D.}\ \bibnamefont {Burkhart}}, \bibinfo {author} {\bibfnamefont {W.}~\bibnamefont {Pfaff}}, \bibinfo {author} {\bibfnamefont {M.}~\bibnamefont {Zhang}}, \bibinfo {author} {\bibfnamefont {K.}~\bibnamefont {Chou}}, \bibinfo {author} {\bibfnamefont {P.}~\bibnamefont {{Campagne-Ibarcq}}}, \bibinfo {author} {\bibfnamefont {P.}~\bibnamefont {Reinhold}}, \bibinfo {author} {\bibfnamefont {L.}~\bibnamefont {Frunzio}}, \bibinfo {author} {\bibfnamefont {S.~M.}\ \bibnamefont {Girvin}}, \bibinfo {author} {\bibfnamefont {L.}~\bibnamefont {Jiang}}, \bibinfo {author} {\bibfnamefont {M.~H.}\ \bibnamefont {Devoret}},\ and\ \bibinfo {author} {\bibfnamefont {R.~J.}\ \bibnamefont {Schoelkopf}},\ }\bibfield  {title} {\bibinfo {title} {On-demand quantum state transfer and entanglement between remote microwave cavity memories},\ }\href {https://doi.org/10.1038/s41567-018-0115-y} {\bibfield  {journal} {\bibinfo  {journal} {Nature Physics}\ }\textbf {\bibinfo {volume} {14}},\ \bibinfo {pages} {705} (\bibinfo {year} {2018})}\BibitemShut {NoStop}%
\bibitem [{\citenamefont {{Campagne-Ibarcq}}\ \emph {et~al.}(2020)\citenamefont {{Campagne-Ibarcq}}, \citenamefont {Eickbusch}, \citenamefont {Touzard}, \citenamefont {{Zalys-Geller}}, \citenamefont {Frattini}, \citenamefont {Sivak}, \citenamefont {Reinhold}, \citenamefont {Puri}, \citenamefont {Shankar}, \citenamefont {Schoelkopf}, \citenamefont {Frunzio}, \citenamefont {Mirrahimi},\ and\ \citenamefont {Devoret}}]{CampagneIbarcq2020}%
  \BibitemOpen
  \bibfield  {author} {\bibinfo {author} {\bibfnamefont {P.}~\bibnamefont {{Campagne-Ibarcq}}}, \bibinfo {author} {\bibfnamefont {A.}~\bibnamefont {Eickbusch}}, \bibinfo {author} {\bibfnamefont {S.}~\bibnamefont {Touzard}}, \bibinfo {author} {\bibfnamefont {E.}~\bibnamefont {{Zalys-Geller}}}, \bibinfo {author} {\bibfnamefont {N.~E.}\ \bibnamefont {Frattini}}, \bibinfo {author} {\bibfnamefont {V.~V.}\ \bibnamefont {Sivak}}, \bibinfo {author} {\bibfnamefont {P.}~\bibnamefont {Reinhold}}, \bibinfo {author} {\bibfnamefont {S.}~\bibnamefont {Puri}}, \bibinfo {author} {\bibfnamefont {S.}~\bibnamefont {Shankar}}, \bibinfo {author} {\bibfnamefont {R.~J.}\ \bibnamefont {Schoelkopf}}, \bibinfo {author} {\bibfnamefont {L.}~\bibnamefont {Frunzio}}, \bibinfo {author} {\bibfnamefont {M.}~\bibnamefont {Mirrahimi}},\ and\ \bibinfo {author} {\bibfnamefont {M.~H.}\ \bibnamefont {Devoret}},\ }\bibfield  {title} {\bibinfo {title} {Quantum error correction of a qubit encoded in grid states of an oscillator},\ }\href {https://doi.org/10.1038/s41586-020-2603-3} {\bibfield  {journal} {\bibinfo  {journal} {Nature}\ }\textbf {\bibinfo {volume} {584}},\ \bibinfo {pages} {368} (\bibinfo {year} {2020})}\BibitemShut {NoStop}%
\bibitem [{\citenamefont {Valadares}\ \emph {et~al.}(2024)\citenamefont {Valadares}, \citenamefont {Huang}, \citenamefont {Chu}, \citenamefont {Dorogov}, \citenamefont {Chua}, \citenamefont {Kong}, \citenamefont {Song},\ and\ \citenamefont {Gao}}]{Valadares2024}%
  \BibitemOpen
  \bibfield  {author} {\bibinfo {author} {\bibfnamefont {F.}~\bibnamefont {Valadares}}, \bibinfo {author} {\bibfnamefont {N.-N.}\ \bibnamefont {Huang}}, \bibinfo {author} {\bibfnamefont {K.~T.~N.}\ \bibnamefont {Chu}}, \bibinfo {author} {\bibfnamefont {A.}~\bibnamefont {Dorogov}}, \bibinfo {author} {\bibfnamefont {W.}~\bibnamefont {Chua}}, \bibinfo {author} {\bibfnamefont {L.}~\bibnamefont {Kong}}, \bibinfo {author} {\bibfnamefont {P.}~\bibnamefont {Song}},\ and\ \bibinfo {author} {\bibfnamefont {Y.~Y.}\ \bibnamefont {Gao}},\ }\bibfield  {title} {\bibinfo {title} {On-demand transposition across light-matter interaction regimes in bosonic {{cQED}}},\ }\href {https://doi.org/10.1038/s41467-024-50201-7} {\bibfield  {journal} {\bibinfo  {journal} {Nature Communications}\ }\textbf {\bibinfo {volume} {15}},\ \bibinfo {pages} {5816} (\bibinfo {year} {2024})}\BibitemShut {NoStop}%
\bibitem [{\citenamefont {Cai}\ \emph {et~al.}(2024)\citenamefont {Cai}, \citenamefont {Mu}, \citenamefont {Wang}, \citenamefont {Zhou}, \citenamefont {Ma}, \citenamefont {Pan}, \citenamefont {Hua}, \citenamefont {Liu}, \citenamefont {Xue}, \citenamefont {Yu}, \citenamefont {Wang}, \citenamefont {Song}, \citenamefont {Zou},\ and\ \citenamefont {Sun}}]{Cai2024}%
  \BibitemOpen
  \bibfield  {author} {\bibinfo {author} {\bibfnamefont {W.}~\bibnamefont {Cai}}, \bibinfo {author} {\bibfnamefont {X.}~\bibnamefont {Mu}}, \bibinfo {author} {\bibfnamefont {W.}~\bibnamefont {Wang}}, \bibinfo {author} {\bibfnamefont {J.}~\bibnamefont {Zhou}}, \bibinfo {author} {\bibfnamefont {Y.}~\bibnamefont {Ma}}, \bibinfo {author} {\bibfnamefont {X.}~\bibnamefont {Pan}}, \bibinfo {author} {\bibfnamefont {Z.}~\bibnamefont {Hua}}, \bibinfo {author} {\bibfnamefont {X.}~\bibnamefont {Liu}}, \bibinfo {author} {\bibfnamefont {G.}~\bibnamefont {Xue}}, \bibinfo {author} {\bibfnamefont {H.}~\bibnamefont {Yu}}, \bibinfo {author} {\bibfnamefont {H.}~\bibnamefont {Wang}}, \bibinfo {author} {\bibfnamefont {Y.}~\bibnamefont {Song}}, \bibinfo {author} {\bibfnamefont {C.-L.}\ \bibnamefont {Zou}},\ and\ \bibinfo {author} {\bibfnamefont {L.}~\bibnamefont {Sun}},\ }\bibfield  {title} {\bibinfo {title} {Protecting entanglement between logical qubits via quantum error correction},\ }\href {https://doi.org/10.1038/s41567-024-02446-8} {\bibfield  {journal} {\bibinfo  {journal} {Nature Physics}\ }\textbf {\bibinfo {volume} {20}},\ \bibinfo {pages} {1022} (\bibinfo {year} {2024})}\BibitemShut {NoStop}%
\bibitem [{\citenamefont {Yang}\ \emph {et~al.}(2025)\citenamefont {Yang}, \citenamefont {Agrenius}, \citenamefont {Usova}, \citenamefont {{Romero-Isart}},\ and\ \citenamefont {Kirchmair}}]{Yang2025}%
  \BibitemOpen
  \bibfield  {author} {\bibinfo {author} {\bibfnamefont {I.}~\bibnamefont {Yang}}, \bibinfo {author} {\bibfnamefont {T.}~\bibnamefont {Agrenius}}, \bibinfo {author} {\bibfnamefont {V.}~\bibnamefont {Usova}}, \bibinfo {author} {\bibfnamefont {O.}~\bibnamefont {{Romero-Isart}}},\ and\ \bibinfo {author} {\bibfnamefont {G.}~\bibnamefont {Kirchmair}},\ }\bibfield  {title} {\bibinfo {title} {Hot {{Schr{\"o}dinger}} cat states},\ }\href {https://doi.org/10.1126/sciadv.adr4492} {\bibfield  {journal} {\bibinfo  {journal} {Science Advances}\ }\textbf {\bibinfo {volume} {11}},\ \bibinfo {pages} {eadr4492} (\bibinfo {year} {2025})}\BibitemShut {NoStop}%
\bibitem [{\citenamefont {Lotkhov}\ \emph {et~al.}(2011)\citenamefont {Lotkhov}, \citenamefont {Bogoslovsky}, \citenamefont {Zorin},\ and\ \citenamefont {Niemeyer}}]{Lotkhov2011BridgeFree}%
  \BibitemOpen
  \bibfield  {author} {\bibinfo {author} {\bibfnamefont {S.~V.}\ \bibnamefont {Lotkhov}}, \bibinfo {author} {\bibfnamefont {S.~A.}\ \bibnamefont {Bogoslovsky}}, \bibinfo {author} {\bibfnamefont {A.~B.}\ \bibnamefont {Zorin}},\ and\ \bibinfo {author} {\bibfnamefont {J.}~\bibnamefont {Niemeyer}},\ }\bibfield  {title} {\bibinfo {title} {Bridge-free technique for fabrication of submicron {{Al}}/{{AlOx}}/{{Al}} tunnel junctions for superconducting quantum circuits},\ }\href {https://doi.org/10.1088/0957-4484/22/31/315302} {\bibfield  {journal} {\bibinfo  {journal} {Nanotechnology}\ }\textbf {\bibinfo {volume} {22}},\ \bibinfo {pages} {315302} (\bibinfo {year} {2011})}\BibitemShut {NoStop}%
\bibitem [{\citenamefont {Vool}\ \emph {et~al.}(2018)\citenamefont {Vool}, \citenamefont {Kou}, \citenamefont {Smith}, \citenamefont {Frattini}, \citenamefont {Serniak}, \citenamefont {Reinhold}, \citenamefont {Pop}, \citenamefont {Shankar}, \citenamefont {Frunzio}, \citenamefont {Girvin},\ and\ \citenamefont {Devoret}}]{vool2018driving}%
  \BibitemOpen
  \bibfield  {author} {\bibinfo {author} {\bibfnamefont {U.}~\bibnamefont {Vool}}, \bibinfo {author} {\bibfnamefont {A.}~\bibnamefont {Kou}}, \bibinfo {author} {\bibfnamefont {W.~C.}\ \bibnamefont {Smith}}, \bibinfo {author} {\bibfnamefont {N.~E.}\ \bibnamefont {Frattini}}, \bibinfo {author} {\bibfnamefont {K.}~\bibnamefont {Serniak}}, \bibinfo {author} {\bibfnamefont {P.}~\bibnamefont {Reinhold}}, \bibinfo {author} {\bibfnamefont {I.~M.}\ \bibnamefont {Pop}}, \bibinfo {author} {\bibfnamefont {S.}~\bibnamefont {Shankar}}, \bibinfo {author} {\bibfnamefont {L.}~\bibnamefont {Frunzio}}, \bibinfo {author} {\bibfnamefont {S.~M.}\ \bibnamefont {Girvin}},\ and\ \bibinfo {author} {\bibfnamefont {M.~H.}\ \bibnamefont {Devoret}},\ }\bibfield  {title} {\bibinfo {title} {Driving forbidden transitions in the fluxonium artificial atom},\ }\href {https://doi.org/10.1103/PhysRevApplied.9.054046} {\bibfield  {journal} {\bibinfo  {journal} {Physical Review Applied}\ }\textbf {\bibinfo {volume} {9}},\ \bibinfo {pages} {054046} (\bibinfo {year} {2018})}\BibitemShut {NoStop}%
\bibitem [{\citenamefont {Zhang}\ \emph {et~al.}(2021)\citenamefont {Zhang}, \citenamefont {Chakram}, \citenamefont {Roy}, \citenamefont {Earnest}, \citenamefont {Lu}, \citenamefont {Huang}, \citenamefont {Weiss}, \citenamefont {Koch},\ and\ \citenamefont {Schuster}}]{Zhang2021universal}%
  \BibitemOpen
  \bibfield  {author} {\bibinfo {author} {\bibfnamefont {H.}~\bibnamefont {Zhang}}, \bibinfo {author} {\bibfnamefont {S.}~\bibnamefont {Chakram}}, \bibinfo {author} {\bibfnamefont {T.}~\bibnamefont {Roy}}, \bibinfo {author} {\bibfnamefont {N.}~\bibnamefont {Earnest}}, \bibinfo {author} {\bibfnamefont {Y.}~\bibnamefont {Lu}}, \bibinfo {author} {\bibfnamefont {Z.}~\bibnamefont {Huang}}, \bibinfo {author} {\bibfnamefont {D.~K.}\ \bibnamefont {Weiss}}, \bibinfo {author} {\bibfnamefont {J.}~\bibnamefont {Koch}},\ and\ \bibinfo {author} {\bibfnamefont {D.~I.}\ \bibnamefont {Schuster}},\ }\bibfield  {title} {\bibinfo {title} {Universal fast-flux control of a coherent, low-frequency qubit},\ }\href {https://doi.org/10.1103/PhysRevX.11.011010} {\bibfield  {journal} {\bibinfo  {journal} {Physical Review X}\ }\textbf {\bibinfo {volume} {11}},\ \bibinfo {pages} {011010} (\bibinfo {year} {2021})}\BibitemShut {NoStop}%
\bibitem [{\citenamefont {Wang}\ \emph {et~al.}(2024)\citenamefont {Wang}, \citenamefont {Wu}, \citenamefont {Wang}, \citenamefont {Ma}, \citenamefont {Zhang}, \citenamefont {Chen}, \citenamefont {Deng}, \citenamefont {Gao}, \citenamefont {Hu}, \citenamefont {Ma}, \citenamefont {Song}, \citenamefont {Xia}, \citenamefont {Ying}, \citenamefont {Zhan}, \citenamefont {Zhao},\ and\ \citenamefont {Deng}}]{Wang2024efficient}%
  \BibitemOpen
  \bibfield  {author} {\bibinfo {author} {\bibfnamefont {T.}~\bibnamefont {Wang}}, \bibinfo {author} {\bibfnamefont {F.}~\bibnamefont {Wu}}, \bibinfo {author} {\bibfnamefont {F.}~\bibnamefont {Wang}}, \bibinfo {author} {\bibfnamefont {X.}~\bibnamefont {Ma}}, \bibinfo {author} {\bibfnamefont {G.}~\bibnamefont {Zhang}}, \bibinfo {author} {\bibfnamefont {J.}~\bibnamefont {Chen}}, \bibinfo {author} {\bibfnamefont {H.}~\bibnamefont {Deng}}, \bibinfo {author} {\bibfnamefont {R.}~\bibnamefont {Gao}}, \bibinfo {author} {\bibfnamefont {R.}~\bibnamefont {Hu}}, \bibinfo {author} {\bibfnamefont {L.}~\bibnamefont {Ma}}, \bibinfo {author} {\bibfnamefont {Z.}~\bibnamefont {Song}}, \bibinfo {author} {\bibfnamefont {T.}~\bibnamefont {Xia}}, \bibinfo {author} {\bibfnamefont {M.}~\bibnamefont {Ying}}, \bibinfo {author} {\bibfnamefont {H.}~\bibnamefont {Zhan}}, \bibinfo {author} {\bibfnamefont {H.-H.}\ \bibnamefont {Zhao}},\ and\ \bibinfo {author} {\bibfnamefont {C.}~\bibnamefont {Deng}},\ }\bibfield  {title} {\bibinfo {title} {Efficient initialization of fluxonium qubits based on auxiliary energy levels},\ }\href {https://doi.org/10.1103/PhysRevLett.132.230601} {\bibfield  {journal} {\bibinfo  {journal} {Physical Review Letters}\ }\textbf {\bibinfo {volume} {132}},\ \bibinfo {pages} {230601} (\bibinfo {year} {2024})}\BibitemShut {NoStop}%
\bibitem [{\citenamefont {Nie}\ \emph {et~al.}(2024)\citenamefont {Nie}, \citenamefont {Bista}, \citenamefont {Chow}, \citenamefont {Pfaff},\ and\ \citenamefont {Kou}}]{nie_parametrically_2024}%
  \BibitemOpen
  \bibfield  {author} {\bibinfo {author} {\bibfnamefont {K.}~\bibnamefont {Nie}}, \bibinfo {author} {\bibfnamefont {A.}~\bibnamefont {Bista}}, \bibinfo {author} {\bibfnamefont {K.}~\bibnamefont {Chow}}, \bibinfo {author} {\bibfnamefont {W.}~\bibnamefont {Pfaff}},\ and\ \bibinfo {author} {\bibfnamefont {A.}~\bibnamefont {Kou}},\ }\bibfield  {title} {\bibinfo {title} {Parametrically controlled microwave-photonic interface for the fluxonium},\ }\href {https://doi.org/10.1103/PhysRevApplied.22.054021} {\bibfield  {journal} {\bibinfo  {journal} {Physical Review Applied}\ }\textbf {\bibinfo {volume} {22}},\ \bibinfo {pages} {054021} (\bibinfo {year} {2024})}\BibitemShut {NoStop}%
\bibitem [{\citenamefont {Groszkowski}\ and\ \citenamefont {Koch}(2021)}]{scqubits_1}%
  \BibitemOpen
  \bibfield  {author} {\bibinfo {author} {\bibfnamefont {P.}~\bibnamefont {Groszkowski}}\ and\ \bibinfo {author} {\bibfnamefont {J.}~\bibnamefont {Koch}},\ }\bibfield  {title} {\bibinfo {title} {Scqubits: A {{Python}} package for superconducting qubits},\ }\href {https://doi.org/10.22331/q-2021-11-17-583} {\bibfield  {journal} {\bibinfo  {journal} {Quantum}\ }\textbf {\bibinfo {volume} {5}},\ \bibinfo {pages} {583} (\bibinfo {year} {2021})}\BibitemShut {NoStop}%
\bibitem [{\citenamefont {Chitta}\ \emph {et~al.}(2022)\citenamefont {Chitta}, \citenamefont {Zhao}, \citenamefont {Huang}, \citenamefont {{Mondragon-Shem}},\ and\ \citenamefont {Koch}}]{scqubits_2}%
  \BibitemOpen
  \bibfield  {author} {\bibinfo {author} {\bibfnamefont {S.~P.}\ \bibnamefont {Chitta}}, \bibinfo {author} {\bibfnamefont {T.}~\bibnamefont {Zhao}}, \bibinfo {author} {\bibfnamefont {Z.}~\bibnamefont {Huang}}, \bibinfo {author} {\bibfnamefont {I.}~\bibnamefont {{Mondragon-Shem}}},\ and\ \bibinfo {author} {\bibfnamefont {J.}~\bibnamefont {Koch}},\ }\bibfield  {title} {\bibinfo {title} {Computer-aided quantization and numerical analysis of superconducting circuits},\ }\href {https://doi.org/10.1088/1367-2630/ac94f2} {\bibfield  {journal} {\bibinfo  {journal} {New Journal of Physics}\ }\textbf {\bibinfo {volume} {24}},\ \bibinfo {pages} {103020} (\bibinfo {year} {2022})}\BibitemShut {NoStop}%
\bibitem [{\citenamefont {Binder}\ \emph {et~al.}(2023)\citenamefont {Binder}, \citenamefont {Ge}, \citenamefont {He}, \citenamefont {Lyu}, \citenamefont {Rossi}, \citenamefont {Coto}, \citenamefont {Minganti}, \citenamefont {Gerardo}, \citenamefont {Ball}, \citenamefont {Zhang}, \citenamefont {{Garc{\'i}a-P{\'e}rez}}, \citenamefont {Pirandola}, \citenamefont {Paternostro}, \citenamefont {Fux}, \citenamefont {Leghtas}, \citenamefont {Clerk}, \citenamefont {Le~Boit{\'e}}, \citenamefont {Fazio}, \citenamefont {Murch}, \citenamefont {{Hacohen-Gourgy}}, \citenamefont {Combes}, \citenamefont {Fiderer}, \citenamefont {Brandner}, \citenamefont {Chenu}, \citenamefont {Chiacchio}, \citenamefont {Xu}, \citenamefont {Scopa}, \citenamefont {Snizhko}, \citenamefont {Arenz}, \citenamefont {Cappellaro}, \citenamefont {Lambert}, \citenamefont {Nori},\ and\ \citenamefont {Nation}}]{qutip3}%
  \BibitemOpen
  \bibfield  {author} {\bibinfo {author} {\bibfnamefont {F.~C.}\ \bibnamefont {Binder}}, \bibinfo {author} {\bibfnamefont {W.}~\bibnamefont {Ge}}, \bibinfo {author} {\bibfnamefont {J.}~\bibnamefont {He}}, \bibinfo {author} {\bibfnamefont {C.}~\bibnamefont {Lyu}}, \bibinfo {author} {\bibfnamefont {M.~A.~C.}\ \bibnamefont {Rossi}}, \bibinfo {author} {\bibfnamefont {R.~S.}\ \bibnamefont {Coto}}, \bibinfo {author} {\bibfnamefont {F.}~\bibnamefont {Minganti}}, \bibinfo {author} {\bibfnamefont {O.}~\bibnamefont {Gerardo}}, \bibinfo {author} {\bibfnamefont {T.}~\bibnamefont {Ball}}, \bibinfo {author} {\bibfnamefont {X.}~\bibnamefont {Zhang}}, \bibinfo {author} {\bibfnamefont {G.}~\bibnamefont {{Garc{\'i}a-P{\'e}rez}}}, \bibinfo {author} {\bibfnamefont {S.}~\bibnamefont {Pirandola}}, \bibinfo {author} {\bibfnamefont {M.}~\bibnamefont {Paternostro}}, \bibinfo {author} {\bibfnamefont {S.}~\bibnamefont {Fux}}, \bibinfo {author} {\bibfnamefont {Z.}~\bibnamefont {Leghtas}}, \bibinfo {author} {\bibfnamefont {A.}~\bibnamefont {Clerk}}, \bibinfo {author} {\bibfnamefont {A.}~\bibnamefont {Le~Boit{\'e}}}, \bibinfo {author} {\bibfnamefont {R.}~\bibnamefont {Fazio}}, \bibinfo {author} {\bibfnamefont {K.}~\bibnamefont {Murch}}, \bibinfo {author} {\bibfnamefont {S.}~\bibnamefont {{Hacohen-Gourgy}}}, \bibinfo {author} {\bibfnamefont {J.}~\bibnamefont {Combes}}, \bibinfo {author} {\bibfnamefont {L.~J.}\ \bibnamefont {Fiderer}}, \bibinfo {author} {\bibfnamefont {K.}~\bibnamefont {Brandner}}, \bibinfo {author} {\bibfnamefont {A.}~\bibnamefont {Chenu}}, \bibinfo {author} {\bibfnamefont {E.~I.}\ \bibnamefont {Chiacchio}}, \bibinfo {author} {\bibfnamefont {X.}~\bibnamefont {Xu}}, \bibinfo {author} {\bibfnamefont {S.}~\bibnamefont {Scopa}}, \bibinfo {author} {\bibfnamefont {K.}~\bibnamefont {Snizhko}}, \bibinfo {author} {\bibfnamefont {C.}~\bibnamefont {Arenz}}, \bibinfo {author} {\bibfnamefont {P.}~\bibnamefont {Cappellaro}}, \bibinfo {author} {\bibfnamefont {N.}~\bibnamefont {Lambert}}, \bibinfo {author} {\bibfnamefont {F.}~\bibnamefont {Nori}},\ and\ \bibinfo {author} {\bibfnamefont {P.~D.}\ \bibnamefont {Nation}},\ }\bibfield  {title} {\bibinfo {title} {{{QuTiP}}: {{An}} open-source {{Python}} framework for the dynamics of open quantum systems},\ }\href {https://doi.org/10.22331/q-2023-01-11-933} {\bibfield  {journal} {\bibinfo  {journal} {Quantum}\ }\textbf {\bibinfo {volume} {7}},\ \bibinfo {pages} {933} (\bibinfo {year} {2023})}\BibitemShut {NoStop}%
\bibitem [{\citenamefont {Blais}\ \emph {et~al.}(2021)\citenamefont {Blais}, \citenamefont {Grimsmo}, \citenamefont {Girvin},\ and\ \citenamefont {Wallraff}}]{blais_qed}%
  \BibitemOpen
  \bibfield  {author} {\bibinfo {author} {\bibfnamefont {A.}~\bibnamefont {Blais}}, \bibinfo {author} {\bibfnamefont {A.~L.}\ \bibnamefont {Grimsmo}}, \bibinfo {author} {\bibfnamefont {S.~M.}\ \bibnamefont {Girvin}},\ and\ \bibinfo {author} {\bibfnamefont {A.}~\bibnamefont {Wallraff}},\ }\bibfield  {title} {\bibinfo {title} {Circuit quantum electrodynamics},\ }\href {https://doi.org/10.1103/revmodphys.93.025005} {\bibfield  {journal} {\bibinfo  {journal} {Reviews of Modern Physics}\ }\textbf {\bibinfo {volume} {93}},\ \bibinfo {pages} {025005} (\bibinfo {year} {2021})}\BibitemShut {NoStop}%
\bibitem [{\citenamefont {Koch}\ \emph {et~al.}(2007)\citenamefont {Koch}, \citenamefont {Yu}, \citenamefont {Gambetta}, \citenamefont {Houck}, \citenamefont {Schuster}, \citenamefont {Majer}, \citenamefont {Blais}, \citenamefont {Devoret}, \citenamefont {Girvin},\ and\ \citenamefont {Schoelkopf}}]{koch_transmon_regime}%
  \BibitemOpen
  \bibfield  {author} {\bibinfo {author} {\bibfnamefont {J.}~\bibnamefont {Koch}}, \bibinfo {author} {\bibfnamefont {T.~M.}\ \bibnamefont {Yu}}, \bibinfo {author} {\bibfnamefont {J.}~\bibnamefont {Gambetta}}, \bibinfo {author} {\bibfnamefont {A.~A.}\ \bibnamefont {Houck}}, \bibinfo {author} {\bibfnamefont {D.~I.}\ \bibnamefont {Schuster}}, \bibinfo {author} {\bibfnamefont {J.}~\bibnamefont {Majer}}, \bibinfo {author} {\bibfnamefont {A.}~\bibnamefont {Blais}}, \bibinfo {author} {\bibfnamefont {M.~H.}\ \bibnamefont {Devoret}}, \bibinfo {author} {\bibfnamefont {S.~M.}\ \bibnamefont {Girvin}},\ and\ \bibinfo {author} {\bibfnamefont {R.~J.}\ \bibnamefont {Schoelkopf}},\ }\bibfield  {title} {\bibinfo {title} {Charge-insensitive qubit design derived from the {{Cooper}} pair box},\ }\href {https://doi.org/10.1103/physreva.76.042319} {\bibfield  {journal} {\bibinfo  {journal} {Physical Review A}\ }\textbf {\bibinfo {volume} {76}},\ \bibinfo {pages} {042319} (\bibinfo {year} {2007})}\BibitemShut {NoStop}%
\bibitem [{\citenamefont {Schreier}\ \emph {et~al.}(2008)\citenamefont {Schreier}, \citenamefont {Houck}, \citenamefont {Koch}, \citenamefont {Schuster}, \citenamefont {Johnson}, \citenamefont {Chow}, \citenamefont {Gambetta}, \citenamefont {Majer}, \citenamefont {Frunzio}, \citenamefont {Devoret}, \citenamefont {Girvin},\ and\ \citenamefont {Schoelkopf}}]{schreier_transmon_regime}%
  \BibitemOpen
  \bibfield  {author} {\bibinfo {author} {\bibfnamefont {J.~A.}\ \bibnamefont {Schreier}}, \bibinfo {author} {\bibfnamefont {A.~A.}\ \bibnamefont {Houck}}, \bibinfo {author} {\bibfnamefont {J.}~\bibnamefont {Koch}}, \bibinfo {author} {\bibfnamefont {D.~I.}\ \bibnamefont {Schuster}}, \bibinfo {author} {\bibfnamefont {B.~R.}\ \bibnamefont {Johnson}}, \bibinfo {author} {\bibfnamefont {J.~M.}\ \bibnamefont {Chow}}, \bibinfo {author} {\bibfnamefont {J.~M.}\ \bibnamefont {Gambetta}}, \bibinfo {author} {\bibfnamefont {J.}~\bibnamefont {Majer}}, \bibinfo {author} {\bibfnamefont {L.}~\bibnamefont {Frunzio}}, \bibinfo {author} {\bibfnamefont {M.~H.}\ \bibnamefont {Devoret}}, \bibinfo {author} {\bibfnamefont {S.~M.}\ \bibnamefont {Girvin}},\ and\ \bibinfo {author} {\bibfnamefont {R.~J.}\ \bibnamefont {Schoelkopf}},\ }\bibfield  {title} {\bibinfo {title} {Suppressing charge noise decoherence in superconducting charge qubits},\ }\href {https://doi.org/10.1103/physrevb.77.180502} {\bibfield  {journal} {\bibinfo  {journal} {Physical Review B}\ }\textbf {\bibinfo {volume} {77}},\ \bibinfo {pages} {180502(R)} (\bibinfo {year} {2008})}\BibitemShut {NoStop}%
\end{thebibliography}
\end{document}